\documentclass[11pt]{article}
\pdfoutput=1
\usepackage{jcapmod}

\usepackage{booktabs}
\usepackage[english]{babel}
\usepackage{amsmath,amssymb,amsbsy,amstext, amsthm, simplewick, amsfonts}
\usepackage{hyperref}
\usepackage{graphicx}
\usepackage[small]{caption}
\usepackage{siunitx}
\usepackage{upgreek}
\usepackage{framed}
\usepackage{wrapfig}
\usepackage{multirow}
\usepackage{bbm}
\usepackage[svgnames,dvipsnames,x11names]{xcolor}
\usepackage[utf8x]{inputenc}
\usepackage{selinput}
\usepackage{bm}
\usepackage{float}
\usepackage{geometry}
\usepackage{yfonts}
\usepackage{caption}
\usepackage{subcaption}
\usepackage{sidecap}
\usepackage{longtable}
\usepackage{anyfontsize}
\usepackage[hang,flushmargin]{footmisc} 
\usepackage{graphicx}
\usepackage{dcolumn}
\usepackage{bm}
\usepackage{amsmath, amssymb, amsfonts, amsthm}
\usepackage{epsfig}
\usepackage{textcomp}
\usepackage{tikz-cd} 
\usepackage{youngtab}
\usepackage{enumitem}
\usepackage{adjustbox}
\usepackage{makecell}

\def\d{\delta}

\def\f{\phi}

\def\L{\mathcal{L}}
\def\m{\mu}

\def\o{\omega}

\def \bfx {\textbf{x}}

\def \bfk {\textbf{k}}
\def \del {\partial}
\usepackage[framemethod=default]{mdframed}
\newmdenv[skipabove=7pt,
skipbelow=7pt,
rightline=false,
leftline=false,
topline=false,
bottomline=false,
backgroundcolor=gray!10,
linecolor=gray,
innerleftmargin=5pt,
innerrightmargin=5pt,
innertopmargin=5pt,
innerbottommargin=5pt,
leftmargin=0cm,
rightmargin=0cm,
linewidth=4pt]{eBox}
\newmdenv[skipabove=7pt,
skipbelow=7pt,
rightline=false,
leftline=false,
topline=false,
bottomline=false,
backgroundcolor=gray!10,
linecolor=gray,
innerleftmargin=5pt,
innerrightmargin=5pt,
innertopmargin=-5pt,
innerbottommargin=5pt,
leftmargin=0cm,
rightmargin=0cm,
linewidth=4pt]{eBox2}

\usepackage{array}

\definecolor{blue3}{RGB}{31, 119, 180}
\definecolor{red3}{RGB}{	214, 39, 40}
\definecolor{orange3}{RGB}{255, 127, 14}
\definecolor{green3}{RGB}{44, 160, 44}

\usepackage{colortbl}
\definecolor{lightgreen}{cmyk}{0.2, 0, 0.2, 0.2}
\definecolor{lightgray}{cmyk}{0.1,0.2,0,0.1}
\definecolor{lightgray2}{cmyk}{0.1,0.1,0,0.1}

\makeatletter
\newlength{\apb@width}
\newcommand{\autoparbox}[2][c]{\settowidth{\apb@width}{#2}\parbox[#1]{\apb@width}{#2}}

\makeatother


\def\d{{\rm d}}

\def\nn{\nonumber}

\def \bfp {\textbf{p}}

\def\beq{\begin{equation}}
\def\eeq{\end{equation}}

\allowdisplaybreaks[1]
\usepackage{physics}

\begin{document}



\begin{titlepage}
\setcounter{page}{1} \baselineskip=15.5pt 
\thispagestyle{empty}

\begin{center}
{\fontsize{18}{18} \bf The Cosmological Phonon: \vspace{0.1cm}
\;

Symmetries and Amplitudes on Sub-Horizon Scales}\\
\end{center}

\vskip 18pt
\begin{center}
\noindent
{\fontsize{12}{18}\selectfont Tanguy Grall\footnote{\tt
			tg418@cam.ac.uk}, Sadra Jazayeri\footnote{\tt
			sj571@dampt.cam.ac.uk} and David Stefanyszyn\footnote{\tt
			d.stefanyszyn@damtp.cam.ac.uk}}
\end{center}

\begin{center}
  \vskip 8pt
\textit{Department of Applied Mathematics and Theoretical Physics, University of Cambridge, Wilberforce Road, Cambridge, CB3 0WA, UK} 

\end{center}

\vspace{0.4cm}
 \begin{center}{\bf Abstract}
 \end{center}
 
 \noindent In contrast to massless spinning particles, scalars are not heavily constrained by unitarity and locality. Off-shell, no gauge symmetries are required to write down manifestly local theories, while on-shell consistent factorisation is trivial. Instead a useful classification scheme for scalars is based on the symmetries they can non-linearly realise. Motivated by the breaking of Lorentz boosts in cosmology, in this paper we classify the possible symmetries of a shift-symmetric scalar that is assumed to non-linearly realise Lorentz boosts as, for example, in the EFT of inflation. Our classification method is algebraic; guided by the coset construction and inverse Higgs constraints. We rediscover some known phonon theories within the superfluid and galileid classes, and discover a new galileid theory which we call the \textit{extended galileid}. Generic galileids correspond to the broken phase of galileon scalar EFTs and our extended galileids correspond to special subsets where each galileon coupling is fixed by an additional symmetry. We discuss the broken phase of theories that also admit a perturbation theory around Poincar\'{e} invariant vacua and we show that the so-called exceptional EFTs, the DBI scalar and special galileon, do not admit such a broken phase. Concentrating on DBI we provide a detailed account of this showing that the scattering amplitudes are secretly Poincar\'{e} invariant when the theory is expanded around the superfluid background used in the EFT of inflation. We point out that DBI is an exception to the common lore that the residue of the total energy pole of cosmological correlators is proportional to the amplitude. We also discuss the inevitability of poles in $2 \rightarrow 2$ scattering amplitudes when boost are spontaneously broken meaning that such theories do not admit Adler zeros and generalisations even in the presence of a shift symmetry.


\end{titlepage}

\restoregeometry

\newpage
\setcounter{tocdepth}{2}
\tableofcontents

\newpage




\section{Introduction and Motivation}
Weakly coupled effective field theories (EFTs) of massless particles with linearly realised Poincar\'{e} symmetries are incredibly constrained. At low energies, interacting spin-$S=1$ particles are described by Yang-Mills, a self-interacting $S=2$ particle is the graviton of General Relativity, particles with $S \geq 5/2$ cannot self-interact and cannot interact with the graviton, while particles with $S \geq 3/2$ cannot have an electromagnetic charge. All of these statements, and more including the inevitability of charge conservation, the equivalence principle and the presence of supersymmetry for theories with a $S=3/2$ particle, can be derived purely using on-shell methods\footnote{The term \textit{on-shell methods} is now widely used to refer to scattering amplitude techniques that bypass Lagrangian formalisms thereby avoiding redundancies such as gauge invariance and field redefinitions. See \cite{ElvangHuang} for a review.} and without ever having to switch on a collider or think about a falling elevator. The proofs use consistency of S-matrices, by which we mean they describe local and unitary physics, ranging from demanding cancellations of spurious poles to consistent factorisation when intermediate particles are taken on-shell \cite{WeinbergBook1,WeinbergBook2,BenincasaCachazo,McGadyRodina,BenincasaReview,CheungReview}. This small list of allowed EFTs for massless spinning particles in Minkowski space is a remarkable triumph in theoretical physics in the last 100 years or so. \\

However, scalar field theories which are prevalent in model building slip through this Poincar\'{e} straitjacket. Unlike for spinning particles with $S \geq 1$, it is very easy to write down an EFT for a single scalar degree of freedom that does not introduce spurious poles or violate locality. In an off-shell way of thinking we understand this since no gauge redundancies are required to write down massless scalar EFTs, while in an on-shell way of thinking consistent factorisation of say $2 \rightarrow 2$ scattering processes is trivial since the only on-shell three-particle amplitudes are constant. For this reason we invoke additional global symmetries, beyond those of the linearly realised Poincar\'{e} symmetries, to distinguish between different scalar EFTs. These symmetries are necessarily non-linearly realised since the Coleman-Mandula theorem tells us that a larger set of linearly realised (bosonic) spacetime symmetries is not possible \cite{ColemanMandula}. Throughout this paper we will use the term \textit{linearly realised} for symmetries that are unbroken by the vacuum and \textit{non-linearly realised} for those that are spontaneously broken by the vacuum. \\

Such a classification is complete and can be arrived at from two complementary methods: an on-shell classification \cite{GeneralisedAdler1,GeneralisedAdler2,GeneralisedAdler3, Probing} and an algebraic classification \cite{RSW1,BraunerBogers}. The former exploits the fact that non-linearly realised symmetries yield particular structures in soft scattering amplitudes which as part of on-shell data allows one to directly construct amplitudes and derive theories. This soft bootstrap method is very powerful and in some cases can fully fix the theory. The latter uses an algebraic analysis, within the framework of the coset construction \cite{Coset1,Coset2,Coset3} and inverse Higgs constraints \cite{InverseHiggs}, to search for consistent Lie-algebras that can be non-linearly realised by a single scalar degree of freedom. Indeed all of the new global symmetries must form a closed algebra with the Poincar\'{e} symmetries. An advantage of this second method is that no particular power counting is assumed for the interactions and the classification remains oblivious to field redefinitions. For linearly realised Poincar\'{e} symmetries we therefore have an excellent understanding of the allowed particle content, interactions and symmetries for all spins\footnote{A complete classification for $S=1/2$ fermions has also been performed in \cite{RSW1,SoftBootstrap}.}.\\

However, although the assumption of linearly realised Poincar\'{e} symmetries is an excellent one for particle physics and collider experiments, cosmology is very different. One of the distinguishing features of cosmological EFTs is that the symmetries of the Poincar\'{e} algebra are non-linearly realised. Indeed, the FRW metric breaks both time translations and Lorentz boosts while maintaining spatial translations and three-dimensional rotations as symmetries of the system. This opens up the possibility that a richer structure for the particle content, interactions and symmetries is possible in cosmology. A word of caution though. We have no reason to believe that Poincar\'{e} symmetries are not ultimately a good symmetry of Nature. In cosmology, and also condensed matter, the breaking of some Poincar\'{e} symmetries is usually taken as spontaneous and therefore expected to be restored and linearly realised at high energies. This is manifest in e.g. the EFT of inflation \cite{Cheung:2007st} where a Goldstone boson is introduced to non-linearly realise the broken time translations. In a theory where time translations are not broken spontaneously, no such Goldstone would be necessary. \\

This motivates us to extend the above described classification of scalar EFTs to theories with less linearly realised symmetry. Our aim in this paper is to construct a classification for scalar EFTs that linearly realise some form of \textit{spacetime} translations and spatial rotations while non-linearly realising Lorentz boosts and other symmetries which have the power to constrain the Wilson couplings. An analysis for spinning particles which makes use of the four-particle test \cite{BenincasaCachazo} will appear elsewhere \cite{Boostless}. Note that we are assuming that some form of time translations is a symmetry of the vacuum. This allows us to maintain conservation of energy such that we have a consistent set-up for scattering processes i.e. we have a conserved Hamiltonian and can define asymptotic states. For $P(X)$ theories a classification was presented in \cite{Pajer:2018egx} and, allowing for a breaking of the shift symmetry a classification for $P(\phi,X)$ theories was performed in \cite{Grall:2019qof}. Here we go beyond leading order in derivatives.   \\

One may worry that these assumptions, and the absence of a $S=2$ particle, makes our results irrelevant for cosmology. However this is not the case. Assume that the scalar fluctuation in a cosmological EFT has a constant shift symmetry. This requires the scalar to be derivatively coupled and therefore scattering processes are dominated by large momenta way above the would-be Hubble scale $H$ but below the cut-off of the theory which we denote as $\Lambda$. We can therefore work in a flat space limit by sending $H \rightarrow 0$ which is accurate up to $\mathcal{O}(H / \Lambda)$ corrections. This removes the cosmological expansion from the problem but does not remove $S=2$ fluctuations. We therefore also work in a decoupling limit by sending $M_{\text{pl}} \rightarrow \infty, \dot{H} \rightarrow 0$ with $M_{\text{pl}}^{2} \dot{H}$ kept fixed. In this limit the $S=0$ and $S=2$ fluctuations decouple. There is therefore a consistent and interesting limit of cosmological EFTs which are purely described by a single scalar EFT with self-interactions in flat space and non-linearly realised Lorentz boosts. See section \ref{sec:recap} for more details. These EFTs are of interest to us in this work and our results are therefore important for cosmological model building motivated by e.g. early universe inflation, field theoretic alternatives to the cosmological constant, the dark matter problem, etc. From now on we refer to our single scalar fluctuation as a \textit{phonon} and denote it by $\pi$. We use $\phi$ for Poincar\'{e} invariant theories. \\

For Poincar\'{e} invariant theories the on-shell classification is built upon the existence of an Adler zero \cite{Adler1,Adler2} for theories enjoying a constant shift symmetry $\phi \rightarrow \phi + c$. This symmetry ensures that scattering amplitudes vanish in the limit where one external momenta is taken soft. The classification centres around generalisations of this soft behaviour where amplitudes vanish more quickly in the soft limit thanks to additional symmetry. However, when Lorentz boosts are spontaneously broken there is no Adler zero even when there is a constant shift symmetry. This is because broken boosts allow for non-trivial cubic vertices which yield poles in amplitudes which ultimately  dominate the soft limit (see section \ref{Amplitudes}). No such cubic vertices exist when boosts are linearly realised\footnote{The only non-trivial cubic vertex is $\phi^3$ which is forbidden by the constant shift symmetry. Any other cubic operators vanish on-shell and can be removed by field redefinitions.} and their absence is crucial in deriving the Adler zero soft theorems. We will show that cubic vertices are actually inevitable when boosts are spontaneously broken: they are either required by symmetry or generated quantum mechanically. Due to the absence of a standard Adler zero we therefore choose to go down the route of an algebraic classification within the framework of the coset construction and inverse Higgs constraints which will both be reviewed in section \ref{sec:recap}. There we will also review the classification of Poincar\'{e} invariant theories. Let us emphasise that the absence of an Adler zero does not imply that there are not other soft theorems that constrain the form of the soft scattering amplitudes, and indeed it would be very interesting to derive these soft theorems and use them to construct theories directly at the level of S-matrix.  \\

The paper is organized as follows: we start in Section \ref{sec:recap} by reviewing background material on the coset construction as well as inverse Higgs trees which lay the foundations of our classification. There we also briefly review the EFT of inflation which is the prototypical cosmological formalism inspiring our work. Our classification is presented in Section \ref{sec:trees_in_the_zoo}. The algebraic method we use encompasses the broken phase of theories that also admit a consistent perturbation theory around a Poincar\'{e} invariant background as well as EFTs that do not. We rediscover known theories such as superfluids, with additional scaling \cite{Pajer:2018egx} or full conformal \cite{Hellerman:2015nra} symmetry, and galileids. We also discover a new theory which we call the \textit{extended galileid} which has all the symmetries of the galileid plus an additional symmetry generated by a scalar generator (see Tables \ref{tab:Extended_galileons.} and \ref{tab:extended_galileids}). This theory is reminiscent of the Poincar\'{e} invariant special galileon, since the symmetry starts with a field-independent term that is quadratic in the coordinates, but it is \textit{not} simply the special galileon expanded around a Lorentz breaking vacuum. Interestingly, we find that the so-called exceptional EFTs of scalar DBI and the special galileon \cite{galileon,SG} do not admit such a broken phase where all of the original non-linearly realised symmetries remain so. We find an interesting example of a broken phase of the special galileon where a symmetry that was non-linearly realised in the Lorentz invariant phase becomes linearly realised on the phonon (see Appendix \ref{sec:special_galileon_broken_phase}). For DBI however there is no broken phase and we outline the consequences for cosmology in Section \ref{sec:DBI_Inflation}. In particular, looking at the three-point function for DBI in the EFT of inflation we notice that the order of the total energy pole is different to that of a generic $P(X)$ theory. This is because in the flat space limit the DBI amplitudes are secretly Poincar\'{e} invariant due to increased symmetry in that limit. We conclude that the residue of the total energy pole in correlators is not always proportional to the flat space amplitude (see \cite{Arkani-Hamed:2018kmz} for a discussion of why this is often the case). Finally, in Section \ref{Amplitudes}, we discuss the inevitability of cubic vertices, soft theorems and weak coupling for theories with spontaneously broken boosts. We show that the scaling and conformal superfluid contain a region of parameter space where the theory is weakly coupled on sub-horizon scales. We conclude and discuss avenues for future work in Section \ref{sec:discussion}.


\section*{Conventions} 
\label{sec:conventions}
We work in $3+1$ spacetime dimensions. Greek lower case letters refer to Minkowski covariant objects with metric $\eta_{\mu \nu}=\text{diag}(-1,1,1,1)$. Lower case Latin letters refer to $SO(3)$ covariant objects with indices raised and lowered by the Kronecker delta $\delta_{ij}=\text{diag}(1,1,1)$. We also use boldface to distinguish $\bfx^2\equiv\delta_{ij}x^i x^j$ from $x^2 \equiv \eta_{\mu \nu}x^\mu x^\nu$. In the Poincar\'{e} invariant phase we use $\{P_\mu, M_{\mu \nu}\}$ as a basis for the Poincar\'e algebra with non-trivial commutation relations:
\begin{align}
 	\comm{P_\mu}{M_{\rho \sigma}}= \eta_{\mu \sigma} P_\rho - \eta_{\mu \rho} P_\sigma\,, && \comm{M_{\mu \nu}}{M_{\rho \sigma}}=\eta_{\mu \rho}M_{\nu \sigma} + \text{anti-symmetric}\,.
 \end{align} 
While in the broken phase we use the basis $\{\bar{P}_0,\bar{P}_i, \bar{M}_{ij}=\epsilon_{ijk}\bar{J}^k\}$. The non-trivial commutation relations in this basis are:
\begin{align} \label{eq:broken_poincare_comm_rel}
\comm{\bar{P}_{i}}{\bar{J}_{j}}&= \epsilon_{ijk}\bar{P}^{k}\,, \qquad 	
\comm{\bar{J}_{i}}{\bar{J}_{j}}= \epsilon_{ijk}\bar{J}^{k}\,.
\end{align}
We will sometimes denote non-linear boosts as $K_{i}$. For equal time in-in correlators we define
\begin{equation}
\langle \pi_{\bfk_1}\pi_{\bfk_2}\pi_{\bfk_3}\rangle=\langle \pi_{\bfk_1}\pi_{\bfk_2}\pi_{\bfk_3}\rangle'\, (2\pi)^3 \delta^3(\bfk_1+\bfk_2+\bfk_3)\,.
\end{equation}



\section{Recap} 
\label{sec:recap}

Before moving onto our classification let us first review the coset construction, inverse Higgs constraints and outline our methods for classifying EFTs. Here we will also put our cosmological motivation on a firmer footing, and review the Poincar\'{e} invariant classification. The reader familiar with these topics may jump directly to Section \ref{sec:trees_in_the_zoo}.
\subsection{The EFT of Inflation} 
\label{sub:the_eft_of_inflation}
Although in this paper we consider theories that live in flat spacetime, our classification is still useful for studying inflationary theories in their high energy limit, where the mixing with gravity is negligible and the background is effectively flat. In particular, a subset of theories we discover can describe the flat space limit of \textit{single-clock} cosmologies, where a single scalar field acquires a time dependent vev (the "clock") and breaks the time diffeomorphism symmetry \cite{Cheung:2007st}. Time translation symmetry can be restored via the \textit{Stueckelberg method} by introducing a Goldstone boson via
\begin{equation}
\phi(t,\bfx)=\bar{\phi} (t+\pi(t,\bfx) )\,,
\end{equation}
where $\bar{\phi}(t)$ is the time dependent vev driving inflation. 
In the decoupling limit, i.e. when $\frac{H}{M_p}, \frac{\dot{H}}{H^2}\to 0$ with the amplitude of the scalar power spectrum $(H^4/f_\pi^4)$ kept finite\footnote{Following \cite{Baumann:2011su} we have defined $f_\pi^4=2 c_s M_p^2|\dot H|$, the Goldstone decay constant normalising the scalar power spectrum.}, the Lagrangian for $\pi$ simplifies to
\begin{equation}
S=\int dt d^3\bfx \,\sqrt{-\bar{g}} \left(M_p^2\dot{H} \bar{g}^{\mu\nu}\del_\mu\pi \del_\nu \pi + \sum\limits_{n>1} \dfrac{M_n(t+\pi)}{n!} (-2\dot{\pi}+\bar{g}^{\mu\nu}\del_\mu\pi \del_\nu \pi )^n+... \right)\,,
\end{equation}
where $\bar{g}_{\mu\nu}$ is the quasi-dS metric, and ... stands for higher derivative terms. Up to this point, $M_n(t+\pi)$ are arbitrary functions, but in order to get a nearly scale invariant spectrum, one demands an approximate shift symmetry for $\pi$ that in combination with dS dilatation enforces all the $M_n$'s to be constant \cite{Finelli:2017fml,Finelli:2018upr}. As a result, in the flat space limit $\bar{g}_{\mu\nu}\to\eta_{\mu\nu}$, the theory for $\pi$ linearly realises time translations. For future references, let us quote the perturbative Lagrangian up to cubic order:
\begin{eqnarray}
\label{EFT}
S &=& \int dt\,d^3\bfx\,a^3(t)\,  \dfrac{f_\pi^4}{2c_s^3}\left[\left(\dot{\pi}^2-c_s^2\dfrac{(\del_i\pi)^2}{a^2}\right)+ (c_s^2-1)\,\left(\dfrac{1}{a^2}\dot{\pi}(\del_i\pi)^2-\left(1-\frac{2}{3}\frac{c_3}{c_s^2}\right)\dot{\pi}^3 \right)\right]\,.
\end{eqnarray}
Notice that in the EFT of inflation the transformation of $\pi$ under boosts is fixed to be
\begin{equation}
\label{boosttra}
\delta_{K_{i}} \pi=x^i+t\del^{i}\pi+x^i\dot{\pi}\,.
\end{equation}
However, in the classification we will provide in Section \ref{sec:trees_in_the_zoo}, we consider the spacetime dependence of the background $\bar{\phi}(t,\bfx)$ to be as generic as possible. Therefore, the transformation of $\pi$ under boosts is allowed to deviate from \eqref{boosttra} in interesting ways. \\

Once we switch back on the Hubble expansion, the symmetries we find could well be broken by any operator that is proportional to $H$, and as such, cosmological correlators will not remain exactly invariant under these symmetries. However, apart from in a very special case which we will discuss in Section \ref{sec:DBI_Inflation}, cosmological correlators encode information about flat space amplitudes in the residue of their total energy pole \cite{Maldacena:2011nz,Raju:2012zr,Arkani-Hamed:2015bza, Arkani-Hamed:2018kmz, Fraser-Goodhew, Enrico}, hence knowing the flat space limit of a theory provides valuable insights into the behaviour of boundary correlators as well\footnote{Regarding the following discussions, we are indebted to Enrico Pajer for kindly sharing his unpublished manuscript on the relation between scattering amplitudes and correlators' total energy poles.}. For example, the three point function generated by the cubic interactions in \eqref{EFT} is given by
\begin{align}
\label{3pt}
\langle \pi_{\bfk_1}\pi_{\bfk_2}\pi_{\bfk_3}\rangle' &=  f_\pi^4\,\dfrac{c_s^2-1}{c_s^3} \,P_{k_1}P_{k_2}P_{k_3} \Bigg[\dfrac{c_s}{H}\dfrac{1}{e_1^3}\Big(-12e_3^2+4e_1 e_2 e_3-11 e_1^3 e_3 \\\nonumber 
& +4e_1^2 e_2^2+3e_1^4 e_2-e_1^6\Big) + \dfrac{c_s^3}{H}\left(1-\dfrac{2 c_3}{3c_s^2}\right)\dfrac{12 e_3^2}{e_1^3} \Bigg]\,,
\end{align}
where we have defined the elementary symmetric polynomials in three variables:
\begin{align} \label{SymmetricPolynomials}
e_1 &= k_1+k_2+k_3\,, \nonumber \\ 
e_2 &= k_1k_2+k_1 k_3+k_2 k_3\,, \nonumber \\ 
e_3 &= k_1 k_2 k_3\,,
\end{align}
and $P_k$ is simply the powers pectrum of $\pi$, i.e. 
\begin{equation}
P_k=\left(\frac{H}{f_\pi}\right)^4 \frac{1}{k^3}\,.
\end{equation}
On the total energy pole, namely $e_1\to 0$ (which can be achieved upon analytical continuation), the three point function simplifies to
\begin{equation}
\label{poleenergy}
\lim_{e_1\to 0} \dfrac{1}{k_1 k_2 k_3\prod P_{k_i}}\langle \pi_{\bfk_1}\pi_{\bfk_2}\pi_{\bfk_3}\rangle'=f_\pi^4\,\dfrac{c_s^2-1}{c_s^2}\,\dfrac{12 e_3}{H\, e_1^3}\left[-1+c_s^2-\dfrac{2}{3} c_3\right]\,.
\end{equation}
As promised, the right hand side is proportional to the three-particle amplitude of the flat space limit of the theory in \eqref{EFT}, i.e. 
\begin{equation}
{\cal A}(k_1,k_2,k_3)=i6\sqrt{2c_s}\dfrac{c_s^{2}}{f_\pi^2}(c_s^2-1) e_3\left[1-c_s^2+\dfrac{2}{3}c_3 \right]\,.
\end{equation}
We use this result in Section \ref{sec:DBI_Inflation} to explore the consequences of our findings for DBI inflation.  \\

Our results will therefore have important applications for cosmological EFTs in their high energy limit, and this is manifest in the total energy pole of the cosmological correlators (see \cite{EnricoDan} for a recent study of exact linearly realised symmetries for cosmological correlators). 


\subsection{Coset Construction} 
\label{sub:coset_construction}

The coset construction allows one to construct the non-linear realisation of a broken symmetry group without knowledge of how the symmetry was spontaneously broken. It was first introduced for internal symmetries in \cite{Coset1,Coset2} and extended to spacetime symmetries in \cite{Coset3}. It has been reviewed in many cases in the literature, e.g. \cite{KRS,WessZuminoGal,Wheel}, so here we merely outline the most important aspects.\\

The starting point is the coset parametrisation
\begin{align} 
\Omega = e^{x^{\mu}P_{\mu}} e^{\f^A G_A} \,, \label{coset-element}
\end{align}
where $P_{\mu}$ are the spacetime translations and $G_{A}$ are the generators that are spontaneously broken with associated Goldstone fields $\f^{A}$. The generators and Goldstones must form (reducible) representations of the unbroken subgroup which contains the generators $T_{i}$. For Poincar\'{e} invariant field theories the unbroken subgroup is Lorentz $SO(1,3)$ while for our interests in this paper the subgroup is that of spatial rotations $SO(3)$. Note that spacetime translations are included in the coset element since they are non-linearly realised on the coordinates. The building blocks of invariant Lagrangians come from the Maurer-Cartan $1$-form which can be written as 
\begin{align} 
\Omega^{-1} d \Omega = \o^\m P_\m + \o^A G_A + \o^i T_i \,.
\end{align} 
The $\o$ are functions of the coordinates and the Goldstones and are referred to as Maurer-Cartan components. After pulling back to spacetime and writing the components as e.g. $\omega^{\mu} = (\o^\m)_\nu dx^{\nu}$, we interpret the $(\o^\m)_\nu$ as vielbeins $e^\m{}_\nu \equiv (\o^\m)_\nu$, while the components $(\o^A)_{\mu}$ can be used to construct covariant derivatives of the Goldstones $\nabla_\m \f^A = (e^{-1})^{\nu}{}_{\mu} (\o^A)_\nu$. The remaining components $ (\o^i)_{\mu}$ are used to construct higher order derivatives of the Goldstones and to couple the Goldstones to matter fields. These are the building blocks of invariant Lagrangians. In addition to invariant Lagrangians one can also consider Lagrangians that shift by a total derivative. These \textit{Wess-Zumino terms} are derived by constructing exact $5$-forms $\beta_{5} = d \beta_{4}$ out of the Maurer-Cartan $1$-forms followed by integrating $\beta_{4}$ over spacetime. Note that here we have combined time and space translations into a single object $P_{\mu} = (P_{0},P_{i})$. When we review the Poincar\'{e} invariant cases below this is because our linearly realised subgroup contains the Lorentz generators and therefore $P_{\mu}$ is a Lorentzian four-vector. However, later on our linearly realised subgroup will simply be the group of spatial rotations $SO(3)$ in which case it is for notational convenience only; there is no sense in which this a is Lorentzian four-vector since there are no Lorentz boosts in the linearly realised subgroup. \\

An important aspect of the coset construction is that we are required to introduce a Goldstone mode for each non-linear generator. Naively, it would seem impossible for a single scalar EFT to realise symmetries beyond the one generated by its corresponding scalar generator. However, inverse Higgs constraints \cite{InverseHiggs}, which we will now review, open up this possibility. 



\subsection{Inverse Higgs Constraints and Inverse Higgs Trees} 
\label{sub:inverse_higgs_constraints_and_inverse_higgs_trees}

When internal symmetries are spontaneously broken, Goldstone's theorem tell us that we have a massless degree of freedom for each non-linear generator in the resulting non-linear realisation. However, when spacetime symmetries are broken Goldstone's theorem does not apply and indeed there can be fewer Goldstone degrees of freedom than non-linear generators. In this case we distinguish between essential and inessential Goldstones. The former are required to non-linearly realise all the symmetries while the latter are not and can be eliminated in favour of the essential ones via \textit{inverse Higgs constraints} \cite{InverseHiggs}. As described above, the coset construction requires us to introduce these inessential modes, but inverse Higgs constraints allows us to eliminate them without losing any symmetries.\\

An inverse Higgs constraint exists when a commutator of the form $[P,G'] \supset G$ appears in the algebra. In this case the would-be Goldstone mode associated with $G'$, say $\phi'$, can be removed in favour of the spacetime derivatives of the Goldstone modes associated with $G$, say $\phi$, by setting to zero the appropriate projection of the covariant derivative $\nabla \phi$. Here we are suppressing any indices and $P$ schematically denotes spacetime translations. Having a commutator of the form $[P,G'] \supset G$ ensures that $\phi'$ appears linearly in $\nabla \phi$ meaning that we can solve for all its components\footnote{The algebra needs to satisfy additional conditions if $\phi'$ is to appear algebraically in $\nabla \phi$ to all orders, as desired \cite{KRS,McArthur}.}. This also guarantees that in the non-linear realisation $\phi'$ is not massless since the most general Lagrangian includes a $(\nabla \phi)^2$ term. This further emphasises that it cannot be an essential part of the low energy EFT since it can be integrated out for energies below its mass which is expected to be near the symmetry breaking scale.\\

We now move onto \textit{inverse Higgs trees} which were first introduced in \cite{RSW1,RSW2}. When classifying scalar EFTs in terms of non-linearly realised symmetries, inverse Higgs constraints are crucial as they allow the EFT to have extra symmetry without extra degrees of freedom. Inverse Higgs trees offer a systematic way of understanding which algebras can be realised by a given number of degrees of freedom. We now review this procedure for a single scalar Poincar\'{e} invariant EFT where the idea of an inverse Higgs tree is very simple. \\

Since we want to work with a single essential scalar Goldstone, the algebra must contain a scalar generator $Q$ in addition to $ISO(1,3)$. Now the only covariant derivative we can set to zero to solve for an inessential Goldstone lives in the vector representation of the Lorentz group since it is the scalar's covariant derivative. We can therefore add an additional vector generator to the algebra as long as it's commutator with spacetime translations contains $Q$ i.e. $[P_{\mu}, V_{\nu}] \supset \eta_{\mu\nu} Q$. Denoting the vector mode as $A_{\mu}$, the scalar's covariant derivative takes the form $\nabla_{\mu} \phi = \partial_{\mu} \phi + A_{\mu} + \ldots$ where the $\ldots$ depend on the full form of all commutators.\\

\begin{wrapfigure}[16]{R}{5cm}
	\begin{tikzpicture} 
	\node at (2, 1.0){\textbf{Lorentz Invariant Scalar tree}};
	\draw [dashed, thin] (0+0.2,0-0.2) -- (1 - 0.2,-1 + 0.2);
	\node at (0,0){$\bullet$};
	\draw [dashed, thin](1 + 0.2, -1 - 0.2) -- (2 -0.2, -2 + 0.2);
	\draw [dashed, thin](1-0.2,-1-0.2) -- (0+0.2,-2+0.2);
	\node at (1,-1) {\tiny $\yng(1)$};
	\node at (0,-2) {$\bullet$};
	\node at (2,-2) {\tiny $\yng(2)$};
	\draw [dashed, thin](0 + 0.2, -2 - 0.2) -- (1 - 0.2, -3 + 0.2);
	\draw [dashed, thin](2 + 0.2, -2 - 0.2) -- (3 - 0.2, -3 + 0.2);
	\draw [dashed, thin](2 - 0.2, -2 - 0.2) -- (1 + 0.2, -3 + 0.2);
	\node at (1,-3) {\tiny $\yng(1)$};
	\node at (3, -3) {\tiny $\yng(3)$};
	\draw [dashed, thin](1 -0.2, -3 - 0.2) -- (0 + 0.2, -4 + 0.2);
	\draw [dashed,  thin](1 +0.2, -3 -0.2) -- (2 -0.2, -4 + 0.2);
	\draw [dashed, thin](3 - 0.2, -3 - 0.2) -- (2 + 0.2, -4 + 0.2);
	\draw [dashed, thin](3 + 0.2,  -3 - 0.2) -- (4 - 0.2, -4 + 0.2); 
	\node at (0,-4){$\bullet$};
	\node at (2, -4){\tiny $\yng(2)$};
	\node at (4, -4){\tiny $\yng(4)$};
	\end{tikzpicture}
	\caption{Inverse Higgs tree for a single essential scalar in a Lorentz invariant theory, from \cite{RSW1}.}
	\label{fig:scalartree}
\end{wrapfigure}
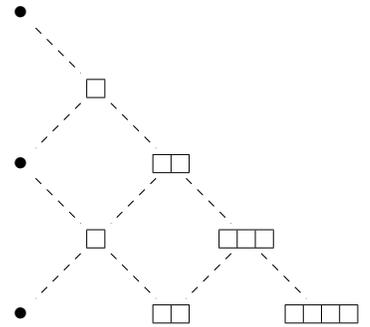

Having added the vector, we have opened up more possibilities as we now have additional covariant derivatives coming from the vector which can also be set to zero to solve for inessential modes. This covariant derivative contains three irreducible representations of the Lorentz group: a scalar, a traceless tensor and a two-form. However, Jacobi identities cannot be satisfied in the presence of the two-form \cite{BraunerBogers} so only two generators can be added. If we denote these generators by $S$ and $S_{\mu\nu}^{T}$ then the required commutators are $[P_{\mu},S] \supset V_{\mu}$ and $[P_{\mu},S_{\rho \sigma}^{T}] \supset \eta_{\mu \rho} V_{\sigma} + \eta_{\mu \sigma} V_{\rho} - \eta_{\rho \sigma}V_{\mu} / 2$. The pattern continues and yields the inverse Higgs tree in Figure \ref{fig:scalartree} where the dashed lines denote connections between generators by spacetime translations and each generator is an irreducible representation of the Lorentz group. All allowed generators live in the Taylor expansion of $\phi(x)$, and the level of a generator in the tree is given by how many acts of $P_{\mu}$ it is away from $Q$. A generator at level-$n$ is denoted $G^{n}$. If a generator has multiple connections to the level below then all connections are required to satisfy Jacobi identities e.g. if we include the vector at level-$3$ then it must be connected, by translations, to both the scalar and the tensor at level-$2$. Note that all commutators between non-linear generators and translations take the schematic form $[P,G^{n}] = G^{n-1} + \text{linear}$. No other non-linear generators can appear on the RHS, this can be guaranteed by basis changes \cite{RSW1}. \\

Now lets assume\footnote{The only other option is that $Q$ generates dilatations in which case there is no Adler zero.} that the symmetry transformation of the essential Goldstone $\phi$ generated by $Q$ is $\delta_{Q}\phi = 1$. Given that the vector generator $V_{\mu}$ at level-$1$ in the tree is related to $Q$ via spacetime translations, it is clear that the symmetry transformation it generates must be of the form $\delta_{V_{\mu}} \phi = x^{\mu} + \ldots$, where $\ldots$ contains field-dependence but no explicit $x^\mu$ dependence. The extension to higher levels in the tree is trivial with all level-$n$ symmetries containing $n$ powers of $x^\mu$. \\

The Adler zero mentioned in the introduction appears thanks to the $n=0$ symmetry which ensures that scattering amplitudes vanish in the limit where one external momentum is taken soft i.e. in the soft limit we can Taylor expand amplitudes as
\begin{align} 
\lim_{p \rightarrow 0}\mathcal{A} = \sum_{q=1}a_{q}p^q,
\end{align} 
where $p$ is the soft momentum and the constant parameters $a_{q}, ~ q \geq 1$ are unconstrained by the shift symmetry. These parameters are related to the Wilson couplings in the EFT. Generalisations of the Adler zero can however constrain the soft amplitudes further \cite{GeneralisedAdler1,GeneralisedAdler2,GeneralisedAdler3}. A theory with finite $n$ symmetries has $a_{q} = 0 ~ \forall ~ q \leq n$ such that the soft amplitude begins at $\mathcal{O}(p^{\sigma})$ where $\sigma = n+1$ is the referred to as the soft degree. We therefore see a direct connection between the level at which the tree terminates, and the soft limit of the resulting scattering amplitudes. The tree nicely encodes the data one provides for soft bootstrap procedures (see e.g. \cite{SoftBootstrap}). Indeed, the generators at level-$0$ tell us what the essential Goldstones are in the EFT, the connections between generators tell us about the linearly realised symmetries, and as we have just explained, the level at which the tree terminates fixes the soft degree of scattering amplitudes.\\

The inverse Higgs tree ensures that Jacobi identities involving two copies of translations and one other generator are satisfied. To complete the classification of possible algebras one therefore needs to satisfy the remaining Jacobi identities. The idea is to write down the most general commutators, consistent with the inverse Higgs tree, and solve the constraints imposed by Jacobi identities. A full classification was performed in \cite{RSW1} assuming the existence of a standard $1/p^2$ propagator. This assumption constrains the tree further and restricts Figure \ref{fig:scalartree} to the far diagonal. The reason for this is simple: the scalar's canonical kinetic term $(\partial \phi)^2$ is the operator with the fewest powers of $\phi$ and so needs to be invariant under the field-independent part of all symmetry transformations. This is only the case if the symmetry parameters are traceless. \\

In the absence of field-dependence the symmetry transformations are referred to as \textit{extended shift symmetries} \cite{ExtendedShifts}\footnote{Extended shift symmetries are relevant for the Goldstone boson equivalence theorem and control the onset of strong coupling for \textit{massive} spinning particles since they dictate the structure of the self-interactions of the longitudinal mode \cite{UVconstraints,BrandoHigherSpin}.} while those with field-dependence are referred to as exceptional shift symmetries and are realised by \textit{exceptional EFTs}. Algebraically the former are Abelian algebras where all commutators between non-linear generators vanish whereas the latter are non-Abelian algebras where at least one commutator between non-linear generators is non-zero. The exceptional EFTs are the natural scalar analogues of gauge and gravity theories\footnote{See \cite{RoestSG} for a recent discussion.} given that some simple properties of their S-matrix can fully fix their interactions. A classification of all single scalar EFTs with such symmetries is complete and is summarised in the table \ref{ScalarSymmetries}\footnote{We are assuming the existence of a shift symmetry and so the dilaton EFT is not included. The dilaton has $\sigma=-1$.}. \\

\begin{table}[h!]
\begin{center}
\begin{tabular}{ |p{3cm}|p{3cm}|p{3.2cm}|}
 \hline
 \multicolumn{3}{|c|}{Full single scalar classification with linearly realised $ISO(1,3)$} \\
 \hline
 Soft degree $\sigma$ & Extended shifts &Exceptional EFTs\\
 \hline
 1   & \checkmark ~~~ $\partial^4 \phi^4$   & $\times$  \\
 2&   \checkmark ~~~ $\partial^6 \phi^4$   &  \checkmark ~~~ $\partial^4 \phi^4$    \\
 3 &\checkmark ~~~ $\partial^{10} \phi^4$ & \checkmark ~~~ $\partial^6 \phi^4$   \\
 4    &\checkmark ~~~ $\partial^{12} \phi^4$  & $\times$  \\
 5&  \checkmark ~~~ $\partial^{16} \phi^4$  &  $\times$  \\
  \vdots&  \vdots   & \vdots  \\
  10&  \checkmark ~~~ $\partial^{30} \phi^4$  &  $\times$  \\
  \vdots&  \vdots   & \vdots  \\
 \hline
\end{tabular}
\caption{The full classification of scalar EFTs with generalised Adler zero soft behaviour with the power counting of leading order four-point vertices. For even $\sigma$ these vertices have $3 \sigma$ derivatives and for odd $\sigma$ they have $3 \sigma +1$ derivatives.}
\label{ScalarSymmetries}
\end{center}
\end{table}
A theory with extended shift symmetries exists for every $\sigma$, while there are only two exceptional EFTs with $\sigma = 2,3$. In each case we have also indicated the schematic power counting for four-point vertices with the fewest derivatives in the corresponding EFT, ignoring those that can be removed by field redefinitions\footnote{As far as we know a full classification of interactions at each order in $\sigma$ is incomplete. This is a non-trivial task since for $\sigma \geq 2$ the leading interactions for the extended shift symmetries are Wess-Zumino terms and are therefore not directly derivable from the coset construction (see \cite{WessZuminoGal} for a derivation of Wess-Zumino terms for $\sigma=2$). The vertices presented in table \ref{ScalarSymmetries} can however be read off from the tuned higher spin potentials in \cite{BrandoHigherSpin}. Indeed, a scalar with a $\sigma$ soft degree is the longitudinal mode of a massive $S = \sigma$ particle. In the high energy limit, the Wess-Zumino interactions arise once all of the individual irreps of the massive spinning particle have been diagonalised at quadratic order. This is familiar for massive $S=2$, see \cite{Aspects} for a review, but extends to higher spins too (see also \cite{PorratiRahman}).}. The two exceptional EFTs are the scalar DBI and special galileon \cite{SG} (the more general galileon EFT \cite{galileon} is the $\sigma=2$ extended shift symmetry). See \cite{dSAdS} for a classification in dS/AdS space. \\






\section{Trees in the Zoo} 
\label{sec:trees_in_the_zoo}
We set out to establish a similar classification for theories which spontaneously break Lorentz boosts.
Our goal is to write down algebras that can be non-linearly realised by a single $SO(3)$-scalar, $\pi(t,\bfx)$, which we take to linearly realise spacetime translations $\bar{P}_{i}, \bar{P}_{0}$ and $SO(3)$ rotations $\bar{J}_i$. In addition we are interested in non-linearly realised symmetries which includes at the very least Lorentz boosts $K_i$. The algebras we will derive are therefore all relativistic meaning that they contain an $ISO(1,3)$ subgroup and all additional generators form representations of the Lorentz group $SO(1,3)$. The realisation of these algebras on $\pi$ however is related to the relativistic algebra by basis changes which are only required to be $SO(3)$ covariant. It is these basis changes that create the non-trivial inverse Higgs constraints which remove the inessential Goldstone modes which now includes the Goldstones of the broken boosts. Crucially, the translation and rotation generators do not have to be the same in the different bases. In the broken phase we use bars to represent the linear generators as
\begin{align}
	\bar P_0\,, \bar P_i\quad  \text{and} \quad \bar{J}_i\, \quad\quad\quad (\text{unbroken})\,.
\end{align}
A familiar example comes from (zero-temperature) superfluids. The non-linearly realised algebra is $ISO(1,3) \times U(1)$ and a basis change $\bar P_0=P_0+\mu Q$, where $\mu$ is order parameter of the symmetry breaking, is required to generate the inverse Higgs constraints to remove the Goldstone bosons of the broken boosts. Such a phenomena is more generally referred to as spontaneous symmetry probing \cite{Nicolis:2011pv}. These linear generators act on $\pi$ as
\begin{align}
\delta_{\bar P^{0}}\pi = - \dot{\pi}, \quad \delta_{\bar P^{i}}\pi = - \partial_{i}\pi, \quad \delta_{\bar J^{i}}\pi = \epsilon_{ijk}x^{j} \partial^{k}\pi.
\end{align}

Let us remark that this work can be seen in continuation of \cite{Nicolis:2015sra} where the authors classified condensed matter systems according to their symmetry breaking pattern. The different symmetry breaking patterns that can arise from basis changes of the Poincar\'e algebra to the subset of linear generators $\{\bar P_0,\bar P_i,\bar J_i\}$\footnote{It is equivalent to listing all the possible change of basis from $\{ P_0, P_i, J_i\}$ to $\{\bar P_0,\bar P_i,\bar J_i\}$ with or without additional linear generators.} yields different physical systems in the IR. This leads to eight classes of theories each with a unique way to realise the symmetries. Within this ``zoology" of condensed matter systems the inverse Higgs trees we consider in this work correspond to the classes that can be realised with a shift-symmetry $Q$ and a single Goldstone mode $\pi(t,\bfx)$. The shift symmetry is simply $\delta_Q\pi=1$.\\

Guided by the Poincar\'{e} invariant classifications outlined in Section \ref{sec:recap}, we will classify different symmetry generators by the level at which they appear in an inverse Higgs tree. A level-$n$ generator $G^{n}_{i_1\dots i_s}$ is defined to act on $\pi$ as 
\begin{equation}
 \delta_{G^{n}_{i_1\dots i_s}}\pi=t^{n-s} x^{i_1}\dots x^{i_s}+\dots, 
 \end{equation} 
where ellipses stand for field-dependent terms\footnote{By extension, the level of a tree refers to its highest non-empty level.}. Such a symmetry generator appears at level-$n$ in a tree since we are required to act with $n$ copies of translations to reach the constant shift symmetry generated by the level-$0$ generator $Q$. Note that this general form includes non-linear boosts. A crucial difference between this Lorentz breaking set-up versus the fully Poincar\'{e} invariant case is that there are no Abelian algebras. Since boosts are non-linearly realised, there will always be commutators between non-linear generators that are non-zero. In that sense \textit{all} of the algebras are non-Abelian and \textit{all} of theories are exceptional. Clearly this is not a wise distinction to make here so we will avoid referring to theories we derive as exceptional. Furthermore, there are infinitely many such algebras, we will explain why below, and therefore it is impossible to perform a full classification. For this reason we truncate our classification of trees to level-2 i.e. we consider symmetries with at most two powers of the spacetime co-ordinates. This means we capture the theories which are most important in the IR. There may be a better distinction between algebras that can yield a full classification, and we leave such a direction for future work.\\

The plan for this section is as follows: in Section \ref{subsub:tree_structure} we describe the organic structure of the inverse Higgs trees and list the various assumptions that we use to derive our classification. In Section \ref{sub:level_1_trees_superfluids} and Section \ref{sub:level_2_trees_galileids_and_superfluids_again} we present the resulting theories associated with the trees truncated 
at level-1 and level-2 respectively. They are comprised of various superfluids and  galileids, including a new theory we call the \textit{extended galileid}. We discuss the broken phase of extended shift symmetric theories in Section \ref{sub:broken_phase_of_ext_shift_theories}. Finally, along the way the reader will notice the surprising absence of the broken phase of Poincar\'{e} invariant exceptional EFTs, namely, scalar DBI and the special galileon. It turns out that these theories do not admit backgrounds breaking Lorentz boosts but preserving space and time translations. We shed light on these peculiarities in Section \ref{sub:obstacles_for_excp_EFTs} before offering a more complete discussion for DBI in Section \ref{sec:DBI_Inflation}.



\subsection{Set-Up} 
\label{subsub:tree_structure}
When Lorentz boosts are spontaneously broken, the resulting long-wavelength theory is no longer built from Lorentz invariant (or covariant) terms: time and space derivatives are treated on an unequal footing. The same is true then for symmetry generators so we parametrise the coset element by
\begin{equation}
	\Omega=e^{t\bar{P}_0}e^{x^i \bar{P}_i}e^{\pi Q}e^{\phi^AG_A}\,,
\end{equation}
where all generators form representations of $SO(3)$. For instance an $SO(3)$ scalar\footnote{Note that this scalar does not need to be a fundamental scalar under $ISO(1,3)$. It could be the zero-component of a vector field, for example.} commutes with $\bar{J}_i$ while an $SO(3)$ vector $V_i$ has the usual commutation relation
\begin{align}
  	\comm{V_i}{\bar{J}_j}=\epsilon_{ijk}V^k\,.
\end{align} 
Since space and time translations are treated separately, the necessary conditions for the existence of inverse Higgs constraints is now different. For a given additional generator $G^n$, the necessary condition for its Goldstone to be inessential is that $\comm{\bar{P}_0}{G^{n}}\supset G^{n-1}$ \emph{or} $\comm{\bar{P}_i}{G^{n}}\supset G^{n-1}$. In many cases both are required. This means that the inverse Higgs trees will be made of two types of branches with connections to lower level generators by $\bar{P}_0$ or $\bar{P}_i$. We use the following notation for these branches:
\begin{figure}[h!]
\centering
	 \begin{tikzpicture}[x=0.4cm,y=0.4cm]
\tikzstyle{Pi} = [black, thick];
\tikzstyle{P0} = [black, dashed, thick];
\tikzstyle{label} = [->];
\draw[style=Pi] (6,4) to (8,4)node [anchor = west, black]{$\comm{\bar{P}_i}{G^{n}}\supset G^{n-1}$};
\draw[style=P0] (6,2) to (8,2)node [anchor = west, black]{$\comm{\bar{P}_0}{G^{n}}\supset G^{n-1}$};
\end{tikzpicture}
\end{figure}
\,\\ 
We take the solid lines as going from north-west to south-east, and the dashed lines as going from north-east to south-west in our figures. As an example see Figure \ref{fig:scaling_superfluid}. By a slight abuse of language we will keep referring to the symmetries generated by the $G^{n}$'s as space-time symmetries.\\

Finally, let us be clear about the assumptions we make to guide us through the woods of inverse Higgs trees:
\begin{enumerate}
\item \emph{SO(1,3) representations}: We want the EFT to describe the Lorentz breaking phase of a fundamentally Lorentz invariant theory so the non-linearly realised algebra must be relativistic up to basis changes (which are only manifestly $SO(3)$ symmetric). By relativistic we mean that all generators form representations of $SO(1,3)$ and the algebra contains the $ISO(1,3)$ algebra as a sub-algebra. These basis changes are necessary to create the required inverse Higgs constraints. \\

Crucially, we will be assuming that only the generators in the inverse Higgs trees are used to form the $SO(1,3)$ irreps. There are other cases, however, where $SO(3)$ generators correspond to linearly realised symmetries yet combine with generators in a tree to form an $SO(1,3)$ irrep. We believe these are rare but provide an example in Appendix \ref{sec:special_galileon_broken_phase}. We believe this provides a neat arena for studying exceptions to the Coleman-Mandula theorem \cite{ColemanMandula} when boosts are not a symmetry of the vacuum.  \\

\item \emph{Canonical Propagator}: We assume a standard kinetic term for the phonon\footnote{We leave the case of theories with different dispersion relations (e.g. the ghost condensate \cite{ghost}) to future work.}, $\dot\pi^2-c_s^2\partial_i\pi \partial^i\pi$, which must be invariant under the field independent part of all symmetries. This is because it is the operator with the fewest powers of $\pi$. Note that, unless otherwise explicitly stated, we will set $c_s=1$ throughout most of this section. For theories on Minkowski space this simply amounts to a rescaling of the spatial coordinate $x^i\to x^i/c_s$.	
\end{enumerate}
For each tree our working strategy is as follows. First, write the most general commutators between generators respecting the symmetries and the inverse Higgs constraints (i.e. the tree structure). Second, use Jacobi identities to place further constraints on the algebra and its coefficients. At this step it is possible (and in fact does happen) that one tree yields several distinct algebras in which case we say it has multiple stems. Finally we resort to assumption 1 above and check whether or not the algebra can be made relativistic after a basis change. In particular it should contain an $SO(3)$ vector with the commutation relations of boosts i.e.
\begin{align}
[K_{i},K_{j}] = - \epsilon_{ijk}J^{k}.
\end{align}
This is in fact a very constraining condition. Indeed, the majority of algebras do not satisfy this property. \\


\subsection{Level-1: Superfluids and Scaling Superfluids} 
\label{sub:level_1_trees_superfluids}
\begin{figure}[h!]
\begin{center}
\subcaptionbox{Superfluid\label{fig:level_1_superfluid}}[.4\linewidth]{
	 \begin{tikzpicture}[x=0.6cm,y=0.6cm]
\tikzstyle{Pi} = [black, thick];
\tikzstyle{P0} = [black, dashed, thick];
\tikzstyle{label} = [->];
\draw (0,0) node [anchor = south, black]{$Q$};
\draw[style=Pi] (0,0) to (2,-2)node [anchor = west, black]{$V_i$};
\end{tikzpicture}}
\subcaptionbox{Scaling Superfluid \label{fig:scaling_superfluid}}[.4\linewidth]{
	 \begin{tikzpicture}[x=0.6cm,y=0.6cm]
\tikzstyle{Pi} = [black, thick];
\tikzstyle{P0} = [black, dashed, thick];
\tikzstyle{label} = [->];
\draw (4,4) node [anchor = south, black]{$Q$};
\draw[style=Pi] (4,4) to (6,2)node [anchor = west, black]{$V_i$};
\draw[style=P0] (4,4) to (2,2)node [anchor = east, black]{$D$};
\end{tikzpicture}}
\caption{Level-1 Inverse Higgs Trees}
\label{fig:level-1_two_trees}
\end{center}
\end{figure}
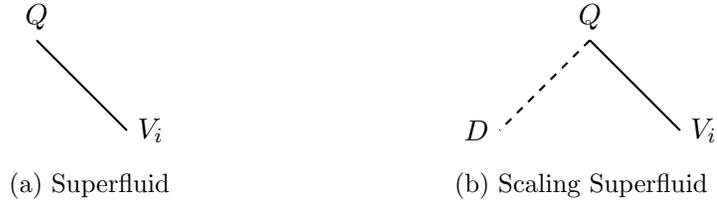

We begin by truncating the trees at level-1. Here we can only add two different generators since the inverse Higgs constraints at our disposal are
\begin{equation}
 	\nabla_t\pi=0\,, \qquad \text{and}\qquad \nabla_i\pi=0\,.
 \end{equation} 
Since these covariant derivatives live in the scalar and vector representation of $SO(3)$ we can add a scalar generator which we denote as $D$ and a vector generator which we denote as $V_i$. In principle this would lead to three different trees but since the algebra must ultimately be relativisitic after a basis change, at least one vector must appear in the tree. There are therefore two possibilities represented in Figure \ref{fig:level-1_two_trees}.
Now since the algebras are so simple in these cases it actually makes sense to work with the relativistic algebra directly and look for basis changes that bring the algebras into the form dictated by the trees. Consider the first case where we add a scalar generator $Q$ to the Poincar\'{e} algebra. Of course $Q$ commutes with the Lorentz generators so we only need to specify it's commutator with translations which takes the general form
\begin{align}
[P_{\mu},Q] = aP_{\mu}.
\end{align} 
All Jacobi identities are satisfied and this is a consistent algebra. To generate the required tree structure such that the Goldstones of boosts are inessential we must define the new time translation generator 
\begin{align} \label{eq:superfluid_basisc_change}
\bar{P}_{0} = P_{0} + \mu Q.
\end{align}
We then have $[\bar{P}_{i},V_{i}] \supset -\mu \delta_{ij} Q$ where $V_{i} = K_i$. However, $\bar{P}_{0}$ must still generate time translations meaning that it must commute with $\bar{P}_{i} = P_{i}$. This is only possible if $a=0$. We then have the algebra of a zero-temperature \emph{superfluid} which in the relativistic phase is simply $ISO(1,3)\times U(1)$ and with $\mu$ the chemical potential \cite{Son:2002zn}. The coset construction for such a symmetry breaking pattern has been performed in \cite{Nicolis:2013lma}. If we parametrise the coset element as 
\begin{align}
\Omega = e^{t\bar{P}_{0}}e^{x^i \bar{P}_{i}}e^{\pi Q}e^{\eta^{i}V_{i}},
\end{align} 
then after the field redefinition 
\begin{equation}
\beta_i\equiv - \frac{\tanh\sqrt{\eta^2}}{\sqrt{\eta^2}}\eta_i, 
\end{equation}
the inverse Higgs constraint allowing us to eliminate the boost Goldstones in favour of the phonon is
\begin{align}
\beta_{i} = - \frac{\partial_i\pi}{\mu+\dot\pi}.
\end{align}
The resulting theory for $\pi$ is neatly written in terms of $\phi = \mu t + \pi$ and to leading order in derivatives takes the $P(X)$ form where $X = (\partial \phi)^2$ \cite{Son:2002zn,Nicolis:2013lma,Nicolis:2011cs}. Indeed, the phonon corresponds in this case to fluctuations around a state of uniform charge density: $\pi=\phi-\expval{\phi}=\phi-\mu t$. Under non-linear boosts the phonon transforms as
\begin{equation} \label{NLBoostsSuperfluid}
\delta_{V_{i}}\pi=  \mu x^i + x^i \dot\pi  + t\partial^i \pi \,,
\end{equation}
as it does in the EFT of inflation. \\

The second tree (Figure \ref{fig:scaling_superfluid}) is also easy to understand. We again simply start with a relativistic algebra, where we add two scalar generators to $ISO(1,3)$, and create the tree through a basis change. Imposing Jacobi identities, the non-trivial commutators are 
\begin{equation}
	\comm{P_\mu}{D}=b P_\mu\,, \qquad \qquad \comm{Q}{D}= -\Delta Q\,,
\end{equation}
with $\Delta$ an arbitrary real parameter and $b \neq 0$ to ensure that the Goldstone associated with $D$ is inessential. Without loss of generality we set $b=1$. As suggested by the notation, $D$ can be seen as the generator of dilatations with $\Delta$ the scaling weight of $Q$. Now to generate the tree we are required to perform the same basis change as above \eqref{eq:superfluid_basisc_change} and then the symmetry algebra in the broken phase has the following non-trivial commutation relations
\begin{align}\nn
	\comm{\bar P_0}{D}&=\bar P_0- \mu (\Delta+1)Q &&  \comm{\bar P_i}{D}=\bar P_i \\
	\label{eq:scaling_superfluid_algebra}
	\comm{\bar P_0}{V_i}&=\bar P_i &&\comm{\bar P_i}{V_j}=\delta_{ij} (\bar P_0- \mu Q)\\\nn
	\comm{Q}{D}&=-\Delta Q && \comm{V_i}{V_j}=-\epsilon_{ijk}\bar{J}^k.
\end{align}
One can see that for the special value $\Delta=-1$ one of the inverse Higgs constraints is lost. This can be understood from the symmetry transformation of $\pi$ which takes the form
\begin{equation}
\delta_D\pi=-\mu(\Delta+1)t-\Delta\pi-t\dot\pi - x^{i} \partial_{i} \pi\,.
\end{equation}
Indeed, when $\Delta=-1$ the symmetry is no longer spontaneously broken, it is linearly realised\footnote{Actually in this case the theory is non-local. This can be seen from the the underlying relativistic theory, at leading order in derivatives, $X$ is scale invariant and so there's no way to compensate for $\d^4x$ measure and construct a scaling invariant action. Then at higher order in derivatives, one would have terms such as $(\partial_\mu \partial_\nu\phi \partial^\mu\partial^\nu\phi)^{-2}$ which make up a scaling invariant action. However these and all such higher derivatives terms are non-local and generate non-local interactions for $\pi$ as well.}.\\

The non-linear realisation of this algebra is the \emph{scaling superluid}. It was introduced along with its derivation via the coset construction in \cite{Pajer:2018egx}. It is a special subset of the general superfluid as can be seen from the fact its algebra contains the superfluid one as a sub-algebra. It follows that under non-linear boosts $\pi$ transforms as \eqref{NLBoostsSuperfluid}. The invariant action can therefore be written in terms of $\phi = \mu t + \pi$ and it takes the  leading order form $P(X)=X^{\alpha}$ where $\alpha=\frac{2}{1+\Delta}$. In terms of $\pi$ the perturbative phonon action is \cite{Pajer:2018egx}
\begin{align}\label{eq:scaling_superfluid_lagrang}
S[\pi]=\int\d^3\bfx\, \d t \Big[ \frac{1}{2}(\dot\pi^2- \partial_i\pi \partial^i\pi)+&\frac{\alpha_1}{\mu^2}\dot\pi^3-\frac{\alpha_2}{\mu^2}\dot\pi \partial^i\pi \partial_i\pi \nonumber \\ +&\frac{\beta_1}{\mu^4}\dot\pi^4-\frac{\beta_2}{\mu^4}\dot\pi^2\partial^i\pi \partial_i\pi+\frac{\beta_3}{\mu^4}(\partial^i\pi \partial_i\pi)^2+\dots\Big]\,,
\end{align}
where $\dots$ stands for quintic and higher order terms. The Wilson couplings $\alpha_i, \beta_i$ are fully fixed in terms of one single parameter, the scaling weight $\Delta$, through the combination
\begin{equation}
c_s^2=\frac{1+\Delta}{3- \Delta}\,,
\end{equation}
which, as the notation suggests, can be identified the speed of sound $c_s$ had we not rescaled the spatial coordinates. These couplings are given by
\begin{align}\label{eq:scaling_superfluid_wilson_coef_alphas}
	\alpha_1=\frac{1}{6}\frac{1-c_s^2}{\sqrt{c_s^3(1+c_s^2)}}\,, \qquad \alpha_2=\frac{1}{2}\frac{1-c_s^2}{\sqrt{c_s^3(1+c_s^2)}}\,,
\end{align}
\begin{align}\label{eq:scaling_superfluid_wilson_coef_betas}
\beta_1=\frac{1}{24}\left(\frac{1-c_s^2}{1+c_s^2}\right)\frac{1-2c_s^2}{c_s^3}\,, \qquad  \beta_2= \frac{1}{4}\left(\frac{1-c_s^2}{1+c_s^2}\right)\frac{1-2c_s^2}{c_s^3}\,, \qquad \beta_3=\frac{1}{8c_s^3}\left(\frac{1-c_s^2}{1+c_s^2}\right)\,.
\end{align}
Absence of pathologies such as gradient instabilities, superluminality or ghosts for this theory requires $\alpha\geq 1 \Leftrightarrow -1< \Delta<1$ \cite{Pajer:2018egx} which also implies the theory satisfies positivity bounds from dispersion relations \cite{Baumann:2015nta,TG_Melville_inprep_positivity}. Also note that, in line with the conjecture of \cite{Baumann:2015nta}, all interactions vanish in the limit $c_{s} = 1$. We will discuss amplitudes and weak coupling in more detail in section \ref{Amplitudes}. In particular, we show that a necessary condition for the theory to remain weakly coupled on sub-horizon scales is $\alpha < 181$. Since all the low energy couplings are fixed in terms of $c_{s}$, this theory is on a similar footing as the exceptional EFTs of the scalar DBI and special galileon. It would be very interesting to construct the S-matrix for this theory directly from the non-linear symmetries. \\

Note that for $\Delta \neq 1$ the theory does not admit a consistent perturbation theory around the Poincar\'{e} invariant vacuum. Indeed there is no kinetic term for the fluctations. So even when the Lagrangian is written in terms of $\phi$ i.e. $\mathcal{L} = X^{\alpha}$ this should always be read as a theory for the fluctuation $\pi$ only. Finally note that both the superfluid and scaling superfluid fall into case $2$ of \cite{Nicolis:2015sra} since we have 
\begin{align} \label{SuperfluidGeneral}
\bar{P}_{0} = P_{0} + \mu Q, \qquad \bar{P}_{i} = P_{i}, \qquad \bar{J}_{i} = J_{i}.
\end{align}



\subsection{Level-2: Galileids, Conformal Superfluid and Extended Galileids} 
\label{sub:level_2_trees_galileids_and_superfluids_again}
At level-2 the transformation rules for the additional symmetries start at quadratic order in the co-ordinates. The possibilities are $t^2,\,t\,x^i,\,x^ix^j$ and so at this level the tree can in principle contain a scalar, a vector and a traceful tensor. It is simple to see that requiring $\dot\pi^2- \partial_i\pi\partial^i\pi$ to be invariant under these symmetry transformations requires the generator of the $t^2$ transformation to be proportional to the trace of the tensor that generates the $x^i\,x^j$ transformation. We therefore have three $SO(3)$ irreps that we can add at level-2 in the tree: a scalar $Z$, a vector $W_i$ and a traceless tensor $S^{T}_{ij}$. This leads to, in principle, seven possible trees each with different generator contents.\\

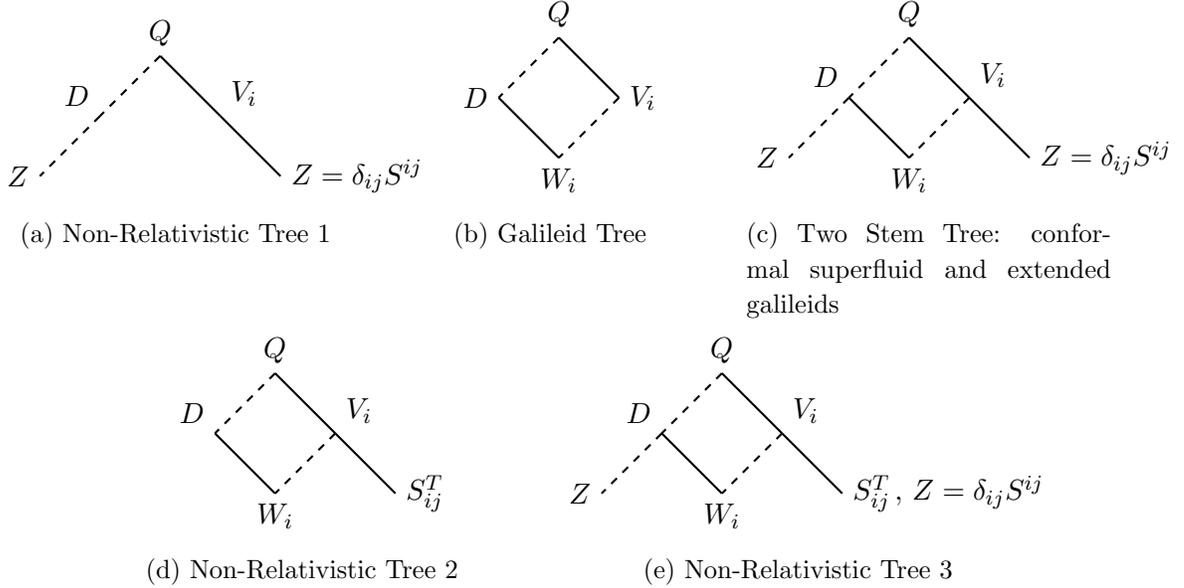
\begin{figure}[h!]
	\begin{center}
	\subcaptionbox{Non-Relativistic Tree 1 \label{subfig:level2_tree1}
}[0.3\textwidth]{
	\begin{tikzpicture}[x=0.4cm,y=0.4cm]
\tikzstyle{Pi} = [black, thick];
\tikzstyle{P0} = [black, dashed, thick];
\tikzstyle{label} = [->];
\draw (4,4) node [anchor = south, black]{$Q$};
\draw[style=Pi] (4,4) to (6,2)node [anchor = south west, black]{$V_i$};
\draw[style=P0] (4,4) to (2,2)node [anchor = south east, black]{$D$};
\draw[style=Pi] (6,2) to (8,0)node [anchor = west, black]{$Z=\delta_{ij}S^{ij}$};
\draw[style=P0] (2,2) to (0,0)node [anchor = east, black]{$Z$};
\end{tikzpicture}}
\subcaptionbox{Galileid Tree \label{subfig:galileid_tree}
}[0.3\textwidth]{
\begin{tikzpicture}[x=0.4cm,y=0.4cm]
\tikzstyle{Pi} = [black, thick];
\tikzstyle{P0} = [black, dashed, thick];
\tikzstyle{label} = [->];
\draw (4,4) node [anchor = south, black]{$Q$};
\draw[style=Pi] (4,4) to (6,2)node [anchor = west, black]{$V_i$};
\draw[style=P0] (4,4) to (2,2)node [anchor = east, black]{$D$};
\draw[style=Pi] (2,2) to (4,0)node [anchor = north, black]{$W_i$};;
\draw[style=P0] (6,2) to (4,0);
\end{tikzpicture}}
\subcaptionbox{Two Stem Tree: conformal superfluid and extended galileids \label{subfig:conf_superfluid_tree}
}[0.3\textwidth]{
	\begin{tikzpicture}[x=0.4cm,y=0.4cm]
\tikzstyle{Pi} = [black, thick];
\tikzstyle{P0} = [black, dashed, thick];
\tikzstyle{label} = [->];
\draw (4,4) node [anchor = south, black]{$Q$};
\draw[style=Pi] (4,4) to (6,2)node [anchor = south west, black]{$V_i$};
\draw[style=P0] (4,4) to (2,2)node [anchor = south east, black]{$D$};
\draw[style=Pi] (2,2) to (4,0)node [anchor = north, black]{$W_i$};;
\draw[style=P0] (6,2) to (4,0);
\draw[style=Pi] (6,2) to (8,0)node [anchor = west, black]{$Z=\delta_{ij}S^{ij}$};
\draw[style=P0] (2,2) to (0,0)node [anchor = east, black]{$Z$};
\end{tikzpicture}}
\\
\subcaptionbox{Non-Relativistic Tree 2 \label{subfig:level2_nonrel_tree2}
}[0.4\textwidth]{
	\begin{tikzpicture}[x=0.4cm,y=0.4cm]
\tikzstyle{Pi} = [black, thick];
\tikzstyle{P0} = [black, dashed, thick];
\tikzstyle{label} = [->];
\draw (4,4) node [anchor = south, black]{$Q$};
\draw[style=Pi] (4,4) to (6,2)node [anchor = south west, black]{$V_i$};
\draw[style=P0] (4,4) to (2,2)node [anchor = south east, black]{$D$};
\draw[style=Pi] (2,2) to (4,0)node [anchor = north, black]{$W_i$};;
\draw[style=P0] (6,2) to (4,0);
\draw[style=Pi] (6,2) to (8,0)node [anchor = west, black]{$S^T_{ij}$};
\end{tikzpicture}}
\subcaptionbox{Non-Relativistic Tree 3 \label{subfig:level2_nonrel_tree3}
}[0.4\textwidth]{
	\begin{tikzpicture}[x=0.4cm,y=0.4cm]
\tikzstyle{Pi} = [black, thick];
\tikzstyle{P0} = [black, dashed, thick];
\tikzstyle{label} = [->];
\draw (4,4) node [anchor = south, black]{$Q$};
\draw[style=Pi] (4,4) to (6,2)node [anchor = south west, black]{$V_i$};
\draw[style=P0] (4,4) to (2,2)node [anchor = south east, black]{$D$};
\draw[style=Pi] (2,2) to (4,0)node [anchor = north, black]{$W_i$};;
\draw[style=P0] (6,2) to (4,0);
\draw[style=Pi] (6,2) to (8,0)node [anchor = west, black]{$S_{ij}^T\,,\, Z=\delta_{ij}S^{ij}$};
\draw[style=P0] (2,2) to (0,0)node [anchor = east, black]{$Z$};
\end{tikzpicture}}
\caption{Level-2 Inverse Higgs Trees}
\label{fig:Level-2_Trees}
\end{center}
\end{figure}

However, not all of these can be related to a relativistic algebra by a basis change. Indeed, the number of generators must be sufficient to form $ISO(1,3)$ representations out of the $SO(3)$ ones. In particular, the presence of a traceless $SO(3)$ tensor requires a $SO(3)$ vector and a $SO(3)$ scalar to form a traceless Lorentz tensor $S_{\mu \nu}^{T}$. It follows that level-2 trees containing $S_{ij}^T$ must also contain $W_{i}$. The level-$1$ vector $V_i$ is always required to be present to satisfy the inverse Higgs relations to remove the would-be Goldstone associated with $S_{ij}^T$. This reduces the number of possible trees from seven to five and they are represented in Figure \ref{fig:Level-2_Trees}. Following the strategy outlined at the end of Section \ref{subsub:tree_structure}, we find that three of these trees (\ref{subfig:level2_tree1}, \ref{subfig:level2_nonrel_tree2} and \ref{subfig:level2_nonrel_tree3}) cannot be related to relativistic algebras after they have been constrained by Jacobi identities. In particular they do not contain a linear combination of the vectors that can be associated with boosts. Among the other two trees (\ref{subfig:galileid_tree} and \ref{subfig:conf_superfluid_tree}) we find three distinct possibilities yielding three classes of phonon EFTs. Two of these have been discussed in the literature before. These are the galileid and the conformal superfluid. The third, however, which we call the extended galileid is a new theory which has not been discussed in the literature before. 

\subsubsection*{Galileid} 
\label{sub:galileid}
First consider tree \ref{subfig:galileid_tree}. After writing down the most general commutation relations consistent with the tree structure and solving all the constraints imposed by Jacobi identities we find the following non-trivial commutation relations
\begin{align}\nn
	\comm{\bar P_0}{D}&= - Q && \comm{\bar P_i}{V_j}=\delta_{ij} Q  \\\nn
	\comm{\bar P_0}{W_i}&= \alpha V_i+\bar P_i && \comm{\bar P_i}{W_j}=\delta_{ij}(\bar P_0 - \alpha D)\\
	\label{eq:galileid_algebra}
	\comm{D}{W_i}&= V_i && \comm{V_i}{W_j}=\delta_{ij}D\\\nn
	\comm{W_i}{W_j}&=-\epsilon_{ijk}\bar{J}^{k}.
\end{align} 
After the basis change
\begin{equation}
\bar P_0= P_0 + \alpha D\,, \qquad \bar P_i=P_i\,,  \qquad \bar J_i=J_i\,,
\label{eq:galileid_change_basis_1}
\end{equation}
and with the identifications $P_\mu=(P_0,P_i)$, $V_\mu:=(D,V_i)$ and $W_i= K_i$, we can bring this algebra into a relativistic form with a $ISO(1,3)$ sub-algebra and 
\begin{align}
\comm{P_\mu}{V_\nu}&=\eta_{\mu \nu}Q\,.
\label{eq:galileon_algebra}
\end{align}
All other commutators, apart from those with $M_{\mu\nu}$ that define the Lorentz representation of each generator, are zero. We recognise this as the galileon algebra $\mathfrak{Gal}(3+1,1)$\cite{Goon:2012dy} which with linearly realised boosts is realised by the well-known galileon EFT \cite{galileon}. The leading Lagrangian in four spacetime dimensions is $\mathcal{L} = \sum_{n=1}^{5} g_{n} \mathcal{L}_{n}$ where
\begin{align} \label{galileonLagrangian}
\mathcal{L}_{1}  &= \phi, \nonumber \\
\mathcal{L}_{2}  &=  (\partial \phi)^{2}, \nonumber \\
\mathcal{L}_{3}  &= (\partial \phi)^{2} \Box \phi,\nonumber \\
\mathcal{L}_{4}  &= (\partial \phi)^{2} ((\Box \phi)^{2} - \partial^{\mu}\partial^{\nu}\phi \partial_{\mu}\partial_{\nu}\phi),  \nonumber \\
\mathcal{L}_{5}  &= (\partial \phi)^{2} ((\Box \phi)^{3} - 3 \Box \phi \partial^{\mu}\partial^{\nu}\phi \partial_{\mu}\partial_{\nu}\phi + 2 \partial^{\mu}\partial^{\nu}\phi \partial_{\nu}\partial_{\rho}\phi \partial^{\rho}\partial_{\mu}\phi).
\end{align}
If we take $\phi$ to have canonical mass dimension and the $g_{i}$ to be dimensionless, we need to add a mass scale $\Lambda$ to $\mathcal{L}_{1}, \mathcal{L}_{3-5}$. Throughout we will work in units where $\Lambda = 1$. To realise a Poincar\'{e} invariant vacuum $\phi=0$ we would fix $g_{1} = 0$. The symmetry generated by $V_{\mu}$ is $\delta_{V_{\mu}} \phi = x^{\mu}$. The special galileon, which enjoys an additional symmetry around the Poincar\'{e} invariant vacuum, has $g_{1} = g_{3} = g_{5} = 0$ \cite{SG}. Note that these operators are invariant under $\delta_{V_{\mu}} \phi = x^{\mu}$ up to a total derivative. Indeed, these are all Wess-Zumino terms \cite{WessZuminoGal} and strictly invariant operators require at least two derivatives per field. Interestingly, these Wess-Zumino operators are not renormalised in perturbation theory \cite{ClassicalAndQuantum,GoonAspects}. This follows from the structure of each term which, ignoring the tadpole, is schematically $(\partial \phi)^{2} (\partial \partial \phi)^n$ where $n \geq 0$ and $(\partial \partial \phi)^n$ are total derivatives. In $D$ spacetime dimensions only the first $D$ of these operators are non-zero and therefore we have a total of $D+1$ Wess-Zumino terms once we include the tadpole. When $n \geq D$ the operators vanish identically. See \cite{Aspects, galileonReview} for more details.  \\

When boosts are spontaneously broken the algebra \eqref{eq:galileid_algebra} is non-linearly realised by the \emph{galileid} which is the galileon EFT expanded around a Lorentz breaking vacuum. Indeed, notice that from (\ref{eq:galileon_algebra}) (plus $ISO(1,3)$) there are several ways to arrive at the galileid algebra (\ref{eq:galileid_algebra}). The most general change of basis is
\begin{equation}
\bar P_0= P_0 + \alpha D\,, \qquad \bar P_i=P_i + \beta V_i\,,  \qquad \bar J_i=J_i\,,
\label{eq:galileid_general_basis_change}
\end{equation}
which amounts to replacing $\alpha \to \alpha- \beta$ in (\ref{eq:galileid_algebra}) with $\alpha \neq \beta$ for the inverse Higgs constraints to be satisfied. From the point of view of the relativistic galileon theory the symmetry breaking patterns associated with the change of basis (\ref{eq:galileid_general_basis_change}) are generated by the background
\begin{equation}
	\expval{\phi(x)}=\frac{1}{2}\left(\beta \bfx^2- \alpha t^2\right),
	\label{eq:galileon_background}
\end{equation}
and indeed, as confirmed by the inverse Higgs tree, when $\alpha= \beta$, $\expval{\phi(x)}\propto x_\mu x^\mu$ is Lorentz invariant and boosts are not broken. Once plugged into the galileon's equation of motion we find a polynomial equation for $\alpha,\beta$ which can be solved for $\alpha = \alpha(\beta)$. Note that this background does not break translations thanks to the galileon symmetry $\delta_{V_{\mu}} \phi = x^{\mu} = \delta_{V_{\mu}} \pi$. The non-linear boosts are generated by $W_{i}$ which appears at level-2 in the tree meaning that its symmetry transformation on $\pi$ differs from \eqref{boosttra}. Indeed we have
\begin{align}
\delta_{W_{i}} \pi = (\beta-\alpha)tx^{i} + t \partial^{i} \pi + x^{i} \dot{\pi}.
\end{align}
When $\alpha = 0$ the theory is referred to as a \textit{type-I galileid} and is in case $3$ of \cite{Nicolis:2015sra} while for $\alpha, \beta \neq 0$ it is referred to as \textit{type-II galileid} in case 5. However, as we have seen here the two algebra's are trivially the same once we redefine the parameters and therefore the phonon symmetries in the two cases are the same. Indeed, only the combination $\alpha -\beta$ appears in the symmetry transformations. So actually there is no difference between type-I and type-II at the level of the phonon EFT. We refer the reader to \cite{Nicolis:2015sra} for further galileid details.


\subsubsection*{Conformal Superfluid} 
\label{sub:conformal_superfluid}
Now consider the tree represented on Figure \ref{subfig:conf_superfluid_tree} which contains five non-linear generators: $Q, D, V_i, Z$ and $W_i$. After writing down the most general commutators consistent with the tree and solving the constraints imposed by Jacobi identities we find two different stems depending on the coefficient of
\begin{equation}
	\label{eq:steming_condition}
	\comm{\bar P_i}{D}=a_1 \bar P_i\,. 
\end{equation}
Consider first the case $a_1\neq0$ where we set $a_{1}=1$ WLOG. We find that the unique algebra which is a basis change away from a relativistic one is
\begin{align}\nn
	\comm{\bar P_0}{D}=&\bar P_0 - \mu Q && \comm{\bar P_i}{D}=\bar P_i\\\nn
	\comm{\bar P_0}{V_i}=&\bar P_i && \comm{\bar P_i}{V_j}=\delta_{ij}(\bar P_0 - \mu Q) \\\nn
	\comm{\bar P_0}{Z}=&-2D && \comm{\bar P_i}{Z}=-2V_i\\
	\label{eq:conf_superfluid_algebra}
	\comm{\bar P_0}{W_i}=&2 V_i && \comm{\bar P_i}{W_j}=2(\delta_{ij} D + \epsilon_{ijk}\bar{J}^{k})\\\nn
	\comm{D}{Z}=&Z  && \comm{D}{W_i}=W_i \\\nn
	\comm{Z}{V_i}=&W_i && \comm{W_i}{V_j}=\delta_{ij}Z\\\nn
	\comm{V_i}{V_j}=&-\epsilon_{ijk}\bar{J}^{k}.
\end{align}
Again here we have only included the non-zero and non-trivial commutators. Note that $Q$ commutes with all generators. This algebra can be brought into a relativistic form by the superfluid basis change \eqref{SuperfluidGeneral} and by identifying $V_i$ with $K_i$ and defining $W_\mu:=(Z,W_i)$. The algebra becomes $SO(2,4) \times U(1)$ i.e. it is the four-dimensional conformal algebra augmented with a single $U(1)$ generator. Here $W_{\mu}$ is the generator of special conformal transformations. In the broken phase this algebra is non-linearly realised by the \emph{conformal superfluid} which is a sub-set of the more general scaling superfluid \eqref{eq:scaling_superfluid_lagrang} with scaling weight $\Delta=0$. Notice a key difference here compared to the galileid: when the algebra is brought into a relativistic form there are not enough inverse Higgs constraints to arrive at a theory of a Poincar\'{e} invariant single scalar. Indeed, when written in terms of a $P(X)$ theory we have $P = X^2$ which is not a perturbative EFT around the $\phi=0$ vacuum. \\

The conformal superfluid first appeared in the literature in the context of large U(1) charge operators in CFTs \cite{Hellerman:2015nra,Monin:2016jmo,Cuomo:2017vzg} and in the context of Holography in \cite{Esposito:2016ria}. The exact Maurer-Cartan form for this symmetry breaking pattern can be found in \cite{Pajer:2018egx}, and once the inverse Higgs constraints are solved to remove the inessential Goldstones the leading order action is given by \eqref{eq:scaling_superfluid_lagrang} with $c_{s}^2 = 1/3$. We will discuss the weak coupling regime of the conformal superfluid in section \ref{Amplitudes} where we show that it can be weakly coupled in the sub-horizon regime of the EFT of inflation.\\

One may wonder how a theory without the dilaton can non-linearly realise the conformal algebra. However, as was explained in \cite{Pajer:2018egx} one can derive the conformal superfluid by taking the dilaton theory and adding a shift symmetric scalar as a matter field. The couplings between this matter field and the dilaton are fixed by conformal symmetry but when the matter Lagrangian is simply $X^2$ the dilaton drops out leaving us with a conformal theory of a shift symmetric scalar that admits a Lorentz breaking phase. More generally, in $D$ spacetime dimensions the leading order conformal superfluid theory is $P(X) = X^{D/2}$.

\subsubsection*{Extended Galileids} 
\label{sub:second_galileids_}
We now take $a_1=0$ in (\ref{eq:steming_condition}). After imposing the constraints from Jacobi identities and asking the algebra to be a basis change away from a relativistic one we find that the non-trivial commutators in the broken phase are
\begin{align}\nn
\comm{\bar P_0}{D}=&-Q && \comm{\bar P_i}{V_j}=\delta_{ij}Q\\\nn
\comm{\bar P_0}{W_i}=&\bar{P}_i+V_i && \comm{\bar P_i}{W_j}=\delta_{ij}(\bar P_0-D)\\\nn
\comm{\bar P_0}{Z}=&(a - b -1)D+b\bar P_0 && \comm{\bar P_i}{Z}=-V_i\\
\comm{D}{W_i}=&V_i && \comm{V_i}{W_j}=\delta_{ij}D \\\nn
\comm{D}{Z}=&(a- b)D+b\bar P_0 && \comm{V_i}{Z}=a V_i+b\bar P_i\\\nn
\comm{Q}{Z}=&a\,Q && \comm{W_i}{W_j}=- \epsilon_{ijk} \bar{J}^{k},
\end{align} 
and after performing the following change of basis 
\begin{equation}
\bar P_0= P_0 + D\,, \qquad \bar P_i=P_i \,,  \qquad \bar J_i=J_i\,,
	\label{eq:extended_galileid_basis_change_1}
\end{equation}
and identifying $P_\mu=(P_0,P_i)$ and $V_\mu=(D,V_i)$, we find the relativistic algebra
\begin{align}\nn
\comm{P_\mu}{V_\nu}=&\eta_{\mu \nu}Q\,, &&\comm{P_\mu}{Z}=-V_\mu\,, \\
\comm{V_\mu}{Z}=&a V_\mu  + b P_\mu  \,, && \comm{Q}{Z}=a\, Q \,.
\label{eq:extended_relativistic_galileon_algebra}
\end{align}
We see that the galileon algebra is a sub-algebra and therefore the theory that non-linearly realises this algebra is expected to be a galileon theory with couplings fixed by the additional symmetry generated by $Z$. The phonon theory, which we call the \textit{extended galileid}, is then obtained by expanding around the galileid background \eqref{eq:galileon_background}. Note that we have performed a basis change to remove a possible $P_{\mu}$ term from the RHS of $[P_{\mu},Z]$. This basis change amounts to the galileon duality \cite{GalDual1,GalDual2} so by writing the algebra in this form we have already accounted for this duality redundancy.\\

The transformation of $\phi$ under $V_{\mu}$ is the galileon transformation $\delta_{V_{\mu}} \phi = x^{\mu}$ and as can be read off from the algebra \eqref{eq:extended_relativistic_galileon_algebra}, the transformation generated by $Z$ is\footnote{We remind the reader that we are working in units where the dimensionful scale $\Lambda=1$. Also note that if we kept a $P_{\mu}$ piece on the RHS of $[P_{\mu},Z]$ we would have a $x^{\mu} \partial_{\mu}\phi$ piece in the transformation rule. This is removed by the galileon duality.}
\begin{equation}
\delta_Z\phi= -\frac{1}{2}x_\mu x^\mu+ a \phi - \frac{1}{2}b (\partial\phi)^2\,.
\label{eq:extended_galileon_symmetry}
\end{equation}
This symmetry is reminiscent of the special galileon symmetry but is indeed different since that symmetry is generated by a traceless tensor $S_{\mu\nu}$ and acts as $\delta_{S_{\mu\nu}} \phi = x^{\mu}x^{\mu} + \partial^{\mu}\phi \partial^{\nu}\phi$ \cite{SG}. We will discuss the special galileon theory further below. Our new galileid algebra was also discussed in \cite{BraunerBogers} but not in the context of Lorentz breaking vacua and the phonon EFT as we will do here. Indeed, in that work the authors were interested in Poincar\'{e} invariant scalar EFTs. Now when we expand the theory around \eqref{eq:galileon_background}, this new symmetry is realised on the phonon as 
\begin{align} \label{DeltaZPi}
\delta_{Z} \pi = \frac{1}{2}(1 - a \alpha + b \alpha^2)t^2 -\frac{1}{2}(1 - a \beta + b \beta^2)\bfx^2 + a \pi - b \alpha t \dot{\pi} -  b \beta x^{i}\partial_{i}\pi - \frac{b}{2} (\partial \pi)^2.
\end{align} 
We now ask that \eqref{eq:extended_galileon_symmetry} is a symmetry of \eqref{galileonLagrangian}. Clearly it has the right structure. Indeed, in the galileon Lagrangian each term differs from the next by one power of the field and two derivatives. If we count explicit co-ordinate dependence as one negative power of the derivatives then we see that \eqref{eq:extended_galileon_symmetry} has precisely the required structure to relate different galileon operators since each term also differs from the next by one power of the field and two derivatives. Now the field-independent part of the symmetry transformation must be a symmetry of the operator with the fewest powers of $\phi$ since there is no other way it can be cancelled, and similarly the part of the transformation with the largest number of fields must be a symmetry of the operator with the largest number of fields. We find that only $\mathcal{L}_{1}$ possess a $\delta \phi = x_\mu x^\mu$ symmetry and only $\mathcal{L}_{5}$ has a $\delta \phi = (\partial \phi)^2$ symmetry. We therefore require $g_{1} \neq 0$ and $g_{5} \neq 0$ (if $b \neq 0$). The reason that $\mathcal{L}_{5}$ is invariant under the last term in the symmetry transformation is that its variation generates the would-be sixth galileon term i.e. an operator of the schematic form $(\partial \phi)^2 (\partial \partial \phi)^{4}$ where $(\partial \partial \phi)^{4}$ is a total derivative. However, as we mentioned above, in four spacetime dimensions such an operator vanishes identically and so $\mathcal{L}_{5}$ is invariant. For more details see \cite{SG}. \\

\begin{table}[h!]
\centering
	  \begin{adjustbox}{max width=\textwidth}
		\begin{tabular}{| c | c | c |}
		\hline
 		\multicolumn{3}{|c|}{Extended Galileons} \\
 		\hline
		Name &  Galileon Lagrangian \eqref{galileonLagrangian} & Extra Symmetry $\delta_Z\phi$ \\
		\hline
		& & \\[-5pt]
		Pure Tadpole & $\L_1$ & $-\frac{1}{2}x_\mu x^\mu$  \\[10pt]
		\hline
		& &  \\[-10pt]
		\makecell{Type-$I_{\pm}$} &\makecell{$\L_1 \pm \L_3 - 3 \L_5$} & $-\frac{1}{2}x_\mu x^\mu\pm6(\partial\phi)^2$ \\[10pt]
		\hline
		& & \\[-10pt]
		\makecell{Type-$II_{\pm}$} & \makecell{$\L_1 \pm \frac{1}{2}\L_2 +2 \L_3 \mp \frac{20}{3} \L_4 + \frac{32}{3}\L_5$} & $-\frac{1}{2}x_\mu x^\mu\mp4\phi+16(\partial\phi)^2$ \\[10pt]
		\hline
		& & \\[-10pt]
		\makecell{Type-$III_{\pm}$} & \makecell{ $\L_1 \pm\frac{1}{2}\L_2 -\frac{1}{4} \L_3 \pm\frac{1}{12}  \L_4 - \frac{1}{48}\L_5$} &$-\frac{1}{2}x_\mu x^\mu\mp4\phi+\frac{5}{2}(\partial\phi)^2$\\[10pt]
		\hline
	\end{tabular}
	\end{adjustbox}
	\caption{Galileon Lagrangians $\L=\sum_n g_n\,\L_n$ with the extra scalar symmetry $\delta_Z\phi=-\frac{1}{2}x_\mu x^\mu+ a\phi - b (\partial\phi)^2/2 $, in units where $\Lambda=1$.}
	\label{tab:Extended_galileons.}
\end{table}

With these constraints we then look for symmetries of the full Lagrangian. We find that there are \textit{seven} possibilities with $({a,b,g_{i}})$ fixed in each case\footnote{Our results are only valid in $3+1$ dimensions but there are indeed other invariant Lagrangians in other dimensions.}. These are summarised in table \ref{tab:Extended_galileons.}. WLOG we have rescaled $\phi$ to set $g_{1} =1$. Similarly, we have rescaled the co-ordinates to fix the magnitude of $g_{2}$ in the type-$II$ and type-$III$ cases and have fixed the magnitude of $g_{3}$ in the type-$I$ case since here $g_{2} = 0$. These rescalings do not affect the transformation rule since we can redefine $(Z,a,b)$ too. We remind the reader that these symmetries were not seen in  \cite{RSW1} and this is precisely due to the presence of the tadpole i.e. the absence of a Poincar\'{e} invariant vacuum\footnote{One may want to expand around the galileid vacuum with $\alpha=\beta$ which is secretly Poincar\'{e} invariant thanks to the galileon symmetry. This eliminates the tadpole but the $\delta_{Z} \pi$ transform does not have a $\pi$-independent part in this case and therefore we do not have all the necessary inverse Higgs constraints and this symmetry does not lead to a generalised Adler zero, consistent with amplitude results \cite{GeneralisedAdler1}.}. \\ 

For type $I_{\pm}-III_{\pm}$ we now constrain $\alpha$ and $\beta$ such that \eqref{eq:galileon_background} is a solution of the theory followed by expanding around this solution to find the quadratic theory for the phonon $\pi$. Note that there is no linear term in $\pi$ since its coefficient is the $\phi$ equation of motion. Our results are summarised in table \ref{tab:extended_galileids} where we present the background solution, condition for absence of ghosts and the speed of sound.\\

\begin{table}[h!]
\centering
	  \begin{adjustbox}{max width=\textwidth}
		\begin{tabular}{| c | c | c | c |}
		\hline
 		\multicolumn{4}{|c|}{Extended Galileids} \\
 		\hline
		Theory & Background Solutions ($\alpha \neq \beta$) & Speed of Sound $c_{s}^2$ & Absence of Ghosts\\
		\hline
		& & &  \\[-5pt]
		\multirow{2}{*}{Type-$I_{\pm}$} & $\alpha \beta = \pm\frac{1}{12}$ & $\pm\frac{1}{36 \beta^2}$ & $ \pm\beta(1\mp12 \beta^2)>0$\\[10pt]
		\cline{2-4}
		& & &  \\[-2pt]
    	& $\beta^2=\pm\frac{1}{12}$ & $\infty$ & Non-Dynamical \\ [10pt]
		\hline
		& & &\\[-5pt]
		\multirow{2}{*}{Type-$II_{\pm}$} & $\beta = \pm\frac{1}{4}$ & $\infty$ &  Non-Dynamical \\[10pt]
		\cline{2-4}
		& & &  \\[-2pt]
    	& $\beta\neq \mp\frac{1}{32}\quad \text{and}\quad \alpha=\frac{1\pm5 \beta}{(32\beta\pm1)}$ & $\frac{9}{(1\pm32 \beta)^2}$ & $\beta > \pm\frac{1}{32}$\\ [10pt]
		\hline
		& & &  \\[-5pt]
		\multirow{2}{*}{Type-$III_{\pm}$} & $\beta = \pm1$ & $\infty$ & Non-Dynamical \\[10pt]
		\cline{2-4}
		& & & \\[-2pt]
    	&  $\alpha=\pm1$ & $0$ & $\beta< \pm1$ \\ [10pt]
    	\hline
	\end{tabular}
	\end{adjustbox}
	\caption{Dynamics of extended galileid phonons around the background solution $\expval{\phi(x)}=\frac{1}{2}\left(\beta|\bfx|^2- \alpha t^2\right)$ with $\alpha \neq \beta$. We show the speed of progagation of the phonons $c_s^2$ around each background together with the condition for the absence of ghosts instabilities.}
	\label{tab:extended_galileids}
\end{table}

We find that all six possible Lagrangians admits two background solutions, and out of these twelve possibilities only three admit fluctuations with a non-zero and non-negative $c_{s}^2$. In those three cases the phonon is therefore dynamical and doesn't have gradient stabilities. Requiring furthermore the absence of ghosts (i.e. a  positive coefficient for $\dot\pi^2$) constrains the allowed value of the background parameter $\beta$ in each case. This extra condition puts bounds on the allowed values of $c_{s}^2$. For example, for Type-$I_{+}$ we see that if $\beta < 0$ we have $c_{s}^2 < 1/3$ while if $\beta > 0$ we have $c_{s}^2 > 1/3$. So in these theories the speed of sound cannot get too large (or small) without propagating ghosts. It would be very interesting to study these theories in more detail especially within the context of galileon inflation \cite{GalInflation1,GalInflation2} and weakly broken galileon symmetry \cite{WeaklyBroken}, and to construct the S-matrices directly from the symmetries. We plan to do this in future work.

\subsection{Broken Phase of Extended Shift Symmetric Theories}
\label{sub:broken_phase_of_ext_shift_theories}
In the Poincar\'{e} invariant cases there are an infinite number of Abelian algebras corresponding to extended shift symmetries. These symmetries are only functions of the coordinates rather than the fields themselves. At the level of the algebra the commutators between non-linear generators are zero. Although we don't have general Lagrangians to work with it is simple to see at the level of the algebras that such theories admit a consistent broken phase with boosts non-linearly realised and some form of spacetime translations linearly realised. The only non-trivial commutators are schematically $[P_{\mu},G^m] \supset G^{m-1}$ for all $m \leq n$ where $n$ is the level at which the tree is tunkated. The galileon we have just discussed is the $n=1$ case. Each $G^m$ is a traceless tensor as explained above with $m$ indices and so the Poincar\'{e} invariant trees are given by Figure \ref{fig:scalartree} reduced to the far diagonal. Now we would like to define new translations $\bar{P}_{0}, \bar{P}_{i}$ such that the boost Goldstones are inessential.   \\

All covariant derivatives other than those corresponding to the top level generator need to be set to zero to solve for inessential Goldstones and so these cannot be used to remove the Goldstones of boosts. The only option is therefore to use the top inessential Goldstone. In all cases the top level generator $G^n$ contains a $SO(3)$ scalar and a $SO(3)$ vector which we call $G^n_{0}$ and $G^n_{i}$ respectively. Now if we define 
\begin{align}
\bar{P}_{0} = P_{0} + \alpha G^n_{0}, \qquad \bar{P}_{i} = P_{i} + \beta G^n_{i},
\end{align}
then $\bar{P}_{0}$ and $\bar{P}_{i}$ still satisfy the commutator relations of spacetime translations and we have 
\begin{align}
[\bar{P}_{0}, K_{i}] \supset G^n_{i}, \qquad  [\bar{P}_{i}, K_{j}] \supset \delta_{ij}G^n_{0},
\end{align}
such that the boost Goldstones are inessential. There are infinitely many of these algebras and each have non-vanishing commutators between non-linear generators since in the relativistic basis the non-linear generators are Lorentzian irreps. This is the primary reason why an exhaustive classification was not possible in this paper, in contrast to the Poincar\'{e} case \cite{RSW1}. However, it would be interesting to consider other distinctions between algebras that could allow for an exhaustive classification. Note that in each case the generators $K_{i}$ appear at the top of the trees. It would be interesting to investigate concrete examples of this further by looking for explicit symmetry breaking vacua within one of these theories. However, each case with $n \geq 2$ is likely to contain ghosts unless new degrees of freedom are included. One possibility might be to IR complete this theories into massive spinning multiplets. See e.g. \cite{BrandoHigherSpin} for details. 

\subsection{Obstacles for Exceptional EFTs}
\label{sub:obstacles_for_excp_EFTs}

Looking back at the theories we have found up to Level-2, one may be surprised that we did not find the exceptional EFTs that are scalar DBI and the special galileon. Indeed, taking DBI for instance, which is a superfluid with $P(X)=-\sqrt{1+X}$ and non-linearly realised $ISO(1,4)$ symmetry, one would expect to find a theory for the phonon $\pi$ on the homogeneous background $\expval{\phi}=\mu t$  which (\emph{seemingly}) breaks boosts. Thus we should find amongst the Level-2 trees above (DBI phonons have two non-linear vector generators) a theory with relativistic symmetry algebra $ISO(1,4)$. In the same vein, one would naively expect to find a special galileid theory corresponding to the special galileon expanded around a background $\expval{\phi(x)}=\frac{1}{2}\left(\beta|\bfx|^2- \alpha t^2\right)$. However, in both cases the background solutions do \emph{not} break boosts while maintaining all non-linearly realised symmetries. The additional non-linearly realised symmetries of these theories protect boosts: the backgrounds are secretly Poincar\'e invariant\footnote{We thank Riccardo Penco for helpful discussions about this point.}.\\

This is actually more general than we have just alluded to. It turns out that \emph{any background} solution to the equations of motion of the leading invariant operators, which preserves some (diagonal) form of space and time translations, of these two exceptional EFTs has Lorentz boosts linearly realised (other than a special case of the special galileon which we discuss in more detail in appendix \ref{sec:special_galileon_broken_phase}). There are two ways to see this:
\begin{enumerate}[label=\roman*)]
\item Look for a vacuum solution which breaks boosts but impose that space and time translations are preserved thanks to the symmetries of the theory.
\item Take the symmetry algebra and look for basis changes that generate the new inverse Higgs constraints such that we can solve for the inessential boosts Goldstones.
\end{enumerate}
Below we follow the first method and show that a broken phase does not exist for these exceptional EFTs. In section \ref{sec:DBI_Inflation} we provide a much more thorough analysis for scalar DBI and outline the consequences for cosmology. 
\subsubsection*{DBI} 
Let us start with scalar DBI. This is a well known $P(X)$ theory with the additional symmetry
\begin{align}
\delta_{V_{\mu}} \phi = x^{\mu} + \phi \partial^{\mu} \phi. 
\label{eq:DBI_symm_transfo}
\end{align}
Along with the Poincar\'{e} symmetries and the shift symmetry $\delta_{Q} \phi = 1$ the full algebra is $ISO(1,4)$. The leading order action is 
\begin{align} \label{DBILag}
\mathcal{L}_{\text{DBI}} = - \sqrt{1 + X},
\end{align}
and higher order corrections can be found in \cite{deRham:2010eu}. Again we have set the dimensionful scale $\Lambda=1$. Now we would like to find a vacuum solution to this theory that breaks Lorentz boosts but preserves some form of spacetime translations thanks to the non-linearly realised symmetries. One might first want to expand the theory around the superfluid vacuum $\expval{\phi} = \mu t$. However this solution does not break boosts. Indeed, under a Lorentz transformation $\phi$ transforms as $\delta_{M_{\mu\nu}} \phi = x^{[\mu} \partial^{\nu]}\phi$ and therefore under a boost we have $\delta_{K_{i}} \expval{\phi} = \mu x^{i}$. However, thanks to \eqref{eq:DBI_symm_transfo} we also have $\delta_{V_{i}} \expval{\phi} =  x^{i}$ and therefore a linear combination of these two transformations leaves the vacuum invariant. We provide more details in section \ref{sec:DBI_Inflation} where we show that the scattering amplitudes are Lorentz invariant. \\

Let's now look at a more general $SO(3)$ invariant solution $\expval{\phi} = f(t,\bfx^2)$. Above we showed that using the shift symmetry to preserve time translations does not yield a Lorentz breaking vacuum so lets now try to use the symmetries in \eqref{eq:DBI_symm_transfo} to preserve some form of translations. In general we have the following transformations of $\expval{\phi}$:
\begin{align}
\delta_{\bar{P}_{0}} \expval{\phi} = - \dot{f}, \quad \delta_{\bar{P}_{i}} \expval{\phi} = - 2 f' x^{i}, \quad  \delta_{Q} \expval{\phi} = 1, \quad \delta_{V_{0}} \expval{\phi} = t - f \dot{f} \quad  \delta_{V_{i}} \expval{\phi} = x^{i}(1 + 2 f f'),
\end{align}
and therefore if a combination of the above is to leave the vacuum invariant under a time translation we require
\begin{align}\label{DBITime}
\dot{f} + a(t - f \dot{f}) + b=0,
\end{align}
and 
\begin{align} \label{DBISpace}
2 f'+ c (1 + 2 f f')=0,
\end{align}
if the vacuum is to be invariant under a spatial translation. Here $a,b,c$ are constants and a $'$ denotes a derivative with respect to $\bfx^2$. First consider \eqref{DBITime}. If $a = 0$ we have 
\begin{align}
f(t,\bfx^2) = -bt + d_{1}(\bfx^2),
\end{align}
while for $a \neq 0$ we have 
\begin{align} \label{fd2}
f(t,\bfx^2) = -\frac{1}{a}\sqrt{d_{2}(\bfx^2) + 2abt + a^2 t^2},
\end{align}
where we have subtracted a constant since it trivially solves the background equation of motion and is Poincar\'{e} invariant. Let's now turn to \eqref{DBISpace}. If $a=0$ then we clearly need $c \neq  0$ since if we also had $c=0$ we would be reduced to the superfluid vacuum which we have already seen is secretly Poincar\'{e} invariant. With $a=0$ and $c \neq 0$ we find that $f(t,\bfx^2)$ becomes
\begin{align}
f(t,\bfx^2) = \frac{1}{c}\sqrt{e-c^2 \bfx^2},
\end{align}
where again we have dropped a constant and $e$ is a new integration constant which must be non-zero such that the equation of motion from \eqref{DBILag} is non-singular. However, this form of $f$ does not solve the background equation of motion which takes the form
\begin{align}
\partial_{\mu}[(1 + X)^{-1/2} \partial^{\mu} \phi] = 0.
\end{align}
We therefore need to take $a \neq 0$. We again have two options corresponding to $c=0$ and $c \neq 0$. Taking first $c = 0$ we see that $d_{2} (\bfx^2)$ must be constant in which case \eqref{fd2} is homogeneous but does not solve the background DBI equation of motion for constant $a$ and $b$. We therefore need to take $c \neq 0$. In this case, starting from (\ref{fd2}) we have
\begin{equation}
f'(t,\bfx^2)=\frac{1}{2}\frac{d_2'(\bfx^2)}{a^2f(t,\bfx^2)}\,,
\end{equation}
and plugging this into (\ref{DBISpace}), and assuming $f(t,\bfx^2)\neq0$, we have 
\begin{equation}	
\label{eq:aneq0cneq0}
f(t,\bfx^2)\left(1+\frac{d_2'}{a^2}\right)+\frac{d_2'}{c\,a^2}=0\,.
\end{equation}
Since $d_2=d_2(\bfx^2)$ there is no way for the time dependence of the first term to be cancelled (when $a\neq0$). So we need $d_2'=-a^2$ but then the equation can't be satisfied: there are no solutions to (\ref{eq:aneq0cneq0}).  We therefore conclude that there are no vacuum solutions to the scalar DBI theory that preserve some form of spacetime translations and rotations but spontaneously break Lorentz boosts.\\

We have also confirmed this using method ii) outlined above. The generator content of the DBI theory suggests it should correspond to the tree in Figure \ref{subfig:galileid_tree}. However, we have checked that it is impossible to bring the algebra into this form by a basis change and therefore it is impossible to realise all the necessary inverse Higgs constraints to reduce to a single scalar. \\

\subsubsection*{Special Galileon} 
The special galileon is the only other exceptional EFT in the Poincar\'{e} invariant case \cite{RSW1}. What makes this galileon theory special with respect to the more general galileon theory \eqref{galileonLagrangian} is that it is invariant under
\begin{align}\label{eq:special_galileon_transfo}
\delta_{S_{\mu\nu}} \phi =x^{\mu}x^{\nu} + \partial^{\mu}\phi \partial^{\nu} \phi,
\end{align}
in addition to the shift symmetry and the usual galileon symmetry $\delta_{V_{\mu}}\phi = x^{\mu}$. The symmetry generator $S_{\mu\nu}$ is symmetric and traceless. This is a symmetry for $g_{2} = -1/2, g_{4} = 1/12, g_{1} = g_{3} = g_{5} = 0$ \cite{SG}. Now is there a vacuum solution to the special galileon theory that breaks Lorentz boosts while preserving some form of spacetime translations?\\

First consider the superfluid background $\expval{\phi} = \mu t$. This is clearly a solution to the theory since it is derivatively coupled. However, the transformation of this background under a boost can be cancelled by $\delta_{V_{i}} \phi$ as we saw for DBI above. Next consider the galileid solution $\expval{\phi(x)}=\frac{1}{2}\left(\beta \bfx^2- \alpha t^2\right)$. Under a boost we have $\delta_{K_{i}}\expval{\phi} = (\beta-\alpha)tx^{i}$ but we also have $\delta_{S_{0i}} \expval{\phi} = (1+\alpha \beta)tx^{i}$. If $\alpha \beta \neq -1$ a linear combination of these leaves the vacuum invariant and so boosts are not broken. If we instead have $\alpha \beta =-1$ (which can indeed solve the background equation of motion) then this part of the special galileon symmetry no longer has a field-independent term. It is for this reason we did not find this theory above and indeed there is no sense in which it would correspond to a special galileid theory: it would have fewer non-linearly realised symmetries. Note that this is an example of the possibility we alluded to in section \ref{subsub:tree_structure} where symmetries outside of the tree structure (i.e. additional linearly realised symmetries) could be used to form $SO(1,3)$ irreps out of $SO(3)$ ones. As we explained there, we believe this deserves further attention as a concrete example of additional linearly realised symmetries when boosts are spontaneously broken. We provide some further details in appendix \ref{sec:special_galileon_broken_phase}. \\

As we did for DBI let's now look at a more general background and ask if boosts can be broken but with some form of spacetime translations preserved. Writing $\expval{\phi} = f(t,\bfx^2)$ we have the following scalar and vector symmetries
\begin{align}
\delta_{\bar{P}_{0}} \expval{\phi} &= -\dot{f}, \quad \delta_{\bar{P}_{i}} \expval{\phi} =- 2 f' x^{i}, \quad  \delta_{Q} \expval{\phi} = 1, \quad   \delta_{V_{0}} \expval{\phi} = t  \quad  \delta_{V_{i}} \expval{\phi} = x^{i}, \\ & \delta_{S} \expval{\phi} = t^2 + \frac{\bfx^2}{3} + \dot{f}^2 + \frac{4 \bfx^2}{3} f'^{2}, \quad \delta_{S_{0i}} \expval{\phi} = t x^{i} - 2\dot{f} f' x^i.
\end{align}
Here we have defined $S = S_{00} = \delta^{ij}S_{ij}$ where the final equality is due to $S_{\mu\nu}$ being traceless. The conditions on $f(t,\bfx^2)$ such that we preserve some form of spacetime translations are therefore
\begin{align}
\dot{f} + a + bt + c \left(t^2 + \frac{\bfx^2}{3} + \dot{f}^2 + \frac{4 \bfx^2}{3}f'^{2}\right) = 0, \label{SpecialGalEq1} \\
2 f' + d + e(t-2 \dot{f} f') = 0, \label{SpecialGalEq2}
\end{align}
where $a, \ldots, e$ are constants. Now it is clear that we need both $\dot{f} \neq 0$ and $f' \neq 0$ otherwise the solution is at most quadratic in the co-ordinates and such solutions do not break boosts, as we have discussed above. If $\dot{f} = 0$ then \eqref{SpecialGalEq2} tells us that $e=0$ and $f(\bfx^2) = -d \bfx^2 / 2 + const$, and if $f' = 0$ \eqref{SpecialGalEq1} tells us $c=0$ and $f(t) = -at - bt^2/2 + const$. So from now on we assume that $f$ is a function of both $t$ and $]bfx^2$. \\

When $e=0$ the second equation is solved by $f(t,\bfx^2)=-\frac{d}{2}\bfx^2+g(t)$ where $g(t)$ is an arbitrary function of time. Pluging this into \eqref{SpecialGalEq1} implies $c=0$ and so we find the general solution
\begin{equation}
	f(t,\bfx^2)=-\frac{1}{2}(d \bfx^2+b t^2)-at\,.
\end{equation}
The linear part is the usual superfluid background while the quadratic part is the galileid background. As discussed above, the only way this background can break boosts is if $d=1/b$ and $a=0$. Again, this case is special (see Appendix \ref{sec:special_galileon_broken_phase}). When $d\neq 1/b$ the background actually does not break boosts. So we need to look for another solution.\\

We turn to the case $e\neq0$, here we found a two parameter $(c_1,c_2)$ solution to (\ref{SpecialGalEq2}):
\begin{equation}
	f(t,\bfx^2)=c_2 + \frac{d+et}{e^2}\pm \frac{1}{e}\sqrt{(d+et)^2\left(1+e^2(\bfx^2-c_1)\right)}\,,
\end{equation}
but this solution cannot solve (\ref{SpecialGalEq1}) at the same time for any choice of parameters. We take this as strong indication that there are no solutions to both (\ref{SpecialGalEq1}) and (\ref{SpecialGalEq2}) with $e\,,c\neq0$. Equations (\ref{SpecialGalEq1}) and (\ref{SpecialGalEq2}) are two non-linear inhomogeneous PDE's: it seems very unlikely that when $e\,,c\neq0$ there could be common solutions to both. That being said, from \eqref{SpecialGalEq2} we can write 
\begin{align} \label{SpecialGalEqInter}
f' = \frac{d+et}{2(e \dot{f} -1)},
\end{align} 
where we have taken $e \dot{f} - 1 \neq 0$ since if it vanished we would need $e=0$ from \eqref{SpecialGalEq2} which would lead to a contradiction. Plugging this expression for $f'$ into \eqref{SpecialGalEq1} we could reduce the two equations to a unique, quartic non-linear inhomogeneous ODE:
\begin{align} \label{SpecialGalEqMaster}
(e \dot{f} -1)^2\left( \dot{f} + a + bt + ct^2 + \frac{c \bfx^2}{3} + c \dot{f}^2\right) + \frac{(d+et)^2 \bfx^2}{3} = 0.
\end{align} 
A complete study of the existence and uniqueness of solutions to this equation is obviously very non-trivial and beyond the scope of our analysis. However, we find it very unlikely that such a complicated solution to this equation would ultimately solve the background equation of motion of the special galileon. \\

One may wonder how the extended galileid can have non-linearly realised boosts when the additional symmetry written in term so $\phi$ \eqref{eq:extended_galileon_symmetry} looks very similar to the special galileon symmetry \eqref{eq:special_galileon_transfo}. The point is that for the extended galileid the additional symmetry is generated by a scalar and therefore in contrast to the special galileon it does not contain a vector component that can be used to compensate for a boost transformation.


\section{DBI with $\mathbf{c_s}<1$: Lorentz Invariance in Disguise}
\label{sec:DBI_Inflation}
As we alluded to in the previous section, the usual linear expectation value for the scalar field $\expval{\phi}=\mu t$, does not break boosts in the DBI theory. This is easily understood by noticing that the variation of the vacuum with respect to the boost in the $i^{th}$ direction precisely cancels against the transformation under the DBI vector symmetry $V_i$ \eqref{eq:DBI_symm_transfo} i.e.
\begin{equation}
\delta_{V^i}\expval{\phi}=\mu^{-1}\, \delta_{M_{0i}}\expval{\phi}\,.
\end{equation}
Using the commutators of DBI symmetry generators, namely 
\begin{equation}
[V_\mu,V_\nu]=M_{\mu\nu}\,,\quad [V_\mu,Q]=-P_\mu\,,\quad [P_\mu, V_\nu]=-\eta_{\mu \nu}\,Q \,, 
\end{equation}
one can easily see that the generator
\begin{equation}
K_i\equiv \dfrac{1}{\sqrt{1-\mu^2}} \left(M_{0i}-\mu V_{i}\right)\,,
\end{equation}
satisfies 
\begin{equation}
[K_i,K_j]=-M_{ij}\,,\quad [\bar{P}_i,K_j]=\dfrac{1}{\sqrt{1-\mu^2}}\bar{P}_0 \delta_{ij}\,,\quad [\bar{P}_0,K_i]=\sqrt{1-\mu^2}\bar{P}_i\,. 
\end{equation}
After rescaling $\bar{P}_0$ to $\sqrt{1-\mu^2}\bar{P}_0$, it becomes evident that $K_i$, which is linearly realised, plays the role of boosts in a Poincar\'e algebra. \\ 

However, from the look of the DBI Lagrangian perturbed around this superfluid vacuum solution, the fact that boosts are not broken is by no means obvious, see e.g.\eqref{pertLag}. For one thing, the fluctuations of the scalar field (or in the language of the EFT for inflation, the Goldstone mode of the broken time translations) acquires a non-trivial speed of sound $c_s\neq 1$ on this background, an avatar of breaking Lorentz symmetry\footnote{Of course what is physically relevant is the ratio between the speed of sound of the phonon and the velocity that appears in boosts (which we have set to unity in this work).}.
Secondly, there are cubic vertices in the Lagrangian, namely $\dot{\pi}^3$ and $\dot{\pi}(\partial_i\pi)^2$,  which are otherwise absent in a Lorentz invariant theory, and they potentially generate $s, t$ and $u$ channel singularities in $2\to2$ scattering amplitudes of the phonons. \\

In this section we elaborate on the secret Lorentz invariance of DBI from two perspectives: firstly, we explicitly show that three-particle amplitudes vanish, $2\to2$ amplitudes are explicitly Lorentz invariant and do not have any poles. Secondly, making use of the underlying geometrical picture of the DBI theory, we work out the non-perturbative field redefinition that maps the theory with $c_s\neq 1$ to the Lorentz invariant theory. Finally, we discuss the cosmological aspects of these subtleties encoded in the cosmological correlators of DBI inflation. 
\subsection{Three and Four Particle Amplitudes}
At constant wrap factor, the Lorentz invariant DBI action is:
\begin{equation}
	\label{origi}
S = -M^4\, \int d^4x\, \sqrt{1+(\del \phi)^2}\,,
\end{equation}
where $M$ is some energy scale. We begin by writing down the perturbative action for $\pi$, defined through $\phi=\mu (t+\pi)$: 
\begin{align}
\label{pertLag}
S&= M^4 \dfrac{1-c_s^2}{2c_s^3}\int d^4x\, \left[ \dot{\pi}^2-c_s^2 (\del_i\pi)^2-\dfrac{(1-c_s^2)}{c_s^2}\dot{\pi}^3+(1-c_s^2)\dot{\pi}(\del_i\pi)^2+\right. \\ \nonumber 
& \qquad \left. +\dfrac{(5-4 c_s^2)(1-c_s^2)}{4c_s^4}\dot{\pi}^4+\dfrac{(1-c_s^2)}{4}(\del_i\pi)^4+\dfrac{(-3+2c_s^2)(1-c_s^2)}{2c_s^2}\dot{\pi}^2(\del_i\pi)^2+... \right]\,,
\end{align}
where have set $\mu=(1-c_s^2)^{1/2}$. In terms of the canonically normalised field $\pi_c$ we have\footnote{In the case where the action \eqref{pertLag} is derived from the EFT of inflation in the flat-space, decoupling limit, the energy scale $M$ is fixed by $f_\pi$ and $c_s$ as $M^4=f_\pi^4/(1-c_s^2)$. However we keep it general in what follows.}
\begin{align}
{\cal L}_2 &=\dfrac{1}{2}\dot{\pi}_c^2-\dfrac{c_s^2}{2}(\del_i\pi_c)^2\,, \\
{\cal L}_3 &=-\dfrac{1}{M^2}\left(\dfrac{1-c_s^2}{c_s}\right)^{1/2}\dot{\pi}_c\, \mathcal L_2,\\
\label{last}
{\cal L}_4 &=\dfrac{1}{2 c_s\, M^4}\, \mathcal L_2\left[\mathcal L_2 +2(1-c_s^2)\dot{\pi}_c^2\right]\,.
\end{align}
The three-particle amplitudes generated by the two cubic vertices are 
\begin{align}
 {\cal A}_{\dot{\pi}^3}(p_1,p_2,p_3) &=-6 i E_1 E_2 E_3\,,\\ \nonumber
{\cal A}_{\dot{\pi}(\del_i\pi)^2} (p_1,p_2,p_3)&=-2i  \left(E_1 \bfp_2\cdot\bfp_3+E_2 \bfp_1\cdot\bfp_3+E_3 \bfp_1\cdot\bfp_2 \right),
\end{align}
where $p_i^\mu=(E_i,\bfp_i)$ are the four momenta of the external phonons, with $E_i=c_s |\bfp_i|$ on-shell and all particles are considered as ingoing, hence $p_1^\mu+p_2^\mu+p_3^\mu=0$. Writing $\bfp_i\cdot\bfp_j$ in terms of the energies and using the conservation of energy and momentum we find
\begin{equation}
{\cal A}_{\dot{\pi}(\del_i\pi)^2}=-6 i c_s^{-2}\, E_1 E_2 E_3\,.
\end{equation}
As a result, the three particle amplitude for DBI vanishes. Of course, this is not an accident since, up to unimportant boundary terms, the cubic part of the DBI Lagrangian is proportional to the free field equation of motion
\begin{equation}
\label{cubic}
{\cal L}_3\propto \dot{\pi}_c \pi_c \left(\ddot{\pi}_c-c_s^2\partial_i\partial^i \pi_c \right)+\text{boundary},
\end{equation}
and therefore does not yield an on-shell three-particle amplitude. \\

For the benefit of computing the $2 \rightarrow 2$ scattering amplitude, let us define
\begin{align}
s &=(E_1+E_2)^2-c_s^2(\bfp_1+\bfp_2)^2=2E_1 E_2-2 c_s^2 (\bfp_1\cdot\bfp_2)\,,\\ \nonumber
t &=(E_1+E_3)^2-c_s^2(\bfp_1+\bfp_3)^2=2E_1 E_3-2 c_s^2 (\bfp_1\cdot\bfp_3)\,,\\ \nonumber
u &=(E_1+E_4)^2-c_s^2(\bfp_1+\bfp_4)^2=2E_1 E_4-2 c_s^2 (\bfp_1\cdot\bfp_4)\,,
\end{align}
and as usual for massless particles we have $s+t+u=0$. The entire contribution to the four-particle amplitude from the exchange diagram simplifies to
\begin{equation}
\label{exch}
i {\cal A}^{\text{exchange}}=\dfrac{-i}{M^4}\left(\dfrac{1-c_s^2}{c_s}\right)\left[(E_1+E_2)^2 s+ (E_1+E_3)^2 t+ (E_1+E_4)^2 u\right]\,.
\end{equation}
As expected, this contribution does not have the common $s$, $t$ and $u$ singularity, precisely because there is no on-shell three particle amplitude. Moreover, \eqref{exch} exactly cancels the contribution to the four-particle amplitude from the $\dot{\pi}_c^2 \mathcal L_2$ part of \eqref{last}. Therefore, the full $2\to2$ amplitude is effectively generated solely by the following contact interaction
\begin{equation}
{\cal L}_{\text{contact}}=\dfrac{1}{8 c_s\, M^4}\left(\dot{\pi}_c^2-c_s^2(\del_i\pi_c)^2\right)^2\,,
\end{equation}
and is given by
\begin{equation}
i {\cal A}_{2\to 2}=-\dfrac{i}{8 c_s M^4}(s^2+t^2+u^2)\,.
\end{equation}
This amplitude entertains a new boost symmetry as it involves the Lorentz contractions of a new set of four momenta
\begin{equation}
\tilde p_i^\mu\equiv (E_i, c_s\bfp_i)\,.
\end{equation}
Notice that our new boosts are different from the non-linearly realised boost of the original Lagrangian \eqref{origi}, as they now leave the velocity of phonons (i.e. $c_s$) invariant, but obviously  they still obey the same commutators. In conclusion, the three and four particle amplitudes are exactly what one finds in a Lorentz invariant theory. 

\subsection{A Field Redefinition to All Orders}
Of course, the miraculous cancellations observed in the three and four particle amplitudes are artefacts of expanding the Lagrangian in terms of variables that hugely obscure the intrinsic Lorentz invariance of the theory. Here we find the non-perturbative field redefinition that allows us to write the Lagrangian in a manifestly Lorentz invariant form. \\

Thanks to the geometrical origin of the DBI theory, it is straightforward to see why $\expval{\phi}=\mu t$ is not a Lorentz violating vacuum and to immediately derive the field redefinition we are aiming for. The DBI Lagrangian \eqref{origi} can be interpreted as a Nambu-Goto action of a 4-dimensional flat brane living in a 5-dimensional Minkowski bulk with action
\begin{equation}
\label{DBI}
S=-M^4\, \int\,d^4x\,\sqrt{-\text{Det}\,\left(\eta_{AB}\dfrac{\del X^A}{\del x^\mu}\dfrac{\del X^B}{\del x^\nu}\right)}\,,\qquad A,B=0,1,...,4\,.
\end{equation}
This action has two sets of symmetries, 
\begin{equation}
ISO(1,4):\,X^A\to \Lambda^A{}_{\,B}X^B+c^A\,,\qquad \text{Diff}(4): X^A(x)\to X^A(x+\xi(x))\,. 
\end{equation}
By using the latter one can gauge fix the Lagrangian by choosing
\begin{equation}
\label{worldvol}
X^A=\Big(t,\,\textbf{x},\,\phi(t,\textbf{x})\Big)\,,
\end{equation}
and re-derive \eqref{origi}. From this standpoint, the $\expval{\phi}=\mu t$ solution corresponds to the brane moving with a uniform velocity in the extra dimension. Obviously, as seen by a boosted observer that is co-moving with the brane, the dynamics of the phonons is relativistic\footnote{In the 5-dimensional picture, all inertial frames are physically equivalent.}. To see this more explicitly, notice that under a boost transformation along the fifth dimension, with Lorentz factor $\gamma^{-1}=\sqrt{1-\mu^2}$, the world-volume \eqref{worldvol} transforms into 
\begin{align}
\bar{X}^A &=\Big(\gamma(t-\mu \phi (t,\textbf{x}) ),\,\textbf{x},\, \gamma(\phi(t,\textbf{x})-\mu t) \Big)\,= \left(\frac{t}{\gamma}-\mu^2\gamma\,\pi(t,\bfx),\,\bfx,\, \gamma \mu\, \pi(t,\bfx)  \right),
\end{align}
where we have set $\phi(t,\bfx)=\mu (t+\pi)$ in the second equality. We further use the diff invariance of \eqref{DBI} to reparametrise the time coordinate on the brane as 
\begin{equation}
\label{timerep}
\tilde{t}\equiv \frac{t}{\gamma}-\mu^2\gamma\,\pi(t,\bfx)\,.
\end{equation}
This relation can be inverted perturbatively in $\pi$ and its time derivatives as
\begin{equation}
t=\gamma\,\tilde{t}+\mu^2 \gamma^2\, \pi(\gamma\tilde{t},\bfx)\left[1+\mu^2 \gamma^2\, \dot{\pi}(\gamma \tilde{t},\bfx)\right]+... \,.
\end{equation} 
Clearly, the $t= \gamma\, \tilde t$ part of the transformation above is the usual time dilation of the moving frame and explains why phonons propagate with velocity
$c_s=\gamma^{-1}=\sqrt{1-\mu^2}<1$ as measured by an observer sitting at rest, whereas for the comoving observers, phonons move on the relativistic light cone. Following \eqref{timerep}, $X^A$ transform into 
\begin{equation}
\tilde{X}^A=\Big(\tilde{t},\,\bfx,\,  \gamma\,\mu\,\pi(t[\tilde{t}],\bfx)\Big)\,,
\end{equation}
where $t[\tilde{t}]$ is a shorthand for the inverse of \eqref{timerep}. Defining a new field through
\begin{equation}
\label{redef}
\tilde{\pi}(\tilde{t},\bfx)\equiv \mu\, \gamma\,\pi(t[\tilde{t}],\bfx)=\dfrac{\mu}{\sqrt{1-\mu^2}}\,\pi(t[\tilde{t}],\bfx)\,,
\end{equation}
it is clear that the Lagrangian for $\tilde{\pi}$ is that of DBI in flat space, i.e. 
$S=-M^4\int d^4 \tilde{x} \sqrt{1+(\del_\mu \tilde{\pi})^2}$. Expanding around $\tilde{\pi}=0$, the inverse of \eqref{redef} will take the following form 
\begin{equation}
\label{perfdef}
\pi (t,\bfx)=\dfrac{c_s}{\sqrt{1-c_s^2}}\tilde{\pi}(c_s t,\bfx)-\tilde{\pi}(c_s t,\bfx)\, \dot{\tilde{\pi}}(c_s t,\bfx)+... \,.
\end{equation}
As a non-trivial check, applying the first two terms of the above field transformation to the quadratic part of the perturbative Lagrangian for $\pi$ \eqref{pertLag} leads to a Lorentz invariant kinetic term for $\tilde{\pi}$, and moreover, it eliminates the cubic vertices and leads to a single Lorentz invariant interaction $\dfrac{1}{8}M^4\,(\del_\mu \tilde{\pi})^4$ at quartic order, in complete agreement with the results for three and four particle amplitudes in the previous section. 

\subsection{Imprints on Cosmological Correlators}
DBI inflation \cite{Alishahiha:2004eh} is one of the most studied model of single-field inflation derived from string theory. It is also a prototypical example in which the scalar fluctuations acquires a speed of sound. In the limit of the EFT of inflation, where an approximate shift symmetry for $\pi$ is restored, the potential and the warp factor in the DBI action become constant and the theory will be effectively described by the following action\footnote{See also \cite{Pajer:2008uy} for a string theoretic model of inflation in which the DBI warped factor is constant.}
\begin{equation}
S=-M^4 \int \sqrt{-\bar{g}}\sqrt{1+\bar{g}^{\mu\nu}\del_\mu \phi \del_\nu \phi}\,.
\end{equation}
In the flat space limit, i.e. $\bar{g}^{\mu\nu} \to \eta^{\mu\nu}$, this action enjoys the $ISO(1,4)$ symmetry \cite{Creminelli:2013xfa} as discussed above. Despite the fact that correlators in the EFT of inflation are not invariant under those symmetries, as we discussed in section \ref{sec:recap}, the total energy pole of inflationary correlators is supposed to reflect the symmetries of the theory in the flat space limit. Motivated by this expectation and as an example, let us inspect the behaviour of the DBI three point function on the total energy pole. In terms of the parametrisations of the EFT action \eqref{EFT}, DBI inflation has cubic coupling:
\begin{equation}
\label{c3}
c_3=\dfrac{3}{2}(c_s^2-1)\,.
\end{equation}
This, in combination with \eqref{poleenergy}, indicates that the residue on the total energy pole of the three point function in DBI inflation vanishes i.e.
\begin{align}
\label{poleDBI}\nonumber
\lim_{e_1\to 0} \dfrac{1}{e_3\prod P_{k_i}}\langle \pi_{\bfk_1}\pi_{\bfk_2}\pi_{\bfk_3}\rangle'_{DBI}&=f_\pi^4\,\dfrac{c_s^2-1}{c_s^2}\,\dfrac{12 e_3}{H\, e_1^3}\left[-1+c_s^2-\dfrac{2}{3} c_3\right]+\mathcal{O}(e_1^{-2})\\
&= 0 + \mathcal{O}(e_1^{-2})\,.
\end{align}
This is simply due to the fact that three particle amplitude vanishes for DBI in flat space. Of course, direct inspection of \eqref{3pt} reveals a sub-leading total energy pole of order $1/e_1^2$. The sub-leading pole can be traced back to the cubic operators that will be generated after performing the perturbative field redefinition
\begin{equation}
\pi \to \pi-\dfrac{1-c_s^2}{c_s^2}\dot{\pi}\pi\,.
\end{equation}
In flat space, this field redefinition eliminates the cubic vertices $\dot{\pi}^3$ and $\dot{\pi}(\del_i\pi)^3$, but in the EFT of inflation it leads to two cubic operators proportional to Hubble and a boundary term, namely\footnote{This boundary term does not modify the correlators of $\pi$, inasmuch as it is real at $\eta=0$, and thereby does not contribute to the real part of the wavefunction of the universe: $\psi \propto  \exp(iS)$. Also notice that the shown field redefinition does not change the correlators on the boundary up to slow-roll corrections simply because its non-linear part is proportional to $\dot{\pi}$.}
\begin{equation}
\Delta S =\int a^3\,dt\,d^3x\, \dfrac{M_p^2 |\dot{H}|}{c_s^4}(1-c_s^2)\Big( 3H \pi \dot{\pi}^2-3H \pi \dfrac{(\del_i\pi)^2}{a^2}+...\Big)+S_{b}\,.
\end{equation}
The newly born cubic terms, namely $\pi \dot{\pi}^2$ and $\pi (\del_i \pi)^2$, have one less derivative with respect to $\dot{\pi}^3$ and $\dot{\pi}(\del_i \pi)^2$ and it is these operators that lead to the now leading $1/e_1^2$ pole, in agreement with the explicit expression \eqref{3pt}. We conclude that the secret Lorentz invariance of DBI inflation at high energy scales is encoded in the order of the total energy pole of correlators. As an illustration, we considered the three point function of $\pi$, where a generic single field theory with a speed of sound $c_s\neq 1$ exhibits a pole of order $3$, and, by contrast, DBI inflation has a pole of order $2$. As such, DBI is a counter-example to the common lore that the residue of a boundary correlator at its total energy pole is proportional to the flat space amplitude of the theory. Of course this is a very weak counter-example since here the amplitude is actually zero. Here we have attributed this to the enhanced symmetry in the flat space limit, and it would be interesting to find other examples of this behaviour. \\

Interestingly, one can derive inflationary correlators by taking exact de Sitter invariant ones (i.e. with unbroken de Sitter boosts) and putting some of the external legs on the background \cite{Arkani-Hamed:2018kmz}. A simple example would be to use the correlators of a $P(X)$ theory. Expanding in powers of the background, at linear order one will find that the three-point inflationary correlator coming from $X^2 = (\partial \phi)^4$ is proportional to the DBI one we have discussed here. Indeed, the cubic operators combine into a single object that is proportional to the quadratic theory. One will therefore find that the order of the total energy pole is 2 whereas the de Sitter invariant four-point function from where this three-point function came from had an order 3 pole \cite{Arkani-Hamed:2018kmz}. This is however only a consequence of expanding to linear order in the background. Indeed, by taking higher order contact terms and putting more legs on the background one finds  that $c_{s} \neq 1$ and the generic cubic order pole for the inflationary three-point function will be recovered.


\section{Scattering Amplitudes of Phonons}\label{Amplitudes}


\subsection{Soft Theorems}
The Poincar\'{e} invariant scattering amplitudes of scalar theories with spontaneously broken symmetries (including a shift symmetry) exhibit non-trivial soft theorems dubbed ``generalised Adler zero's". The derivation of these soft theorems resembles that of a Nambu-Goldstone boson of a spontaneously broken $U(1)$ symmetry. Consider for example the following non-linearly realised symmetry of degree $n$ in a Poincar\'e invariant theory
\begin{equation}
\label{nlsym}
\delta\phi=a_{\mu_1 ... \mu_n}x^{\mu_1}... x^{\mu_n}+f(\phi, \del^m \phi,x)\,,
\end{equation}
in which $f(\phi,\del^m \phi,x)$ is a function of $\phi$ and its derivatives and vanishes for $\phi=0$. For a Poincar\'e invariant theory $a_{\mu_1... \mu_n}$ is a $SO(1,3)$ irrep i.e. it is a symmetric traceless rank-$n$ tensor. The Noether current associated with \eqref{nlsym} is given by
\begin{equation}
\label{current}
J^\mu=\left[-a_{\mu_1 ... \mu_n}x^{\mu_1}...x^{\mu_n}\del^\mu +n\, a^\mu_{\,\, \mu_2 ...\mu_n}x^{\mu_2}...x^{\mu_n}\right]\phi+K^\mu(\phi,\del^m\phi,x)\,,
\end{equation}
in which $K^\mu$ starts at quadratic or higher order in $\phi$ and its derivatives. The soft theorem generated by this current can be derived from the following Ward identity
\begin{equation}
\int d^4x\, \exp(-iq\cdot x) \del_\mu \langle \beta, \text{out}| J^\mu (x)| \alpha, \text{in}\rangle=0. 
\end{equation}
Above, $|\alpha, \text{in}\rangle$ and $|\beta, \text{out}\rangle$ are ``in" and ``out" multi-particle states consisting of hard modes in a scattering process, and $q$ is some on-shell soft momentum, i.e. $q^2=0$. Inserting the first term of \eqref{current} in the Ward identity and using the LSZ formula we find 
\begin{align}
&-\int d^4x \exp(-i\,q\cdot x)\,a_{\mu_1 ... \mu_n}x^{\mu_1}...x^{\mu_n} \langle \beta, \text{out}| \Box \phi(x)| \alpha, \text{in}\rangle \nonumber \\ =&-i^{n+1}\,a_{\mu_1...\mu_n}\dfrac{\del}{\del q_{\mu_1}}... \dfrac{\del}{\del q_{\mu_n}}\langle \beta + q^\mu , \text{out}| \alpha, \text{in}\rangle\,.
\end{align}
In computing the contribution from the second term in \eqref{current}, one is allowed to perform integration by part inasmuch as $K^\mu$ is quadratic in the field and the interactions are asymptotically switched off\footnote{See Section 2.4 of \cite{Mirbabayi:2016xvc}}, hence the absence of boundary terms. We thus find
\begin{equation}
\label{ward}
i^{n}\,a_{\mu_1...\mu_n}\dfrac{\del}{\del q_{\mu_1}}... \dfrac{\del}{\del q_{\mu_n}}\langle \beta + q^\mu , \text{out}| \alpha, \text{in}\rangle=q_\mu\, \langle \beta, \text{out}|K^{\mu}(-q)| \alpha, \text{in}\rangle\,.
\end{equation} 
In the soft limit, the right hand side vanishes if and only if the matrix element $\langle \beta, \text{out}|K^{\mu}(-q)| \alpha, \text{in}\rangle$ is regular at $q=0$. In a Poincar\'{e} invariant, shift symmetric theory this is always the case since all cubic vertices can be removed be a field redefinition. Indeed, the most general three-particle amplitudes for scalars are constant and this constant must vanish if there is a shift symmetry. The subsequent soft theorem reads
\begin{equation}
\label{softth}
\lim \limits_{q\to 0} a_{\mu_1...\mu_n}\dfrac{\del}{\del q_{\mu_1}}... \dfrac{\del}{\del q_{\mu_n}}\langle \beta + q^\mu , \text{out}| \alpha, \text{in}\rangle=0\,.
\end{equation}
This, in conjunction with all the other non-linearly realised symmetries at lower degrees (i.e. $0,1,\dots, n-1$), implies that $\langle \beta + q^\mu , \text{out}| \alpha, \text{in}\rangle$ should vanish at order ${\cal O}(q^{n+1})$ in the soft limit. See \cite{Cheung:2016drk} for further details. \\

\begin{figure}[t!]
\begin{center}
\tikzset{every picture/.style={line width=0.6pt}} 

\begin{tikzpicture}[x=0.75pt,y=0.75pt,yscale=-1,xscale=1]

\draw    (189.88,88.05) -- (266.21,129.26) ;
\draw    (180,159.92) -- (263.52,159.92) ;
\draw    (200.52,227) -- (270.56,183.88) ;
\draw   (262.62,153.21) .. controls (262.62,118.81) and (288.75,90.93) .. (320.99,90.93) .. controls (353.23,90.93) and (379.36,118.81) .. (379.36,153.21) .. controls (379.36,187.61) and (353.23,215.5) .. (320.99,215.5) .. controls (288.75,215.5) and (262.62,187.61) .. (262.62,153.21) -- cycle ;
\draw    (372.18,122.55) -- (440.43,84.22) ;
\draw    (380.26,158.96) -- (453,158.96) ;
\draw    (371.28,184.5) -- (441.33,222.83) ;
\draw   (365.7,58.98) .. controls (364.72,62.42) and (363.79,65.71) .. (365.18,67.83) .. controls (366.57,69.95) and (369.9,70.32) .. (373.39,70.71) .. controls (376.88,71.09) and (380.21,71.47) .. (381.6,73.59) .. controls (382.99,75.71) and (382.06,79) .. (381.07,82.44) .. controls (380.09,85.87) and (379.16,89.16) .. (380.55,91.28) .. controls (381.94,93.4) and (385.27,93.78) .. (388.76,94.16) .. controls (392.25,94.55) and (395.58,94.92) .. (396.97,97.04) .. controls (398.36,99.17) and (397.43,102.45) .. (396.44,105.89) .. controls (396.06,107.22) and (395.69,108.53) .. (395.46,109.75) ;

\draw (185.2,130.54) node [anchor=north west][inner sep=0.75pt]  [font=\normalsize]  {$\textcolor[rgb]{0,0,0}{\alpha }$};
\draw (430.61,132.39) node [anchor=north west][inner sep=0.75pt]  [font=\normalsize]  {$\beta $};
\draw (392.87,60.06) node [anchor=north west][inner sep=0.75pt]  [font=\normalsize]  {$\vec{q}$};

\end{tikzpicture}
\end{center}
\caption{The soft limit of the scattering of phonons is dominated by diagrams within which the soft momentum is attached to an external leg. \label{soft}}
\end{figure}
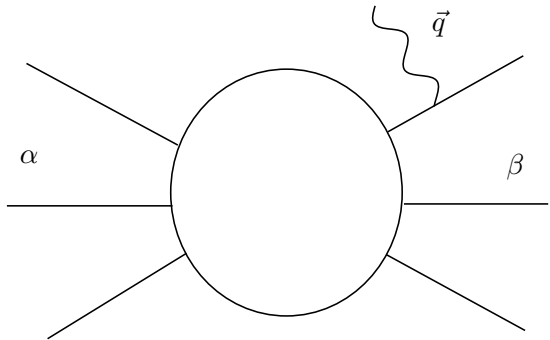

As we pointed out above, the key ingredient in concluding \eqref{softth} was the absence of cubic vertices, or in other words, the absence of non-zero on-shell three-particle amplitudes. When present, the matrix element $\langle \beta, \text{out}|K^{\mu}(-q)| \alpha, \text{in}\rangle$ generically has a $1/q^0$ pole in the soft limit and \eqref{softth} does not hold. This is simple to understand for four-particle amplitudes since when the intermediate particle is taken on-shell in an exchange process, unitarity tells us that the amplitude should factorisation into a product of on-shell three-particle amplitudes. The residue is therefore only non-zero if these three-particle amplitudes are non-zero. To see that this a problem when boosts are non-linearly realised, consider a superfluid theory with a cubic vertex $\dot{\pi}^3$. Similar to the scattering amplitudes with soft photons or gravitons \cite{Weinberg:1964ew}, the soft behaviour of the scattering of phonons in a process such as $\alpha \to \beta +\textbf{q}$ is dominated by diagrams inside which the soft momentum is attached to an external leg, see Figure \ref{soft}. We then find
\begin{align}
\lim_{q\to 0}\langle \beta + q^\mu , \text{out}| \alpha, \text{in}\rangle &\propto \langle \beta , \text{out}| \alpha, \text{in}\rangle\, \sum_{a} \dfrac{\, p_a^2 q}{c_s^2(\eta_a p_a+q)^2-c_s^2 (\eta_a \textbf{p}_a+\textbf{q})^2}\\ \nonumber
&= \langle \beta , \text{out}| \alpha, \text{in}\rangle \dfrac{1}{c_s^2}\sum_{a} \dfrac{\eta_a p_a}{1-\hat{p}_q\cdot\hat{q}}\,.
\end{align}
Here $\eta_a$ equals +1 or -1 for outgoing and ingoing particles, respectively, and $p_a$'s are the external momenta. Now, as an example, consider the current associated with the broken boosts of a superfluid theory
\begin{equation}
J^\mu_{K_i}=(-x^i\del^\mu+\eta^{\mu i})\phi+K^\mu\,.
\end{equation}
 Comparing the above soft limit with the schematic form of the Ward identity in \eqref{ward} indicates that $K^\mu$ must have a pole at $q=0$. The same happens for any other broken symmetry when cubic vertices are present. \\
 
Although in this section we focus on S-matrix soft theorems, we should emphasise that in cosmology there is no roadblock to finding single soft theorems for non-linearly realised symmetries for correlators, even if cubic vertices are present \cite{Creminelli:2012ed, Hinterbichler:2013dpa}\footnote{For recent developments of soft theorems for setups with additional internal symmetries see also \cite{Finelli:2018upr, Finelli:2017fml, Pajer:2019jhb}.}. In our flat space setting, it is also possible to derive similar soft theorems for in-in correlators in Minkowski space. We leave this and the derivation of cosmological soft theorems (associated with non-linear symmetries) to future work. In the second case, one must first perform a classification of symmetries around cosmological backgrounds and we outline our plans in this direction in Section \ref{sec:discussion}. 

\subsection{Inevitability of Cubic Vertices}

It turns out that for the superfluid and galileid theories we have discussed in this paper non-zero three-particle amplitudes, and therefore four-particle amplitudes with poles, are actually inevitable. In Poincar\'{e} invariant theories, non-perturbative three-particle amplitudes are fixed by the little group scaling of the external particles leaving only an undetermined coupling constant, see e.g. \cite{CheungReview} for a review. Scalars do not have helicity and therefore do not transform under the little group. As such the most general three-particle amplitude is a constant. When Lorentz boosts are broken, the non-perturbative three-particle amplitudes have the same structure as their Poincar\'{e} invariant counterparts, but now the previously constant coupling can be an arbitrary function of the three energies of the particles \cite{Boostless}. For identical scalars this function should be a symmetric polynomial in the energies and can therefore be written in terms of the three elementary symmetric polynomials \eqref{SymmetricPolynomials}. However, on-shell conservation of energy dictates $e_{1} = 0$ (we take all particles as incoming) and therefore the three-particle amplitudes are a function of $e_{2}$ and $e_{3}$ only \cite{Boostless}. \\ 

Consider first the superfluid action $\mathcal{L} = P(X)$ with $\phi = \mu t +\pi$ and $X=g^{\mu\nu}\del_\mu\phi \del_\nu\phi$. By expanding the action, canonically normalising the phonon and rescaling the spatial coordinates such that the quadratic action is Lorentz invariant the Lagrangian becomes
\begin{align}
\mathcal{L}_{\pi} = -\frac{1}{2} (\partial_{\mu} \pi)^2 + \alpha_{1} \dot{\pi}^3 + \alpha_{2} \dot{\pi}(\partial_{i} \pi)^2,
\end{align}
where 
\begin{align}
\alpha_{1} &=\dfrac{\mu}{\sqrt{2}}\dfrac{P''-\frac{2}{3}\mu^2 P'''}{\left(P' (P'-2\mu^2 P'')\right)^{\frac{3}{4}}},\\
\alpha_{2} &=-\dfrac{\mu}{\sqrt{2}} \dfrac{P'' (2\mu^2 P''-P')^{\frac{1}{4}}}{(-P')^{\frac{7}{4}}},
\end{align}
with $P' = \partial P / \partial X$. Now if we write $\dot{\pi} (\partial_{i} \pi)^2 = \dot{\pi} (\partial_{\mu} \pi)^2  + \dot{\pi}^3$ we again get two interaction terms but only the $\dot{\pi}^3$ vertex contributes to an on-shell $3$-particle amplitude since as we explained in section \ref{sec:DBI_Inflation} the other term is proportional to $\Box \pi$ after integration by parts and therefore vanishes on-shell. The three-particle amplitude is therefore proportional to $(\alpha_{1} + \alpha_{2}) e_{3}$ and so we require $\alpha_{1} + \alpha_{2} = 0$ if it is to vanish. This is equivalent to
\begin{align}
\label{dbieq}
3 P''^2 - P' P''' = 0,
\end{align}  
where we have taken $2P'' \mu^2 - P' \neq 0$ and $P' \neq 0$ such that $\pi$ is dynamical and interacting. The only non-trivial solution to this equation is $P(X) = -\sqrt{1+X}$ i.e. the scalar DBI theory which we have already shown is actually secretly Lorentz invariant\footnote{Another solution to \eqref{dbieq} is the cuscuton theory $P=\sqrt{X}$ \cite{Cuscuton}, but in this case the phonons are non-dynamical.}. So any superfluid theory, where boosts are actually spontaneously broken and therefore non-linearly realised, has a non-vanishing on-shell three-particle amplitude. This in turn ensures that the $2 \rightarrow 2$ scattering amplitude has a pole. Furthermore, this tells us that DBI is the only $P(X)$ theory that has the softer order 2 total energy pole in its inflationary three-point function. All other $P(X)$ theories have an order 3 pole since they have a non-zero three-particle amplitude. \\ 

Now consider galileids. When the leading Wess-Zumino terms of the galileon theory \eqref{galileonLagrangian} are expanded around the galileid background \eqref{eq:galileon_background} the power counting is unchanged i.e. cubic vertices have four derivatives and these must all be time derivatives if the three-particle amplitude is to have a chance of being non-zero. However, given that $\pi$ is shift symmetric, the only such operator is $\dot{\pi}^2 \ddot{\pi}$ which is a total derivative and therefore does not contribute to the three-particle amplitude. Therefore the leading order galileid vertices do not yield poles in four-particle amplitudes. However, \eqref{galileonLagrangian} is not the full story for galileons. Indeed, quantum corrections generate high order operators with at least two derivatives per field \cite{ClassicalAndQuantum}. This was to be expected since such operators satisfy the symmetries of the EFT and so must be included in the derivative expansion. Take, for example, $(\partial_{\mu}\partial_{\nu} \phi \partial^{\mu} \partial^{\nu} \phi)^3$. When expanded around \eqref{eq:galileon_background} this operator yields a $(\alpha - \beta)^3 \ddot{\pi}^3$ term which contributes to the on-shell three-particle amplitude as long as $\alpha \neq \beta$ which is precisely the condition for boosts to be spontaneously broken. So for galileids non-trivial cubic vertices are not automatic in the tree-level Lagrangian, but they will be generated quantum mechanically and yield poles in four-particle amplitudes. \\

We therefore conclude that for superfluids and galileids, cubic vertices are inevitable meaning that these theories do not have vanishing amplitudes in the limit of one external momentum being taken soft and therefore the they do not admit Adler zero and generalisations. As we mentioned before, there will indeed be other soft theorems due to the non-linear symmetries we have discussed here and it would be very interesting to use these to construct the S-matrices directly from a small amount of on-shell data.

\subsection{Weak Coupling Regime} 
\label{sub:perturbative_unitarity_and_strong_coupling_scales}
To conclude our discussion of scattering amplitudes for the EFTs we have found in Section \ref{sec:trees_in_the_zoo}, we discuss in this section their weakly coupled regime. Perturbative unitarity bounds on scattering amplitudes for theories with spontaneously broken boosts have recently been worked out in \cite{Grall:2020tqc}. We recall their main results here and apply them to the scaling and conformal superfluids. In particular we show necessary conditions for which these theories describe weakly coupled phonons on sub-horizon scales during inflation. We leave the case of the galileid theories for future work.\\

Because of the broken Lorentz symmetry, the amplitudes have fewer kinematical constraints than their Lorentz invariant counterparts. Indeed, $2 \rightarrow 2$ amplitudes $\mathcal{A}(p_1p_2\to p_3p_4)$ now not only depend on the internal energy of the particles $E_s=E_1+E_2$ (e.g. in the center of mass frame $E_s=\sqrt{s}$), but also on the motion of the overall system of particles $\mathbf p_s=\mathbf p_1+\mathbf p_2$ with respect to the background, as well as three scattering angles:
\begin{equation}
	\mathcal{A}(p_1,p_2\to p_3,p_4)=\mathcal A (E_s,|\mathbf{p}_s|,\theta_1,\theta_3,\phi_1-\phi_3),
\end{equation}
where $\theta_i,\,\phi_i$ are the longitudinal and azimuthal angles between $\mathbf{p}_i$ and $\mathbf {p}_s$ respectively. In usual Poincar\'{e} invariant scattering processes the ``momentum of the system" is boosted away to be $\mathbf p_s=0$ (in the center-of-mass frame) but in these theories this would yield a loss of generality: different inertial frames are physically distinct\footnote{In single-clock models of inflation like the EFT of inflation, this is easily understood by the presence of the cosmic frame. The ``clock" driving inflation defines a preferred reference frame in which the expansion is isotropic e.g. set by the vector $n^{\mu}\propto\partial^\mu \expval{\phi(t)}$.}. Partial wave unitary bounds used to probe the strong coupling scale of EFTs must therefore be adapted accordingly. It was proposed in \cite{Grall:2020tqc} to expand the amplitude in terms of spherical harmonics, instead of the usual Legendre polynomials, as
 \begin{equation}
 	\mathcal A(E_s,|\mathbf{p}_s|,\theta_1,\theta_3,\phi_1-\phi_3) = \frac{16\pi^2 E_s^2(1-\rho_s^2)}{E_1E_3}\sum_{\ell_1,\ell_3}\sum_{\substack{-\ell_1 < m_1 < \ell_1 \\ -\ell_3 < m_3 < \ell_3}}  a_{\ell_1\ell_3}^{m_1m_3}(E_s,\rho_s) Y^{m_1 *}_{\ell_1}(\hat{\mathbf p}_1) Y^{m_3}_{\ell_3}(\hat{\mathbf p}_3)\,,
 \end{equation}
 where $\rho_s=c_s|\mathbf{p}_s|E_s^{-1}$ and the spherical wave coefficients satisfy the selection rule $a_{\ell_1\ell3}^{m_1m_3}=\delta^{m_1m_3}(a^{m_1})_{\ell_1\ell_3}$. Perturbative unitarity is then the usual constraint that spherical wave coefficients,
\begin{equation}
	a^{m_1 m_3}_{\ell_1 \ell_3} (  E_s,  \rho_s  ) =  \frac{ 1}{ 1- \rho_s^2} \int \frac{d^2 {\bf p}_1}{4 \pi} \frac{d^2 {\bf p}_3}{4 \pi} \, Y^{m_1}_{\ell_1} ( \hat{\bf p}_1 ) Y^{m_3*}_{\ell_3} ( \hat{\bf p}_3  ) \;  \frac{E_1}{E_s}  \, \frac{E_3}{E_s} \mathcal{A} ( p_1 p_2 \to p_3 p_4 ) \, ,
\end{equation}
are bounded above as\footnote{\label{fn:stronger_bounds}As discussed in \cite{Grall:2020tqc}, there exist in fact stronger unitarity bounds on the $a_{\ell \ell}^{m}$'s, namely $a_{\ell \ell}^{m}>\sum_j|a_{\ell j}^{m}|^2$, which yields stronger bounds on the strong coupling scales. For the sake of clarity we only consider in this work the simplest bounds \eqref{eq:pert_unitarity_bound} and leave an analysis of the full bounds to future work.}
 \begin{equation}
 	\label{eq:pert_unitarity_bound}
 	\Big|\text{Re} \, a_{\ell\ell}^{m} (E_s,\rho_s)\Big|<1/2\,.
 \end{equation}

 Most importantly, when boosts are spontaneously broken, there are two physical scales controlling unitarity of perturbation theory: the internal energy of scattering $E_s$ as well as the momentum of the system $|\mathbf{p}_s|$ (or equivalently the velocity $\rho_s$ of the system). The EFT can break down at either large energy or large momenta. Equivalently one can also think in terms the Mandelstam variable away from the center-of-mass frame $s=E_s^2-c_s^2|\mathbf{p}_s|^2$ with strong coupling attained at some $s_{\text{max}}$ and $|\mathbf{p}_s|_{\text{max}}$ \cite{Grall:2020tqc}.\\

\begin{figure}[h!]
 \centering
 	\includegraphics[width=\linewidth]{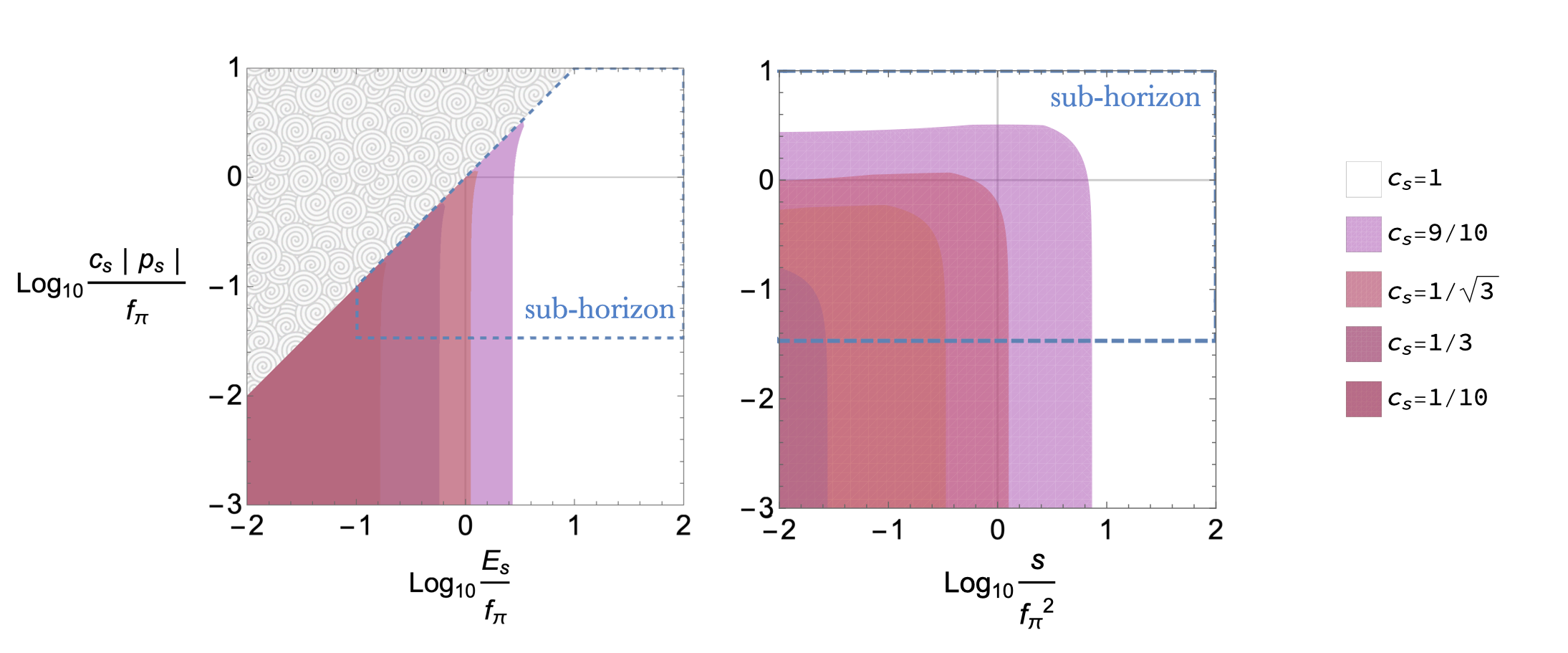}
 	\caption{Bounds on the perturbative unitary regions of the scaling superfluid for spherical wave coefficient up to $\ell=2$. Each colored region corresponds a different value of $c_s$ with the conformal superfluid corresponding to $c_s=1/\sqrt{3}$. All colored regions extend to arbitrarily low $|\mathbf{p}_s|$, $E_s$ and $s$, as spherical wave coefficients get smaller. The dashed enclosed region represents the subhorizon regime of scattering. \textit{Left}: Weakly-coupled regions for various scattering energies $E_s$ and momenta $|\mathbf{p}_s|$. The patterned region is forbidden kinematically. \textit{Right}: Same weakly-coupled regions, this time in terms of the Mandelstam variable $s$ and momentum $|\mathbf{p}_s|$. For a given $c_s$, the maximum $s=s_{\text{max}}$ is reached at low $|\mathbf{p}_s|$ while $|\mathbf{p}_s|_{\text{max}}$ is reached at an intermediate $s$.}
 	\label{fig:conformal_superfluid_unitarity}
 \end{figure}

Let us see how this works out in the case of the scaling superfluid \eqref{eq:scaling_superfluid_lagrang}. The couplings of the theory are uniquely fixed in terms of the speed of sound $c_s$ (or equivalently the scaling exponent $\alpha$). Therefore by using the bound \eqref{eq:pert_unitarity_bound} we can access the unitary region in the $(E_s, |\mathbf{p}_s|)$ (or $(s, |\mathbf{p}_s|)$) plane solely in terms of $c_s$. 
 Furthermore, by setting $c_s=1/\sqrt{3}\Leftrightarrow \alpha=2$, full conformal symmetry is recovered: there are no more free parameters in the action and we can determine the full weakly-coupled region of the conformal superfluid. Bounds on the weakly-coupled regime of the scaling superfluid are shown in Figure \ref{fig:conformal_superfluid_unitarity} for spherical wave coefficients up to $\ell=2$ and for different values of the speed of sound\footnote{Note that as $c_s\to1$, the theory becomes free and the bounds are trivial while as $c_s\to0$ the couplings blow-up and the perturbative unitary region shrinks to zero.}. \\

 Our primary motivation for studying these superfluids was their application to the EFT of inflation to  describe scalar fluctuations on sub-horizon scales (c.f. Section \ref{sub:the_eft_of_inflation}). Naively, since the only relevant energy scale in the decoupling, flat space limit of the EFT is the Goldstone decay constant $f_\pi=(2c_sM_P^2|\dot H|)^{1/4}$, one would take the action (\ref{eq:scaling_superfluid_lagrang}) and set $\mu\to f_\pi$. However this is not the right thing to do. The reason is that for such $P(X)$ theories whose couplings are fully fixed by $c_s$, fixing the background solution to be FLRW (away from the flat space limit) necessarily involves the speed of sound. For the scaling superfluid this can be seen from the Friedmann equation
\begin{equation}
	M_{\text{pl}}^2|\dot H|=X P_{,X}=\mu^4 \alpha=\mu^4\frac{c_s^2+1}{2c_s^2}\,.
\end{equation}
It follows simply then that correct map from the superfluid (\ref{eq:scaling_superfluid_lagrang}) to the flat space limit of the EFT of inflation is 
\begin{equation}
	\mu^2=\sqrt{\frac{c_s}{1+c_s^2}}f_\pi^2\,.
\end{equation}
Observations of the primordial power spectrum fix $f_\pi$ in terms of the Hubble parameter in inflation, $f_\pi=(58.64\pm0.33)H$ \cite{Aghanim:2018eyx}. We can thus deduce from the perturbative unitarity bounds \eqref{eq:pert_unitarity_bound} the conditions for the scaling superfluid to be weakly-coupled on sub-horizon scales. This region is shown on Figure \ref{fig:conformal_superfluid_unitarity}. For most of the $0\leq c_s\leq1$ range, the scaling superfluid is weakly coupled on these scales and the EFT provides a useful description of sub-horizon physics. However for small enough $c_s$, or equivalently large enough exponent $\alpha=\frac{1}{2c_s^2}(c_s^2+1)$, the theory becomes strongly coupled on sub-horizon. Together with the stability constraints (absence of ghosts, gradient instabilities etc.) this effectively bounds the allowed value for the scaling superluid exponent. From our analysis we find
\begin{equation}
	\label{eq:alpha_weak_coupling}
	1\leq \alpha < 181\,,
\end{equation}
which in terms of the speed of sound translates to $\frac{1}{19}<c_s\leq1$. Scaling superfluids with $\alpha \geq181$ or $c_s\leq1/19$ do not describe weakly coupled EFTs on sub-horizon scales during Inflation. On the other hand, the conformal superfluid $(\alpha =2)$ is shown to be weakly coupled on sub-horizon scales up to energies around $f_\pi$ i.e. more than an order of magnitude above $H$.

Finally let us recall that there exists additional bounds on the weak coupling region of these theories, coming from spherical wave coefficients beyond $\ell=2$, as well as improved unitarity bounds (see footnote \ref{fn:stronger_bounds}). From the analysis of \cite{Grall:2020tqc} we can estimate up to which $\ell$ we need to compute the spherical waves to get the strongest possible bounds. Indeed, at low $s$, the $\alpha_1$ coupling dominates the amplitude (since it controls the $1/s$ singularity) and the spherical wave coeficients can be solved analytically for any $a_{\ell_1\ell_3}^{m_1m_3}$. We then have good control over the low $\ell$ approximation. In the case at hand, at small $c_s$, when the scaling superfluid becomes strongly coupled on all sub-horizon scales, the $\ell=2$ bounds we use here to derive \eqref{eq:alpha_weak_coupling} are strongest. On the other hand at large $c_s$, and low $s$, we know there should exist stronger bounds on $|\bf{p}_s|$ from $\ell>2$. 
Although this does not alter the conclusion that the conformal superfluid is weakly-coupled on sub-horizon scales and up to energies $s\approx f_\pi^2$, it would certainly be relevant to anyone wishing to study the theory in more details. We leave a full analysis of these bounds for future work and refer the interested reader to \cite{Grall:2020tqc} for more details.

\section{Discussion and Future Work}
\label{sec:discussion}
Although scalar field theories are not heavily constrained by locality and unitarity in the same way spinning particles are, scalar EFTs can be neatly classified and constrained by symmetries they non-linearly realise. For Poincar\'{e} invariant theories there are two complementary ways of classifying and looking for new symmetries. The first relies on the existence of an Adler zero for soft scattering amplitudes which states that in the limit where one external momentum is taken soft the amplitude vanishes. Asking that the amplitude vanishes more quickly constrains the couplings of the theory and implies the presence of extra symmetry. These techniques have been used in \cite{GeneralisedAdler1,GeneralisedAdler2,GeneralisedAdler3, Probing}. The second uses an algebraic classification, guided by the coset construction and the novel inverse Higgs constraints, to classify the algebras that can be non-linearly realised on a single scalar. This technique has been used in \cite{RSW1,BraunerBogers} and such a classification is now complete and summarised in Table \ref{ScalarSymmetries}. \\

In this paper we have considered a similar classification for systems with less linearly realised symmetry, primarily motivated by cosmology where Lorentz boosts are spontaneously broken. We have considered the possible symmetries of a self-interacting, shift-symmetric phonon which we assume non-linearly realises Lorentz boosts. Such scalar theories arise in the decoupling limit of cosmological EFTs e.g. the EFT of inflation. \\ 

We have shown in section \ref{Amplitudes} that an Adler zero does not exist when Lorentz boosts are spontaneously broken. This is because cubic vertices are now a necessary part of the EFT, even in the presence of a shift symmetry, and therefore four-particle amplitudes do not vanish in the soft limit since there is an exchange diagram with a $1/p_{1} \cdot p_{2}$ pole and we would like to take $p_{1}$ or $p_{2}$ soft. For Poincar\'{e} invariant theories there are no cubic vertices when there is a shift symmetry and so all amplitudes are regular in the soft limit. This does not mean that the shift symmetry for the phonon does not yield non-trivial consequences for amplitudes; it is just that the soft theorem is not an Adler zero. For this reason we have chosen to pursue an algebraic classification in this paper which, in comparison to the on-shell approach, still allows us to avoid redundancies such as field redefinitions. \\

Our classification was presented in section \ref{sec:trees_in_the_zoo}. We have rediscovered some well-known phonon theories such as superfluids which can be further classified into scaling or conformal superfluids, and galileids which correspond to the broken phase of the galileon whose role in cosmology ranges from the longitudinal mode of a massive graviton \cite{Aspects,dRGT} to a method of driving inflation \cite{GalInflation1,GalInflation2} or studying its alternatives \cite{Genesis}. Within the broad class of galileid theories we have uncovered a special subset which we have dubbed the \textit{extended galileid}. We found six choices of the galileon couplings (c.f. Table \ref{tab:Extended_galileons.}) that lead to a new symmetry, thanks to an additional scalar generator, which when expanded around the galileid vacuum leads to twelve theories for the phonon (Table \ref{tab:extended_galileids}). Six of these have an infinite speed of sound, two have a vanishing speed of sound while the other four have a finite speed of sound. For three of these, there is a wide range of parameters that can satisfy $0 < c_{s} \leq 1$ thereby avoiding IR instabilities. Clearly these new extended galileid theories deserve further attention. \\

Let us emphasise that the extended galileid theory is not simply the special galileon \cite{SG} expanded around a Lorentz breaking vacuum solution. Indeed, we have shown that the exceptional EFTs of the scalar DBI and the special galileon do not admit such a broken phase where all of the defining non-linear symmetries remain so. Their defining symmetries are so powerful that if some form of spacetime translations are preserved by the background, some form of Lorentz boosts are too. We have presented a detailed account of this for scalar DBI and have outlined the consequences for cosmology in Section \ref{sec:DBI_Inflation}. The primary result there is that the order of the total energy pole for cosmological correlators in DBI inflation is different to what one finds for a generic superfluid theory. This follows directly from the fact that the DBI theory, when expanded around the superfluid vacuum, actually has Lorentz invariant scattering amplitudes and therefore a vanishing three-particle amplitude. On the other hand, for the special galileon we found a Lorentz breaking vacuum solution where one of the non-linear symmetries of the original theory becomes linearly realised. We provide further details in Appendix \ref{sec:special_galileon_broken_phase}. \\

Finally, in section \ref{sub:perturbative_unitarity_and_strong_coupling_scales} we have discussed the weakly coupled regime of these EFTs. When boosts are spontaneously broken, it was shown in \cite{Grall:2020tqc}, that EFTs have different resolving power in space and in time. Equivalently, perturbative unitarity of scattering amplitudes can break down at either large internal energies or internal momentum. We have computed these partial-wave bounds and derived weakly-coupled regions in the energy-momentum cut-off plane for the scaling and conformal superfluids. Doing so we have shown that, when applied in the context of the EFT of inflation, a necessary condition for the scaling superfluid to describe weakly-coupled sub-horizon dynamics of scalar fluctuations during inflation is that $1\leq \alpha< 181$ or equivalently $1/19<c_s\leq1$. It would be interesting to extend this analysis to other theories such as the galileid and the extended galileids.\\

There are many avenues for future research directions and below we summarise what we think are the most interesting ones:
\begin{itemize}
\item \emph{Including gravity:} are there theories where both the phonon (inflaton) and the graviton transform in non-trivial ways under some symmetry group such that even at finite $M_{\text{pl}}$ and away from the decoupling limit symmetries remain exact? We have seen in this paper that unitarity, in the form of a two-derivative kinetic term for the phonon, constrains the possible symmetry groups in powerful ways. Presumably demanding this for the graviton too will be very constraining. For any of these such putative symmetry breaking patterns one could construct a generalised set-up for inflationary EFTs where the goldstone mode of the broken shift symmetry has different transformations than the usual ones.
\item \emph{Non-linear symmetries of cosmological correlators:} can we extend this analysis to symmetries of correlators in, say, de Sitter space? The space of linearly realised symmetries is very constrained \cite{EnricoDan} in much the same way the linear symmetries of Poincar\'{e} invariant amplitudes are constrained by the Coleman-Mandula theorem \cite{ColemanMandula}. It will also be interesting to investigate symmetries of amplitudes and correlators when the de Sitter boosts are weakly broken as in inflation. 
\item \emph{Other dispersion relations:} in this paper we have assumed that the dispersion relation of the phonon is of the usual linear form. However, there are interesting theories that fall outside of this class. Most notably there is the ghost condensate \cite{ghost} and it would be interesting if interactions there could also be constrained by additional symmetry. 
\item \emph{More degrees of freedom:} in this work we have assumed a single essential Goldstone mode. However, as explained in \cite{Nicolis:2015sra} there are other symmetry breaking patterns (that involve the breaking of boosts) leading to EFTs with more essential Goldstones. A well-known example is that of solids \cite{SolidInflation} which involves three scalars combined into an internal $SO(3)$ multiplet. Our analysis here using inverse Higgs trees could be easily be extended to cases with additional essential Goldstones and therefore additional generators at level-$0$ in the trees.  
\item \emph{(Linear) symmetries of the S-matrix when boosts are broken:} we have discovered that the special galileon, when expanded around a particular background solution, can develop an additional linearly realised symmetry. It would be interesting to explore the space of allowed linearly realised symmetries of scattering amplitudes when boosts are spontaneously broken. Such an analysis has recently been performed for cosmological correlators in \cite{EnricoDan}.
\end{itemize}

\section*{Acknowledgements}
We are grateful to Henry Fraser Goodhew, Mehrdad Mirbabayi, Scott Melville, Enrico Pajer, Riccardo Penco, Diederik Roest and Jakub Supe\l \;for useful discussions. In particular we thank Diederik Roest and Enrico Pajer for comments on a draft, and thank Enrico Pajer for sharing with us an unpublished manuscript on cosmological correlators. The Mathematica package xAct \cite{xact} was used extensively in this work, T.G. is grateful to Garrett Goon for sharing some of his notebooks. T.G. is supported by the Cambridge Trust and thanks IDB University for their hospitality during the completion of this work. S.J. and D.S. are supported in part by the research program VIDI with Project No. 680-47-535, which is (partly) financed by the Netherlands
Organisation for Scientific Research (NWO).




\appendix
\section{Special Galileon Broken Phase} 
\label{sec:special_galileon_broken_phase}
In this appendix we provide more details on the special galileon expanded around the background \eqref{eq:galileon_background} with $\beta=-1/\alpha$. We show that this background does break boosts spontaneously while preserving some form of space and time translations. However it falls out of our classification for an interesting reason: it has an additional linearly realised symmetry that combines with the non-linear symmetries to form a Lorentz covariant generator in the Lorentz invariant phase. \\

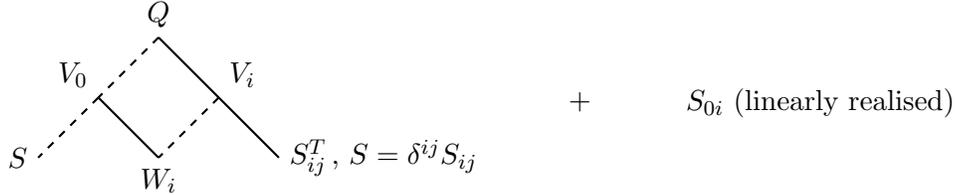
\begin{figure}[h!]
	\begin{center}
		\begin{tikzpicture}[x=0.4cm,y=0.4cm]
\tikzstyle{Pi} = [black, thick];
\tikzstyle{P0} = [black, dashed, thick];
\tikzstyle{label} = [->];
\draw (4,4) node [anchor = south, black]{$Q$};
\draw[style=Pi] (4,4) to (6,2)node [anchor = south west, black]{$V_i$};
\draw[style=P0] (4,4) to (2,2)node [anchor = south east, black]{$V_0$};
\draw[style=Pi] (2,2) to (4,0)node [anchor = north, black]{$W_i$};;
\draw[style=P0] (6,2) to (4,0);
\draw[style=Pi] (6,2) to (8,0)node [anchor = west, black]{$S_{ij}^T\,,\, S=\delta^{ij}S_{ij}$};
\draw[style=P0] (2,2) to (0,0)node [anchor = east, black]{$S$};
\draw (26,1) node [anchor = south, black]{$S_{0i}$ (linearly realised)};
\draw (18,1.2) node [anchor = south, black]{+};
\end{tikzpicture}
\caption{Inverse Higgs tree for the special galileon broken phase. This theory is not included in our classification because there is an additional linearly realised generator $S_{0i}$.}
\label{fig:spe_gal_tree}
	\end{center}
\end{figure}

We have seen in Section \ref{sub:obstacles_for_excp_EFTs} that when $\beta=-1/\alpha$ the vector part of the special galileon symmetry:
\begin{equation}
	\delta_{S_{0i}}\phi=tx^i - \dot\phi \partial^i\phi\,,
\end{equation}
leaves the background
\begin{equation}
	\expval{\phi}=-\frac{1}{2 \alpha}\left(\bfx^2+ \alpha^2 t^2\right)
\end{equation}
exactly invariant. Hence it cannot compensate for the boosts transformation and thus boosts are indeed spontaneously broken. Now the phonon transforms \emph{linearly} under the symmetry generated by $S_{0i}$. Indeed we have\footnote{Note that this transformation contains a piece that is quadratic in $\pi$. As we mentioned in the introduction, we call all symmetries that are a symmetry of the vacuum linearly realised and those that are broken by the vacuum non-linearly realised. So even though this symmetry is different to the usual linearly realised symmetries e.g. a translation or a Lorentz transformation that contain only a linear term in $\pi$, it is still on the same footing and much different to those with a field-independent term which are indeed broken by the vacuum.}
\begin{equation}
	\label{eq:spe_gal_linear_symmetry}
	\delta_{S_{0i}}\pi=\alpha t \partial^i\pi+\frac{1}{\alpha}x^i\dot\pi-\dot\pi \partial^i\pi\,.
\end{equation}
The remaining $SO(3)$ generators that form the covariant $S_{\mu \nu}$ are still non-linearly realised as
\begin{align}
	\delta_{S}\pi&= (1+\alpha^2)t^2+ \frac{4+\alpha^2}{3 \alpha^2}x_\mu x^\mu - 2 \alpha t \dot\pi -\frac{4}{3 \alpha}x^i\partial_i\pi+\dot\pi^2+\frac{1}{3}\partial^i\pi \partial_i\pi\,,\\
	\delta_{S_{ij}}\pi&=x^ix^j \left(1+\frac{4}{\alpha^2}\right)+2 \beta x^{(i}\partial^{j)}\pi+ \partial^i\pi \partial^j\pi\,,
\end{align}
where recall that $S\equiv S_{00}= \delta^{ij}S_{ij}$. As expected, these two symmetry generators do satisfy the required inverse Higgs constraints such that they can be realised by a single phonon.  We have
\begin{align}
	\comm{P_0}{S}&=2 \alpha P_0 + 2 (1+ \alpha^2)V_0\,, &&\comm{P_i}{S}=\frac{4}{3 \alpha}P_i + \frac{4+ \alpha^2}{3 \alpha^2}V_i\,,\\
	\comm{P_0}{S_{ij}}&=0\,, && \comm{P_i}{S_{jk}}=\left(1+ \frac{4}{\alpha^2}\right)(\delta_{i(j}V_{k)})-\frac{2}{\alpha}\delta_{i(j}P_{k)}\,,\\
	\comm{P_0}{S_{0i}}&=- \alpha P_i\,, && \comm{P_i}{S_{0j}}=-\frac{1}{\alpha}\delta_{ij}P_0\,.
\end{align}
Hence this theory would correspond to a level-2 inverse Higgs tree  but with an additional linearly realised generator as shown in Figure \ref{fig:spe_gal_tree}.
By making the change of basis 
\begin{align}
	\bar P_0&= P_0 + \alpha V_0\,,\\
	\bar P_i&= P_i + \frac{1}{\alpha}V_i\,,
\end{align}
we also verify that space and time translations are indeed linearly realised.\\

Now it is interesting to take a look at this phonon theory to see if IR instabilities can be avoided. First we solve the background equation of motion of the special galileon to constrain $\alpha$ in the background \eqref{eq:galileon_background}. We find the following four solutions
\begin{align}
	\alpha_{++}&=1+\sqrt{2}\,, && \alpha_{- -}=-1-\sqrt{2}\,,\\
	\alpha_{+-}&=1-\sqrt{2}\,, && \alpha_{-+}=-1+\sqrt{2}\,.
\end{align}
The quadratic action for $\pi$ is then
\begin{equation}
	S^{(2)}_\pi=\int \d^3x\d t \left(\frac{3}{\alpha^2}-1\right)\left(\dot\pi^2+c_s^2 (\partial_i\pi)^2\right)\,,
\end{equation}
where the speed of sound is
\begin{equation}
	c_s^2=3+ \frac{8}{\alpha^2-3}\,.
\end{equation}
For $\alpha_{++}$ and $\alpha_{- -}$, the kinetic term has the wrong sign and the theory is sick, while for $\alpha_{-+}$ and $\alpha_{+-}$ the sign is positive and the speed of sound is $c_s^2=3-2\sqrt{2}\simeq 0.17$. Therefore, at least at the basic level of the quadratic action, there a solution of the theory that has a healthy boosts breaking phase.

\bibliographystyle{JHEP}
\bibliography{papers}

\providecommand{\href}[2]{#2}\begingroup\raggedright\begin{thebibliography}{10}

\bibitem{ElvangHuang}
H.~Elvang and Y.-t. Huang, \emph{{Scattering Amplitudes}},
  \href{https://arxiv.org/abs/1308.1697}{{\ttfamily 1308.1697}}.

\bibitem{WeinbergBook1}
S.~Weinberg, \emph{{The Quantum theory of fields. Vol. 1: Foundations}}.
\newblock Cambridge University Press, 2005.

\bibitem{WeinbergBook2}
S.~Weinberg, \emph{{The quantum theory of fields. Vol. 2: Modern
  applications}}.
\newblock Cambridge University Press, 2013.

\bibitem{BenincasaCachazo}
P.~Benincasa and F.~Cachazo, \emph{{Consistency Conditions on the S-Matrix of
  Massless Particles}},  \href{https://arxiv.org/abs/0705.4305}{{\ttfamily
  0705.4305}}.

\bibitem{McGadyRodina}
D.~A. McGady and L.~Rodina, \emph{{Higher-spin massless $S$-matrices in
  four-dimensions}},
  \href{http://dx.doi.org/10.1103/PhysRevD.90.084048}{\emph{Phys. Rev.}
  {\bfseries D90} (2014) 084048},
  [\href{https://arxiv.org/abs/1311.2938}{{\ttfamily 1311.2938}}].

\bibitem{BenincasaReview}
P.~Benincasa, \emph{{New structures in scattering amplitudes: a review}},
  \href{http://dx.doi.org/10.1142/S0217751X14300051}{\emph{Int. J. Mod. Phys.}
  {\bfseries A29} (2014) 1430005},
  [\href{https://arxiv.org/abs/1312.5583}{{\ttfamily 1312.5583}}].

\bibitem{CheungReview}
C.~Cheung, \emph{{TASI Lectures on Scattering Amplitudes}},  in
  \emph{{Proceedings, Theoretical Advanced Study Institute in Elementary
  Particle Physics : Anticipating the Next Discoveries in Particle Physics
  (TASI 2016): Boulder, CO, USA, June 6-July 1, 2016}}, pp.~571--623, 2018.
\newblock \href{https://arxiv.org/abs/1708.03872}{{\ttfamily 1708.03872}}.
\newblock \href{http://dx.doi.org/10.1142/9789813233348_0008}{DOI}.

\bibitem{ColemanMandula}
S.~R. Coleman and J.~Mandula, \emph{{All Possible Symmetries of the S Matrix}},
  \href{http://dx.doi.org/10.1103/PhysRev.159.1251}{\emph{Phys. Rev.}
  {\bfseries 159} (1967) 1251--1256}.

\bibitem{GeneralisedAdler1}
C.~Cheung, K.~Kampf, J.~Novotny and J.~Trnka, \emph{{Effective Field Theories
  from Soft Limits of Scattering Amplitudes}},
  \href{http://dx.doi.org/10.1103/PhysRevLett.114.221602}{\emph{Phys. Rev.
  Lett.} {\bfseries 114} (2015) 221602},
  [\href{https://arxiv.org/abs/1412.4095}{{\ttfamily 1412.4095}}].

\bibitem{GeneralisedAdler2}
C.~Cheung, K.~Kampf, J.~Novotny, C.-H. Shen and J.~Trnka, \emph{{On-Shell
  Recursion Relations for Effective Field Theories}},
  \href{http://dx.doi.org/10.1103/PhysRevLett.116.041601}{\emph{Phys. Rev.
  Lett.} {\bfseries 116} (2016) 041601},
  [\href{https://arxiv.org/abs/1509.03309}{{\ttfamily 1509.03309}}].

\bibitem{GeneralisedAdler3}
C.~Cheung, K.~Kampf, J.~Novotny, C.-H. Shen and J.~Trnka, \emph{{A Periodic
  Table of Effective Field Theories}},
  \href{http://dx.doi.org/10.1007/JHEP02(2017)020}{\emph{JHEP} {\bfseries 02}
  (2017) 020}, [\href{https://arxiv.org/abs/1611.03137}{{\ttfamily
  1611.03137}}].

\bibitem{Probing}
A.~Padilla, D.~Stefanyszyn and T.~Wilson, \emph{{Probing Scalar Effective Field
  Theories with the Soft Limits of Scattering Amplitudes}},
  \href{http://dx.doi.org/10.1007/JHEP04(2017)015}{\emph{JHEP} {\bfseries 04}
  (2017) 015}, [\href{https://arxiv.org/abs/1612.04283}{{\ttfamily
  1612.04283}}].

\bibitem{RSW1}
D.~Roest, D.~Stefanyszyn and P.~Werkman, \emph{{An Algebraic Classification of
  Exceptional EFTs}},
  \href{http://dx.doi.org/10.1007/JHEP08(2019)081}{\emph{JHEP} {\bfseries 08}
  (2019) 081}, [\href{https://arxiv.org/abs/1903.08222}{{\ttfamily
  1903.08222}}].

\bibitem{BraunerBogers}
M.~P. Bogers and T.~Brauner, \emph{{Lie-algebraic classification of effective
  theories with enhanced soft limits}},
  \href{http://dx.doi.org/10.1007/JHEP05(2018)076}{\emph{JHEP} {\bfseries 05}
  (2018) 076}, [\href{https://arxiv.org/abs/1803.05359}{{\ttfamily
  1803.05359}}].

\bibitem{Coset1}
S.~R. Coleman, J.~Wess and B.~Zumino, \emph{{Structure of phenomenological
  Lagrangians. 1.}},
  \href{http://dx.doi.org/10.1103/PhysRev.177.2239}{\emph{Phys. Rev.}
  {\bfseries 177} (1969) 2239--2247}.

\bibitem{Coset2}
C.~G. Callan, Jr., S.~R. Coleman, J.~Wess and B.~Zumino, \emph{{Structure of
  phenomenological Lagrangians. 2.}},
  \href{http://dx.doi.org/10.1103/PhysRev.177.2247}{\emph{Phys. Rev.}
  {\bfseries 177} (1969) 2247--2250}.

\bibitem{Coset3}
D.~V. Volkov, \emph{{Phenomenological Lagrangians}}, {\emph{Fiz. Elem. Chast.
  Atom. Yadra} {\bfseries 4} (1973) 3--41}.

\bibitem{InverseHiggs}
E.~A. Ivanov and V.~I. Ogievetsky, \emph{{The Inverse Higgs Phenomenon in
  Nonlinear Realizations}},
  \href{http://dx.doi.org/10.1007/BF01028947}{\emph{Teor. Mat. Fiz.} {\bfseries
  25} (1975) 164--177}.

\bibitem{SoftBootstrap}
H.~Elvang, M.~Hadjiantonis, C.~R.~T. Jones and S.~Paranjape, \emph{{Soft
  Bootstrap and Supersymmetry}},
  \href{http://dx.doi.org/10.1007/JHEP01(2019)195}{\emph{JHEP} {\bfseries 01}
  (2019) 195}, [\href{https://arxiv.org/abs/1806.06079}{{\ttfamily
  1806.06079}}].

\bibitem{Cheung:2007st}
C.~Cheung, P.~Creminelli, A.~L. Fitzpatrick, J.~Kaplan and L.~Senatore,
  \emph{{The Effective Field Theory of Inflation}},
  \href{http://dx.doi.org/10.1088/1126-6708/2008/03/014}{\emph{JHEP} {\bfseries
  03} (2008) 014}, [\href{https://arxiv.org/abs/0709.0293}{{\ttfamily
  0709.0293}}].

\bibitem{Boostless}
E.~Pajer, D.~Stefanyszyn and J.~Supel, \emph{{The Boostless Bootstrap, to
  appear}}.

\bibitem{Pajer:2018egx}
E.~Pajer and D.~Stefanyszyn, \emph{{Symmetric Superfluids}},
  \href{http://dx.doi.org/10.1007/JHEP06(2019)008}{\emph{JHEP} {\bfseries 06}
  (2019) 008}, [\href{https://arxiv.org/abs/1812.05133}{{\ttfamily
  1812.05133}}].

\bibitem{Grall:2019qof}
T.~Grall, S.~Jazayeri and E.~Pajer, \emph{{Symmetric Scalars}},
  \href{http://dx.doi.org/10.1088/1475-7516/2020/05/031}{\emph{JCAP} {\bfseries
  05} (2020) 031}, [\href{https://arxiv.org/abs/1909.04622}{{\ttfamily
  1909.04622}}].

\bibitem{Adler1}
S.~L. Adler, \emph{{Consistency conditions on the strong interactions implied
  by a partially conserved axial vector current}},
  \href{http://dx.doi.org/10.1103/PhysRev.137.B1022}{\emph{Phys. Rev.}
  {\bfseries 137} (1965) B1022--B1033}.

\bibitem{Adler2}
S.~L. Adler, \emph{{Consistency conditions on the strong interactions implied
  by a partially conserved axial-vector current. II}},
  \href{http://dx.doi.org/10.1103/PhysRev.139.B1638}{\emph{Phys. Rev.}
  {\bfseries 139} (1965) B1638--B1643}.

\bibitem{Hellerman:2015nra}
S.~Hellerman, D.~Orlando, S.~Reffert and M.~Watanabe, \emph{{On the CFT
  Operator Spectrum at Large Global Charge}},
  \href{http://dx.doi.org/10.1007/JHEP12(2015)071}{\emph{JHEP} {\bfseries 12}
  (2015) 071}, [\href{https://arxiv.org/abs/1505.01537}{{\ttfamily
  1505.01537}}].

\bibitem{galileon}
A.~Nicolis, R.~Rattazzi and E.~Trincherini, \emph{{The Galileon as a local
  modification of gravity}},
  \href{http://dx.doi.org/10.1103/PhysRevD.79.064036}{\emph{Phys.\ Rev.\ D}
  {\bfseries 79} (2009) 064036},
  [\href{https://arxiv.org/abs/0811.2197}{{\ttfamily 0811.2197}}].

\bibitem{SG}
K.~Hinterbichler and A.~Joyce, \emph{{Hidden symmetry of the Galileon}},
  \href{http://dx.doi.org/10.1103/PhysRevD.92.023503}{\emph{Phys. Rev.}
  {\bfseries D92} (2015) 023503},
  [\href{https://arxiv.org/abs/1501.07600}{{\ttfamily 1501.07600}}].

\bibitem{Arkani-Hamed:2018kmz}
N.~Arkani-Hamed, D.~Baumann, H.~Lee and G.~L. Pimentel, \emph{{The Cosmological
  Bootstrap: Inflationary Correlators from Symmetries and Singularities}},
  \href{http://dx.doi.org/10.1007/JHEP04(2020)105}{\emph{JHEP} {\bfseries 04}
  (2020) 105}, [\href{https://arxiv.org/abs/1811.00024}{{\ttfamily
  1811.00024}}].

\bibitem{Baumann:2011su}
D.~Baumann and D.~Green, \emph{{Equilateral Non-Gaussianity and New Physics on
  the Horizon}},
  \href{http://dx.doi.org/10.1088/1475-7516/2011/09/014}{\emph{JCAP} {\bfseries
  09} (2011) 014}, [\href{https://arxiv.org/abs/1102.5343}{{\ttfamily
  1102.5343}}].

\bibitem{Finelli:2017fml}
B.~Finelli, G.~Goon, E.~Pajer and L.~Santoni, \emph{{Soft Theorems For
  Shift-Symmetric Cosmologies}},
  \href{http://dx.doi.org/10.1103/PhysRevD.97.063531}{\emph{Phys. Rev. D}
  {\bfseries 97} (2018) 063531},
  [\href{https://arxiv.org/abs/1711.03737}{{\ttfamily 1711.03737}}].

\bibitem{Finelli:2018upr}
B.~Finelli, G.~Goon, E.~Pajer and L.~Santoni, \emph{{The Effective Theory of
  Shift-Symmetric Cosmologies}},
  \href{http://dx.doi.org/10.1088/1475-7516/2018/05/060}{\emph{JCAP} {\bfseries
  05} (2018) 060}, [\href{https://arxiv.org/abs/1802.01580}{{\ttfamily
  1802.01580}}].

\bibitem{Maldacena:2011nz}
J.~M. Maldacena and G.~L. Pimentel, \emph{{On graviton non-Gaussianities during
  inflation}}, \href{http://dx.doi.org/10.1007/JHEP09(2011)045}{\emph{JHEP}
  {\bfseries 09} (2011) 045},
  [\href{https://arxiv.org/abs/1104.2846}{{\ttfamily 1104.2846}}].

\bibitem{Raju:2012zr}
S.~Raju, \emph{{New Recursion Relations and a Flat Space Limit for AdS/CFT
  Correlators}},
  \href{http://dx.doi.org/10.1103/PhysRevD.85.126009}{\emph{Phys. Rev.}
  {\bfseries D85} (2012) 126009},
  [\href{https://arxiv.org/abs/1201.6449}{{\ttfamily 1201.6449}}].

\bibitem{Arkani-Hamed:2015bza}
N.~Arkani-Hamed and J.~Maldacena, \emph{{Cosmological Collider Physics}},
  \href{https://arxiv.org/abs/1503.08043}{{\ttfamily 1503.08043}}.

\bibitem{Fraser-Goodhew}
H.~Fraser~Goodhew, S.~Jazayeri and E.~Pajer, \emph{in preparation}, .

\bibitem{Enrico}
E.~Pajer, \emph{{Unpublished Manuscript}}.

\bibitem{EnricoDan}
D.~Green and E.~Pajer, \emph{{On the Symmetries of Cosmological
  Perturbations}},  \href{https://arxiv.org/abs/2004.09587}{{\ttfamily
  2004.09587}}.

\bibitem{KRS}
R.~Klein, D.~Roest and D.~Stefanyszyn, \emph{{Spontaneously Broken Spacetime
  Symmetries and the Role of Inessential Goldstones}},
  \href{http://dx.doi.org/10.1007/JHEP10(2017)051}{\emph{JHEP} {\bfseries 10}
  (2017) 051}, [\href{https://arxiv.org/abs/1709.03525}{{\ttfamily
  1709.03525}}].

\bibitem{WessZuminoGal}
G.~Goon, K.~Hinterbichler, A.~Joyce and M.~Trodden, \emph{{Galileons as
  Wess-Zumino Terms}},
  \href{http://dx.doi.org/10.1007/JHEP06(2012)004}{\emph{JHEP} {\bfseries 06}
  (2012) 004}, [\href{https://arxiv.org/abs/1203.3191}{{\ttfamily 1203.3191}}].

\bibitem{Wheel}
L.~V. Delacrétaz, S.~Endlich, A.~Monin, R.~Penco and F.~Riva,
  \emph{{(Re-)Inventing the Relativistic Wheel: Gravity, Cosets, and Spinning
  Objects}}, \href{http://dx.doi.org/10.1007/JHEP11(2014)008}{\emph{JHEP}
  {\bfseries 11} (2014) 008},
  [\href{https://arxiv.org/abs/1405.7384}{{\ttfamily 1405.7384}}].

\bibitem{McArthur}
I.~N. McArthur, \emph{{Nonlinear realizations of symmetries and unphysical
  Goldstone bosons}},
  \href{http://dx.doi.org/10.1007/JHEP11(2010)140}{\emph{JHEP} {\bfseries 11}
  (2010) 140}, [\href{https://arxiv.org/abs/1009.3696}{{\ttfamily 1009.3696}}].

\bibitem{RSW2}
D.~Roest, D.~Stefanyszyn and P.~Werkman, \emph{{An Algebraic Classification of
  Exceptional EFTs Part II: Supersymmetry}},
  \href{http://dx.doi.org/10.1007/JHEP11(2019)077}{\emph{JHEP} {\bfseries 11}
  (2019) 077}, [\href{https://arxiv.org/abs/1905.05872}{{\ttfamily
  1905.05872}}].

\bibitem{ExtendedShifts}
K.~Hinterbichler and A.~Joyce, \emph{{Goldstones with Extended Shift
  Symmetries}}, \href{http://dx.doi.org/10.1142/S0218271814430019}{\emph{Int.
  J. Mod. Phys.} {\bfseries D23} (2014) 1443001},
  [\href{https://arxiv.org/abs/1404.4047}{{\ttfamily 1404.4047}}].

\bibitem{UVconstraints}
S.~Melville, D.~Roest and D.~Stefanyszyn, \emph{{UV Constraints on Massive
  Spinning Particles: Lessons from the Gravitino}},
  \href{http://dx.doi.org/10.1007/JHEP02(2020)185}{\emph{JHEP} {\bfseries 02}
  (2020) 185}, [\href{https://arxiv.org/abs/1911.03126}{{\ttfamily
  1911.03126}}].

\bibitem{BrandoHigherSpin}
B.~Bellazzini, F.~Riva, J.~Serra and F.~Sgarlata, \emph{{Massive Higher Spins:
  Effective Theory and Consistency}},
  \href{http://dx.doi.org/10.1007/JHEP10(2019)189}{\emph{JHEP} {\bfseries 10}
  (2019) 189}, [\href{https://arxiv.org/abs/1903.08664}{{\ttfamily
  1903.08664}}].

\bibitem{RoestSG}
D.~Roest, \emph{{The Special Galileon as Goldstone of Diffeomorphisms}},
  \href{https://arxiv.org/abs/2004.09559}{{\ttfamily 2004.09559}}.

\bibitem{Aspects}
K.~Hinterbichler, \emph{{Theoretical Aspects of Massive Gravity}},
  \href{http://dx.doi.org/10.1103/RevModPhys.84.671}{\emph{Rev. Mod. Phys.}
  {\bfseries 84} (2012) 671--710},
  [\href{https://arxiv.org/abs/1105.3735}{{\ttfamily 1105.3735}}].

\bibitem{PorratiRahman}
M.~Porrati and R.~Rahman, \emph{{A Model Independent Ultraviolet Cutoff for
  Theories with Charged Massive Higher Spin Fields}},
  \href{http://dx.doi.org/10.1016/j.nuclphysb.2009.02.010}{\emph{Nucl. Phys.}
  {\bfseries B814} (2009) 370--404},
  [\href{https://arxiv.org/abs/0812.4254}{{\ttfamily 0812.4254}}].

\bibitem{dSAdS}
J.~Bonifacio, K.~Hinterbichler, A.~Joyce and R.~A. Rosen, \emph{{Shift
  Symmetries in (Anti) de Sitter Space}},
  \href{http://dx.doi.org/10.1007/JHEP02(2019)178}{\emph{JHEP} {\bfseries 02}
  (2019) 178}, [\href{https://arxiv.org/abs/1812.08167}{{\ttfamily
  1812.08167}}].

\bibitem{Nicolis:2011pv}
A.~Nicolis and F.~Piazza, \emph{{Spontaneous Symmetry Probing}},
  \href{http://dx.doi.org/10.1007/JHEP06(2012)025}{\emph{JHEP} {\bfseries 06}
  (2012) 025}, [\href{https://arxiv.org/abs/1112.5174}{{\ttfamily 1112.5174}}].

\bibitem{Nicolis:2015sra}
A.~Nicolis, R.~Penco, F.~Piazza and R.~Rattazzi, \emph{{Zoology of condensed
  matter: Framids, ordinary stuff, extra-ordinary stuff}},
  \href{http://dx.doi.org/10.1007/JHEP06(2015)155}{\emph{JHEP} {\bfseries 06}
  (2015) 155}, [\href{https://arxiv.org/abs/1501.03845}{{\ttfamily
  1501.03845}}].

\bibitem{ghost}
N.~Arkani-Hamed, H.-C. Cheng, M.~A. Luty and S.~Mukohyama, \emph{{Ghost
  condensation and a consistent infrared modification of gravity}},
  \href{http://dx.doi.org/10.1088/1126-6708/2004/05/074}{\emph{JHEP} {\bfseries
  05} (2004) 074}, [\href{https://arxiv.org/abs/hep-th/0312099}{{\ttfamily
  hep-th/0312099}}].

\bibitem{Son:2002zn}
D.~T. Son, \emph{{Low-energy quantum effective action for relativistic
  superfluids}},  \href{https://arxiv.org/abs/hep-ph/0204199}{{\ttfamily
  hep-ph/0204199}}.

\bibitem{Nicolis:2013lma}
A.~Nicolis, R.~Penco and R.~A. Rosen, \emph{{Relativistic Fluids, Superfluids,
  Solids and Supersolids from a Coset Construction}},
  \href{http://dx.doi.org/10.1103/PhysRevD.89.045002}{\emph{Phys. Rev.}
  {\bfseries D89} (2014) 045002},
  [\href{https://arxiv.org/abs/1307.0517}{{\ttfamily 1307.0517}}].

\bibitem{Nicolis:2011cs}
A.~Nicolis, \emph{{Low-energy effective field theory for finite-temperature
  relativistic superfluids}},
  \href{https://arxiv.org/abs/1108.2513}{{\ttfamily 1108.2513}}.

\bibitem{Baumann:2015nta}
D.~Baumann, D.~Green, H.~Lee and R.~A. Porto, \emph{{Signs of Analyticity in
  Single-Field Inflation}},
  \href{http://dx.doi.org/10.1103/PhysRevD.93.023523}{\emph{Phys.\ Rev.\ D}
  {\bfseries 93} (2016) 023523},
  [\href{https://arxiv.org/abs/1502.07304}{{\ttfamily 1502.07304}}].

\bibitem{TG_Melville_inprep_positivity}
T.~Grall and S.~Melville, \emph{{Positivity Bounds with Spontaneously Broken
  Boosts}}.
\newblock In Preparation.

\bibitem{Goon:2012dy}
G.~Goon, K.~Hinterbichler, A.~Joyce and M.~Trodden, \emph{{Galileons as
  Wess-Zumino Terms}},
  \href{http://dx.doi.org/10.1007/JHEP06(2012)004}{\emph{JHEP} {\bfseries 06}
  (2012) 004}, [\href{https://arxiv.org/abs/1203.3191}{{\ttfamily 1203.3191}}].

\bibitem{ClassicalAndQuantum}
A.~Nicolis and R.~Rattazzi, \emph{{Classical and quantum consistency of the DGP
  model}}, \href{http://dx.doi.org/10.1088/1126-6708/2004/06/059}{\emph{JHEP}
  {\bfseries 06} (2004) 059},
  [\href{https://arxiv.org/abs/hep-th/0404159}{{\ttfamily hep-th/0404159}}].

\bibitem{GoonAspects}
G.~Goon, K.~Hinterbichler, A.~Joyce and M.~Trodden, \emph{{Aspects of Galileon
  Non-Renormalization}},
  \href{http://dx.doi.org/10.1007/JHEP11(2016)100}{\emph{JHEP} {\bfseries 11}
  (2016) 100}, [\href{https://arxiv.org/abs/1606.02295}{{\ttfamily
  1606.02295}}].

\bibitem{galileonReview}
C.~Deffayet and D.~A. Steer, \emph{{A formal introduction to Horndeski and
  Galileon theories and their generalizations}},
  \href{http://dx.doi.org/10.1088/0264-9381/30/21/214006}{\emph{Class. Quant.
  Grav.} {\bfseries 30} (2013) 214006},
  [\href{https://arxiv.org/abs/1307.2450}{{\ttfamily 1307.2450}}].

\bibitem{Monin:2016jmo}
A.~Monin, D.~Pirtskhalava, R.~Rattazzi and F.~K. Seibold, \emph{{Semiclassics,
  Goldstone Bosons and CFT data}},
  \href{http://dx.doi.org/10.1007/JHEP06(2017)011}{\emph{JHEP} {\bfseries 06}
  (2017) 011}, [\href{https://arxiv.org/abs/1611.02912}{{\ttfamily
  1611.02912}}].

\bibitem{Cuomo:2017vzg}
G.~Cuomo, A.~de~la Fuente, A.~Monin, D.~Pirtskhalava and R.~Rattazzi,
  \emph{{Rotating superfluids and spinning charged operators in conformal field
  theory}}, \href{http://dx.doi.org/10.1103/PhysRevD.97.045012}{\emph{Phys.\
  Rev.\ D} {\bfseries 97} (2018) 045012},
  [\href{https://arxiv.org/abs/1711.02108}{{\ttfamily 1711.02108}}].

\bibitem{Esposito:2016ria}
A.~Esposito, S.~Garcia-Saenz and R.~Penco, \emph{{First sound in holographic
  superfluids at zero temperature}},
  \href{http://dx.doi.org/10.1007/JHEP12(2016)136}{\emph{JHEP} {\bfseries 12}
  (2016) 136}, [\href{https://arxiv.org/abs/1606.03104}{{\ttfamily
  1606.03104}}].

\bibitem{GalDual1}
C.~de~Rham, M.~Fasiello and A.~J. Tolley, \emph{{Galileon Duality}},
  \href{http://dx.doi.org/10.1016/j.physletb.2014.03.061}{\emph{Phys. Lett. B}
  {\bfseries 733} (2014) 46--51},
  [\href{https://arxiv.org/abs/1308.2702}{{\ttfamily 1308.2702}}].

\bibitem{GalDual2}
K.~Kampf and J.~r. Novotny, \emph{{Unification of Galileon Dualities}},
  \href{http://dx.doi.org/10.1007/JHEP10(2014)006}{\emph{JHEP} {\bfseries 10}
  (2014) 006}, [\href{https://arxiv.org/abs/1403.6813}{{\ttfamily 1403.6813}}].

\bibitem{GalInflation1}
C.~Burrage, C.~de~Rham, D.~Seery and A.~J. Tolley, \emph{{Galileon inflation}},
  \href{http://dx.doi.org/10.1088/1475-7516/2011/01/014}{\emph{JCAP} {\bfseries
  1101} (2011) 014}, [\href{https://arxiv.org/abs/1009.2497}{{\ttfamily
  1009.2497}}].

\bibitem{GalInflation2}
T.~Kobayashi, M.~Yamaguchi and J.~Yokoyama, \emph{{G-inflation: Inflation
  driven by the Galileon field}},
  \href{http://dx.doi.org/10.1103/PhysRevLett.105.231302}{\emph{Phys. Rev.
  Lett.} {\bfseries 105} (2010) 231302},
  [\href{https://arxiv.org/abs/1008.0603}{{\ttfamily 1008.0603}}].

\bibitem{WeaklyBroken}
D.~Pirtskhalava, L.~Santoni, E.~Trincherini and F.~Vernizzi, \emph{{Weakly
  Broken Galileon Symmetry}},
  \href{http://dx.doi.org/10.1088/1475-7516/2015/09/007}{\emph{JCAP} {\bfseries
  1509} (2015) 007}, [\href{https://arxiv.org/abs/1505.00007}{{\ttfamily
  1505.00007}}].

\bibitem{deRham:2010eu}
C.~de~Rham and A.~J. Tolley, \emph{{DBI and the Galileon reunited}},
  \href{http://dx.doi.org/10.1088/1475-7516/2010/05/015}{\emph{JCAP} {\bfseries
  1005} (2010) 015}, [\href{https://arxiv.org/abs/1003.5917}{{\ttfamily
  1003.5917}}].

\bibitem{Alishahiha:2004eh}
M.~Alishahiha, E.~Silverstein and D.~Tong, \emph{{DBI in the sky}},
  \href{http://dx.doi.org/10.1103/PhysRevD.70.123505}{\emph{Phys. Rev. D}
  {\bfseries 70} (2004) 123505},
  [\href{https://arxiv.org/abs/hep-th/0404084}{{\ttfamily hep-th/0404084}}].

\bibitem{Pajer:2008uy}
E.~Pajer, \emph{{Inflation at the Tip}},
  \href{http://dx.doi.org/10.1088/1475-7516/2008/04/031}{\emph{JCAP} {\bfseries
  04} (2008) 031}, [\href{https://arxiv.org/abs/0802.2916}{{\ttfamily
  0802.2916}}].

\bibitem{Creminelli:2013xfa}
P.~Creminelli, R.~Emami, M.~Simonovi? and G.~Trevisan, \emph{{ISO(4,1) Symmetry
  in the EFT of Inflation}},
  \href{http://dx.doi.org/10.1088/1475-7516/2013/07/037}{\emph{JCAP} {\bfseries
  1307} (2013) 037}, [\href{https://arxiv.org/abs/1304.4238}{{\ttfamily
  1304.4238}}].

\bibitem{Mirbabayi:2016xvc}
M.~Mirbabayi and M.~Simonovi\'c, \emph{{Weinberg Soft Theorems from Weinberg
  Adiabatic Modes}},  \href{https://arxiv.org/abs/1602.05196}{{\ttfamily
  1602.05196}}.

\bibitem{Cheung:2016drk}
C.~Cheung, K.~Kampf, J.~Novotny, C.-H. Shen and J.~Trnka, \emph{{A Periodic
  Table of Effective Field Theories}},
  \href{http://dx.doi.org/10.1007/JHEP02(2017)020}{\emph{JHEP} {\bfseries 02}
  (2017) 020}, [\href{https://arxiv.org/abs/1611.03137}{{\ttfamily
  1611.03137}}].

\bibitem{Weinberg:1964ew}
S.~Weinberg, \emph{{Photons and Gravitons in $S$-Matrix Theory: Derivation of
  Charge Conservation and Equality of Gravitational and Inertial Mass}},
  \href{http://dx.doi.org/10.1103/PhysRev.135.B1049}{\emph{Phys. Rev.}
  {\bfseries 135} (1964) B1049--B1056}.

\bibitem{Creminelli:2012ed}
P.~Creminelli, J.~Noreña and M.~Simonovi\'c, \emph{{Conformal consistency
  relations for single-field inflation}},
  \href{http://dx.doi.org/10.1088/1475-7516/2012/07/052}{\emph{JCAP} {\bfseries
  07} (2012) 052}, [\href{https://arxiv.org/abs/1203.4595}{{\ttfamily
  1203.4595}}].

\bibitem{Hinterbichler:2013dpa}
K.~Hinterbichler, L.~Hui and J.~Khoury, \emph{{An Infinite Set of Ward
  Identities for Adiabatic Modes in Cosmology}},
  \href{http://dx.doi.org/10.1088/1475-7516/2014/01/039}{\emph{JCAP} {\bfseries
  01} (2014) 039}, [\href{https://arxiv.org/abs/1304.5527}{{\ttfamily
  1304.5527}}].

\bibitem{Pajer:2019jhb}
S.~Jazayeri, E.~Pajer and D.~van~der Woude, \emph{{Solid Soft Theorems}},
  \href{http://dx.doi.org/10.1088/1475-7516/2019/06/011}{\emph{JCAP} {\bfseries
  06} (2019) 011}, [\href{https://arxiv.org/abs/1902.09020}{{\ttfamily
  1902.09020}}].

\bibitem{Cuscuton}
N.~Afshordi, D.~J.~H. Chung and G.~Geshnizjani, \emph{{Cuscuton: A Causal Field
  Theory with an Infinite Speed of Sound}},
  \href{http://dx.doi.org/10.1103/PhysRevD.75.083513}{\emph{Phys. Rev.}
  {\bfseries D75} (2007) 083513},
  [\href{https://arxiv.org/abs/hep-th/0609150}{{\ttfamily hep-th/0609150}}].

\bibitem{Grall:2020tqc}
T.~Grall and S.~Melville, \emph{{Inflation in Motion: Unitarity Constraints in
  Effective Field Theories with Broken Lorentz Symmetry}},
  \href{https://arxiv.org/abs/2005.02366}{{\ttfamily 2005.02366}}.

\bibitem{Aghanim:2018eyx}
{\scshape Planck} collaboration, N.~Aghanim et~al., \emph{{Planck 2018 results.
  VI. Cosmological parameters}},
  \href{https://arxiv.org/abs/1807.06209}{{\ttfamily 1807.06209}}.

\bibitem{dRGT}
C.~de~Rham, G.~Gabadadze and A.~J. Tolley, \emph{{Resummation of Massive
  Gravity}},
  \href{http://dx.doi.org/10.1103/PhysRevLett.106.231101}{\emph{Phys. Rev.
  Lett.} {\bfseries 106} (2011) 231101},
  [\href{https://arxiv.org/abs/1011.1232}{{\ttfamily 1011.1232}}].

\bibitem{Genesis}
P.~Creminelli, A.~Nicolis and E.~Trincherini, \emph{{Galilean Genesis: An
  Alternative to inflation}},
  \href{http://dx.doi.org/10.1088/1475-7516/2010/11/021}{\emph{JCAP} {\bfseries
  11} (2010) 021}, [\href{https://arxiv.org/abs/1007.0027}{{\ttfamily
  1007.0027}}].

\bibitem{SolidInflation}
S.~Endlich, A.~Nicolis and J.~Wang, \emph{{Solid Inflation}},
  \href{http://dx.doi.org/10.1088/1475-7516/2013/10/011}{\emph{JCAP} {\bfseries
  10} (2013) 011}, [\href{https://arxiv.org/abs/1210.0569}{{\ttfamily
  1210.0569}}].

\bibitem{xact}
J.~Mart\'in-Garc\'ia, ``{xAct, Efficient tensor computer algebra for the
  Wolfram Language}.''

\end{thebibliography}\endgroup

\end{document}